\documentclass[twocolumn,superscriptaddress,aps,prb,longbibliography]{revtex4-2}
\usepackage[utf8]{inputenc}
\usepackage[T1]{fontenc}
\usepackage{newtxtext,newtxmath}
\usepackage{graphicx}
\usepackage{hyperref}
\usepackage{bm}
\usepackage{microtype}
\usepackage{physics}
\usepackage{xcolor}
\usepackage[caption=false]{subfig}

\begin{document}
\title{Factors affecting the topological Hall effect in strongly correlated layered magnets: spin of the magnetic atoms, polar and azimuthal angle subtended by the spin texture}
\author{Kaushal Kumar Kesharpu}
\email{kesharpu@theor.jinr.ru}
\affiliation{Bogoliubov Laboratory of Theoretical Physics, Joint Institute for Nuclear Research, Dubna 141980, Russia}

\date{\today}

\begin{abstract}

  The Hamiltonian of a two dimensional (2D) magnetic material in the strong correlation regime with a spin texture, for which both azimuthal and polar angle changes, is solved using $su(2)$ path integral method. The dependence of the Chern number on the atomic spin ($S$), azimuthal angle ($\vec{q}_{1}$) and polar angle ($\vec{q}_{2}$) modulation vector of the spin texture on a bipartite honeycomb lattice is found. For $S \leq 3$ it was found that Chern number depends strongly on $\vec{q}_{2}$ and $S$. We discuss applicability of the model to several van der Waals magnets. Experimentally, it is expected that, with increase in spin modulation vector the sign of the topological Hall conductivity changes, $+\sigma_{xy}^{THE} \to -\sigma_{xy}^{THE}$ or vice-versa, when $S$ is constant. We also propose several heterostrucures for experimental realization of this effect.
\end{abstract}

\maketitle
\section{Introduction}
Motion of electrons on an adiabatically changing chiral spin texture in the strong coupling regime gives rise to the topological Hall effect (THE) \cite{bruno_2004_TopologicalHall_PhysRevLett}. For the strong coupling case the spin of the electrons follows the local direction of the magnetization. If the magnetization varies in a closed loop, then the electrons acquire a geometric phase in the parameter space, which in turn gives rise to the THE \cite{zhang_2020_RealspaceBerry_PhysRevB,kimbell_2022_ChallengesIdentifying_CommunMater,wang_2022_TopologicalHall_ProgressInMaterialsScience}. A large number of spin textures --- e.g. skyrmions \cite{raju_2019_EvolutionSkyrmions_NatCommun,soumyanarayanan_2017_TunableRoomtemperature_NatureMater,shao_2019_TopologicalHall_NatElectron,wang_2018_FerroelectricallyTunable_NatureMater}, conical \cite{shiomi_2012_EmergenceTopological_PhysRevB,ghimire_2020_CompetingMagnetic_SciAdv,afshar_2021_SpinSpiral_PhysRevB}, hedgehog \cite{fujishiro_2019_TopologicalTransitions_NatCommun}, magnetic bubble \cite{vistoli_2019_GiantTopological_NaturePhys} to name a few \cite[for a complete list see Tab. 1 of Ref.][]{wang_2022_TopologicalHall_ProgressInMaterialsScience} --- generating THE had already been observed experimentally in 2D layered magnetic materials \cite{wang_2022_TopologicalHall_ProgressInMaterialsScience}. Microscopically, the Dzyloshinskii-Moriya interaction (DMI) \cite{wiesendanger_2016_NanoscaleMagnetic_NatRevMater,fert_2017_MagneticSkyrmions_NatRevMater}, the dipolar interaction\cite{ezawa_2010_GiantSkyrmions_PhysRevLett}, frustrated chirality \cite{batista_2016_FrustrationChiral_RepProgPhys,karube_2018_DisorderedSkyrmion_SciAdv}, out-of-plane anisotropy \cite{leonov_2020_StabilityInplane_PhysRevB,preissinger_2021_VitalRole_NpjQuantumMater} and Fermi surface curvatures \cite{ozawa_2017_ZeroFieldSkyrmions_PhysRevLett} are responsible for THE. Due to the involvement of these different microscopic effects understanding the competition between them \cite{komatsu_2021_PhaseDiagram_PhysRevB,li_2020_CompetingMagnetic_PhysRevLett,sun_2021_ManipulationMagnetic_AdvFunctMater,bernand-mantel_2020_UnravelingRole_PhysRevB} and ways to manipulate them is important from the point of view of the fundamental physics as well as applied spintronics \cite{sierra_2021_VanWaals_NatNanotechnol}. Recently, Van der Waals (vdW) magnets have emerged as one of the promising class of materials for investigation of these effects \cite{burch_2018_MagnetismTwodimensional_Nature}, due to the possibility of changing their properties through intrinsic means (chemical doping \cite{yang_2021_VanWaals_AdvSci}, stacking order and twist of the monolayers \cite{tong_2019_MagneticProximity_PhysRevAppl,burch_2018_MagnetismTwodimensional_Nature,jiang_2019_StackingTunable_PhysRevB,behura_2021_MoirePhysics_EmergentMater,tran_2020_MoireTransition_2DMater}, free carrier doping) as well as extrinsic means (electric field, magnetic field, strain, pressure
\cite{mak_2019_ProbingControlling_NatRevPhys}).

\begin{figure}
  \centering
  \includegraphics[width=0.35\textwidth]{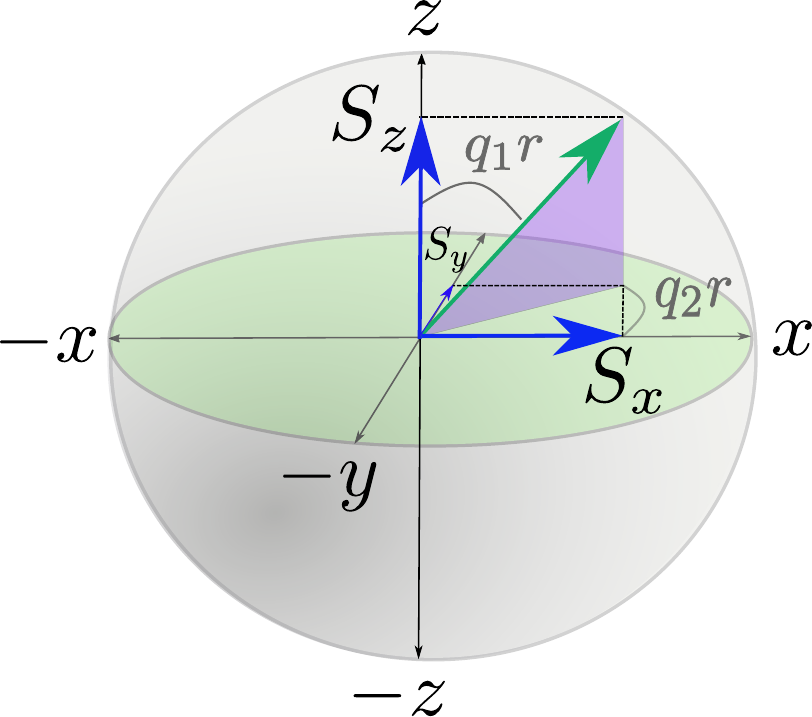}
  \caption{Representation of spin texture given in Eq. (\ref{eq:spin-config-int}) on a Bloch sphere. $S_{x}$, $S_{y}$, and $S_{z}$ represent the $x$, $y$, and $z$ component of the spin. $\vec{q}_{1}$ is the polar angle modulation vector and $\vec{q}_{2}$ is the azimuthal angle modulation vector.}
  \label{fig:spin-sphere}
\end{figure}
\begin{figure}
  \centering
  \subfloat[][Square]{\includegraphics[width=0.25\textwidth]{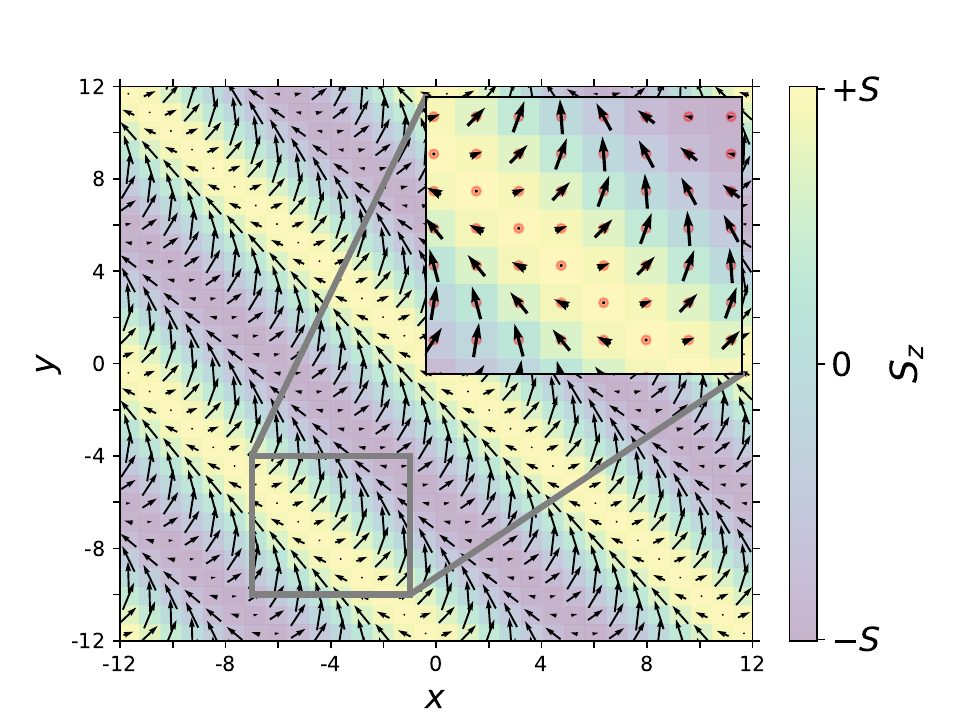}\label{fig:spin-struct-square}}
  \subfloat[][Triangular]{\includegraphics[width=0.25\textwidth]{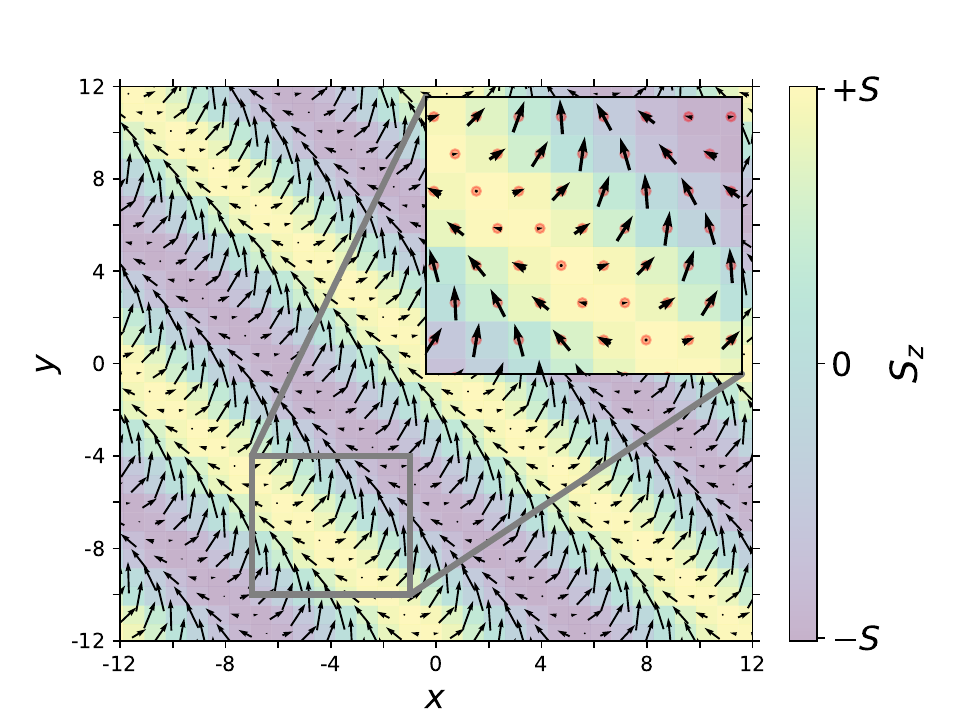}\label{fig:spin-struct-triangle}}\\
  \subfloat[][Honeycomb]{\includegraphics[width=0.25\textwidth]{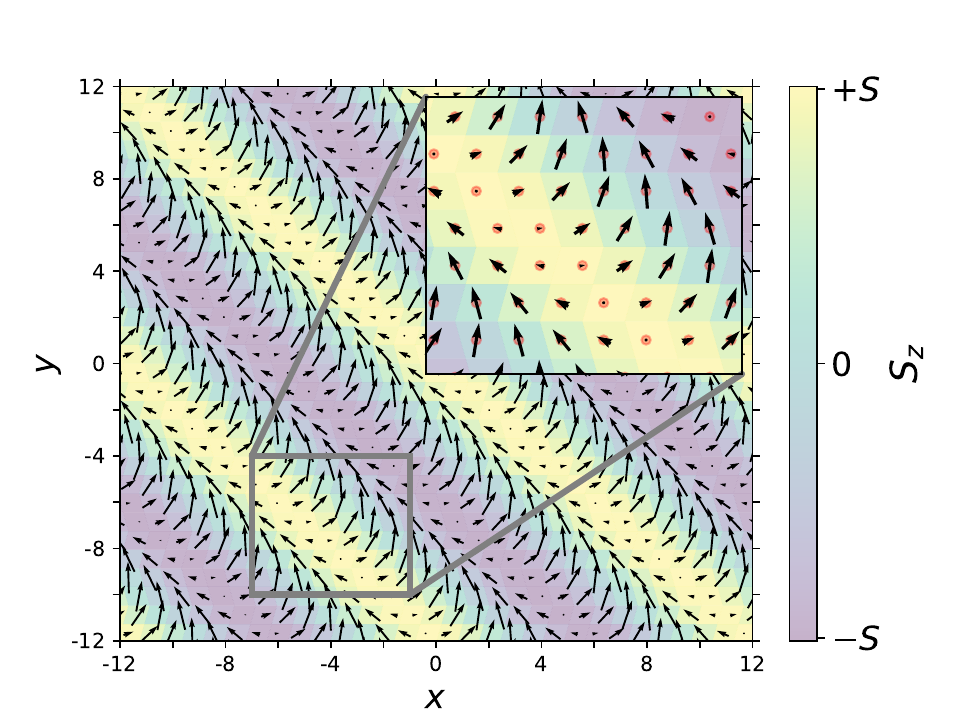}\label{fig:spin-struct-honeycomb}}
  \subfloat[][Kagome]{\includegraphics[width=0.25\textwidth]{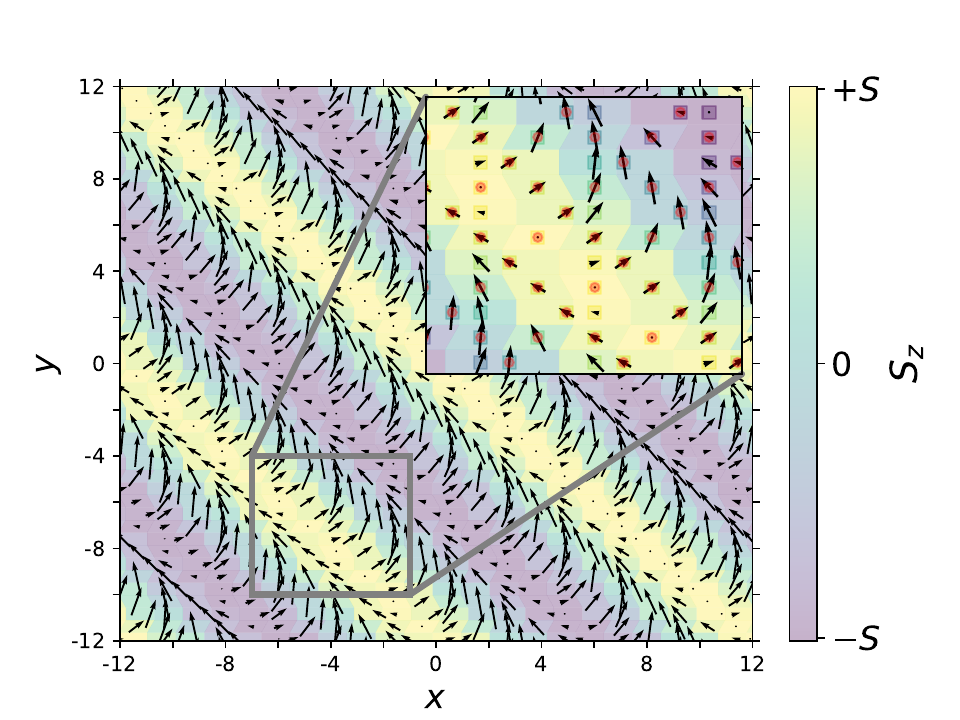}\label{fig:spin-struct-kagome}}\\
  \caption{Spin texture generated from Eq. (\ref{eq:spin-config-int}) for $\vec{q}_{1}= \vec{q}_{2} =\left( \pi/6, \pi/6 \right)$ on (a) square, (b) triangular, (c) honeycomb, and (d) Kagome lattice. The direction of the arrow represents the angle extended by $S_{x}$ and $S_{y}$ on $xy$ plane, defined as: $\arctan \left( S_{y}/S_{x} \right)$. The color represents the values of the $S_{z}$.}
  \label{fig:spin-struct}
\end{figure}

The vdW magnets are primarily divided into five different family of compounds \cite{yang_2021_VanWaals_AdvSci,wang_2022_MagneticGenome_ACSNano}: (i) transition metal halides, (ii) transition metal phosphorous tri-chalcogenides, (iii) transition metal di-chalcogenides, (iv) ternary iron based tellurides, (v) transition metal oxyhalides. In these compounds itinerant ferromagnetism, Kondo lattice bheaviour, and Mott insulating phase \cite{wang_2022_MagneticGenome_ACSNano,kurebayashi_2022_MagnetismSymmetry_NatRevPhys,zhang_2023_MottnessTwodimensional_PhysRevB,corasaniti_2020_ElectronicCorrelations_PhysRevB,sarkar_2020_ElectronicCorrelation_PhysRevMater,zhu_2016_ElectronicCorrelation_PhysRevB,ghosh_2023_UnravelingEffects_NpjComputMater} are observed due to the strong electronic correlation. Keeping this in mind, to investigate the electronic properties of these 2D materials with localized spin moments a low-energy effective theory using $su(2)$ path integral method was proposed recently \cite{kesharpu_2023_TopologicalHall_PhysRevB}. It was shown that the THE for a material with honeycomb bipartite lattice and strongly coupled electron spin to the background high spin-$S$ conical spin texture --- here $S$ is the effective magnetic moment of the magnetic atoms --- depends only on the: (i) atomic spin $S$ , (ii) and the spin modulation vector of the spin texture. In the investigated conical spin texture only the azimuthal angle of the spin projection (on \emph{xy} plane) changed through neighboring sites. Hence naturally the question arises, how the topological properties of the materials changes if both the azimuthal angle as well as the polar angle of the spin texture changes? To answer this question in this work we analyzed the spin texture: 
\begin{equation}
  \label{eq:spin-config-int}
  \vec{S}_{i} \equiv
  \begin{bmatrix}
    S_{x}\\
    S_{y}\\
    S_{z}\\
  \end{bmatrix}
  = S
  \begin{bmatrix}
    \sin \vec{q}_{1} \vec{r}_{i} \quad \cos \vec{q}_{2}\vec{r}_{i}\\
    \sin \vec{q}_{1} \vec{r}_{i} \quad \sin \vec{q}_{2}\vec{r}_{i}\\
    \cos \vec{q}_{1}\vec{r}_{i}
  \end{bmatrix}
\end{equation}
Here $S_{x}$, $S_{y}$, and $S_{z}$ are the $x$, $y$, and $z$ component of the localized spin momentum $\vec{S}_{i}$. $\vec{q}_{1}=\left( q_{1x}, q_{1y} \right)$ and $\vec{q}_{2}=\left( q_{2x}, q_{2y} \right)$ are the two spin modulating wave vectors on a 2D plane. $\vec{r}_{i}=\left( x,y \right)$ is the position vector on the 2D plane. If $\vec{S}_{i}$ is represented on the Bloch sphere, where north pole corresponds to the $S_{z} = \ket{+S}$ state and south pole corresponds to the $S_{z} = \ket{-S}$ state, then $\vec{q}_{1}$ is the polar angle modulating vector and $\vec{q}_{2}$ is the azimuthal angle modulating vectors as shown in Fig. \ref{fig:spin-sphere}. $S=1/2,1,3/2,\dots$ is the highest spin state of the background magnetic atoms. The spin texture on a 2D square, triangular, honeycomb, and kagome lattice is shown in \ref{fig:spin-struct}. Here the $S_{z}$ component is shown through color, and the $S_{x}$, $S_{y}$ components are shown through direction of the arrows defined as $\arctan \left( S_{y}/S_{x} \right)$.

\begin{figure}
  \centering
  \subfloat[][Fe$_{3}$GeTe$_{2}$]{\includegraphics[width=0.45\textwidth]{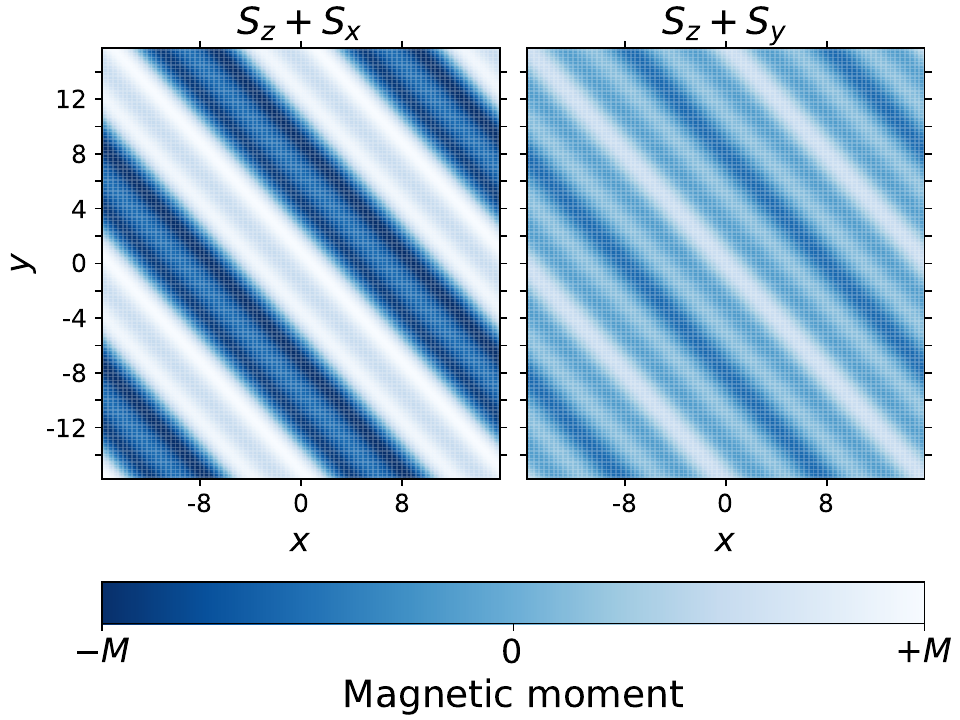}\label{fig:meijer_compound}}\\
  \subfloat[][Cr$_{2}$Ge$_{2}$Te$_{6}$]{\includegraphics[width=0.45\textwidth]{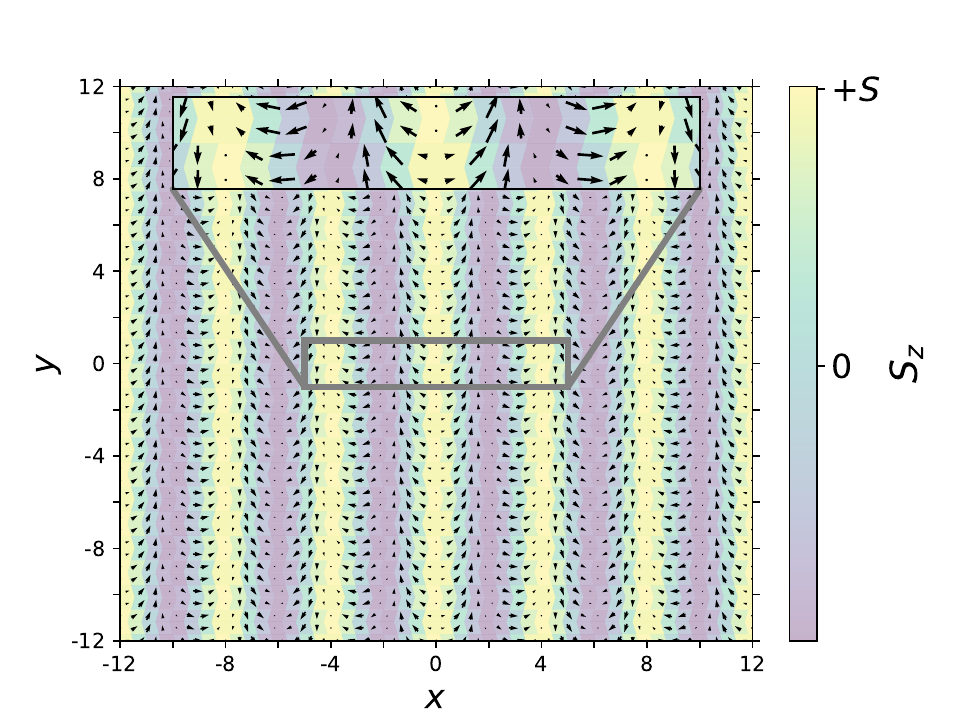}\label{fig:myung}}
  \caption{(a) Generation of the qualitative spin texture in Fe$_{3}$GeTe$_{2}$ experimentally observed in Fig. 1 of Ref. \cite{meijer_2020_ChiralSpin_NanoLett} using Eq. (\ref{eq:spin-config-int}). On the left hand side the summation of magnetization along $z$ and $x$ axis ($S_{z}+S_{x}$) is plotted. On the right the summation of magnetization along $z$ and $y$ axis ($S_{z}+S_{y}$) is plotted. We observe more pronounced magnetic texture for $S_{z}+S_{x}$ compared to $S_{z}+S_{y}$; the same was observed in experiment. (b) Generation of the experimentally observed spin texture in Cr$_{2}$Ge$_{2}$Te$_{6}$ in Fig. 2c of Ref. \cite{han_2019_TopologicalMagneticSpin_NanoLett}, using Eq. (\ref{eq:spin-config-int}), with $\vec{q}_{1}=\left( \pi/2,0 \right)$ and $\vec{q}_{2}=\left( \pi/3,0 \right)$.}
  \label{fig:exper-spin-text}
\end{figure}
Experimentally analogous spin texture was observed in vdW ferromagnet Fe$_{3}$GeTe$_{2}$ (F3GT). It is a conductor with itinerant ferromagnetism \cite{fei_2018_TwodimensionalItinerant_NatureMater} and Kondo lattice behavior \cite{zhang_2018_EmergenceKondo_SciAdv}. In F3GT using scanning electron microscopy with polarization analysis a modulated spin spiral on $xz$ plane (Neel order), as well as on $xy$ plane was observed \cite{meijer_2020_ChiralSpin_NanoLett}. The same spin texture was also observed through Lorentz transmission electron microscopy (LTEM) and micromagnetic simulations \cite{ding_2022_TuningDensity_NPGAsiaMater,peng_2021_TunableNeel_AdvFunctMater,gao_2022_ManipulationTopological_PhysRevB}.
In Fig. \ref{fig:meijer_compound} using Eq. (\ref{eq:spin-config-int}) with $\vec{q}_{1}= \left( \pi/8, \pi/8 \right)$ and $\vec{q}_{2}=\left( \pi/4, \pi/4 \right)$ we plot qualitatively same spin texture as was experimentally observed \cite[see Fig. 1 of Ref.][]{meijer_2020_ChiralSpin_NanoLett}. We say qualitatively, as one can see the pronounced peak and dip of the magnetization for $S_{z}+S_{x}$ in comparison to $S_{z}+S_{y}$, as was observed in experiment. In another sister compound Cr$_{2}$Ge$_{2}$Te$_{6}$ (CGT) through LTEM analogous spin texture was detected \cite{han_2019_TopologicalMagneticSpin_NanoLett}. In Fig. \ref{fig:myung} using Eq. (\ref{eq:spin-config-int}) for $\vec{q}_{1}=\left( \pi/2,0 \right)$ and $\vec{q}_{2}=\left( \pi/3,0 \right)$ we reproduce the experimentally observed spin texture \cite[see Fig. 2c of Ref.][]{han_2019_TopologicalMagneticSpin_NanoLett}. Physically this spin texture can be thought of as Neel spin order sandwiched in between two Bloch domain walls. On a side note, such magnetic texture was also observed in heterostructures of the multiple ferromagnetic monolayers. In these materials combined effect of the perpendicular magnetic anisotropy (PMA) --- due to the dipole interaction between the layers --- and the interfacial DMI give rise to the spin texture \cite{lemesh_2018_TwistedDomain_PhysRevB}. In [Co/Ni]$_{n}$/Ir/Pt(111) heterostructure depending on the thickness of the magnetic multilayer stack [Co/Ni]$_{n}$ and the Ir layer, either Bloch-type or Neel-type domain walls were observed, however, for some specific thickness of both these layers one can find both Bloch and Neel domain walls \cite{chen_2013_TailoringChirality_NatCommun}. It is in this region one can find the spin texture represented by Eq. (\ref{eq:spin-config-int}). The same is true for Co/Pd \cite{garlow_2019_QuantificationMixed_PhysRevLett} and Fe/Ni/Cu(001) \cite{chen_2013_NovelChiral_PhysRevLett} multilayers.

In this work we solve the Hamiltonian of the 2D magnetic materials in the strong electron correlation limit with spin texture given by Eq. (\ref{eq:spin-config-int}). It is assumed that the system has localized spin. The model can be applied to the arbitrary 2D crystal structure with magnetic atoms having arbitrary high spin-$S$. The high localized spin treatment of the problem is necessary, as in 2D materials due to reduced coordination number of the surface atoms the localized electronic bands at surface become narrower compared to the bulk; the narrow band favors the localization, exchange splitting and higher magnetic moments \cite{s_1947_Ferromagnetism_RepProgPhys}. Besides, the studied van der Waals magnets have high correlation which favours narrower bands. It was found that, for spin $S \leq 3$ in a bipartite honeycomb lattice for polar spin modulation vector $\vec{q}_{1}=\left(q_{1x},0 \right)$ and azimuthal spin modulation vector $\vec{q}_{2}=\left( q_{2x},0 \right)$, the Chern number depends strongly on $\vec{q}_{2}$, weakly on $\vec{q}_{1}$; for higher spins the effect of $\vec{q}_{1}$ also increases [see Fig. \ref{fig:chern-numb-all-spin}]. The general expression of Chern number, Eq. (\ref{eq:chern-large-S}), is analogous to the Chern number found in Ref. \cite{ohgushi_2000_SpinAnisotropy_PhysRevB}; it represents the amount of effective magnetic flux penetrating the single unit sublattice plaquette. We also found that the Chern number depends in an analogous manner as the Haldane model \cite{haldane_1988_ModelQuantum_PhysRevLett} for large $S$ and $q_{2x}$ [see Fig. \ref{fig:chern-num-haldane}].

The article is structured as follows, in Sec. \ref{sec:method-calculation} we give the resulting Hamiltonian on a bipartriate lattice. The complete theory, and the mathematical procedure to solve the Hamiltonian is described in App. \ref{sec:app-th} and \ref{sec:solut-hamilt-bipartr} respectively. In Sec. \ref{sec:integer-spin-1} and \ref{sec:half-integer-spins} we find the Chern number for integer and half-integer spins. In Sec. \ref{sec:discussion} we discuss different aspects of the topological properties of the result, and possible experimental setups.

\section{Method and Calculation}
\label{sec:method-calculation}
\label{sec:integer-spin}
\begin{figure}
  \centering
  \includegraphics[width=0.45\textwidth]{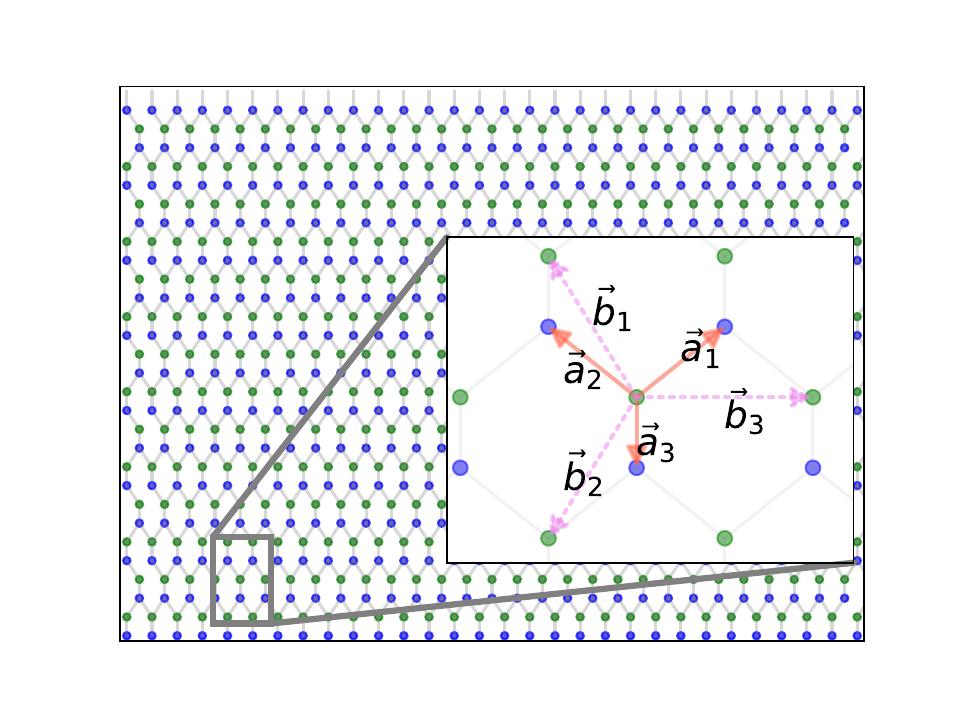}
  \caption{The bipartite honeycomb lattice. $\vec{a}_{1}$, $\vec{a}_{2}$, and $\vec{a}_{3}$ represent the three nearest neighbor vectors. $\vec{b}_{1}$, $\vec{b}_{2}$, and $\vec{b}_{3}$ represents the three next nearest neighbor vectors.}\label{fig:hc-lat}
\end{figure}

To model the vdW materials we start from the Kondo lattice system where the localized magnetic moments on each lattice sites behave as spinless scattering sites \footnote{Kondo lattice model is derived from the periodic Anderson model in the strong correlation regime \cite{tsunetsugu_1997_GroundstatePhase_RevModPhys,gulacsi_2006_KondoLattice_PhilosophicalMagazine}. Periodic Anderson model takes into account the hybridization of the localized $f$ electrons.}. At large Kondo coupling, i.e. when the coupling between the conduction electron spins and the localized spin momentum is strong, the Kondo lattice model is identical to the Hubbard model with large electron correlation \cite{ivantsov_2022_StrongCorrelation_AnnalsOfPhysics}. The necessary theory for solving this model using $su(2)$ path integral method was given recently \cite{kesharpu_2023_TopologicalHall_PhysRevB}. Basically the idea is to represent the Kondo lattice model in strong coupling regime (the Hubbard model in strong correlation regime) through the Hubbard $X^{pq}$ operators \footnote{ The Hubbard operators $X_{i}^{pq} \equiv \bra{p} \ket{q}$ describes the transition at site from $\ket{p}$ state to $\ket{q}$ state. Under strong correlation there are three different states possible, i.e. states with up-spin $\ket{\uparrow}$, down-spin $\ket{\downarrow}$ and empty site $\ket{0}$. Hence, for example $X_{i}^{\uparrow 0}$ represent the transition from empty state to the up-spin state at $i$-th site. In terms of usual electron creation ($c_{i\sigma}^{\dagger}$), annihilation ($c_{i\sigma}$), and number ($n_{i\sigma} = c_{i \sigma}^{\dagger} c_{i \sigma}$) operators, the Hubbard operators are represented as:
  \begin{equation*}
    \label{eq:fn:Hubbard-opr}
    \begin{aligned}
      &X_{i}^{0\sigma} = c_{i\sigma} (1-n_{i \sigma'}),
        \quad X_{i}^{\sigma 0} =(1-n_{i \sigma'})c_{i\sigma}^{\dagger}\\
      &X_{i}^{0 0} = 1 - \sum\limits_{\sigma} \left( 1 - n_{i \sigma'} \right) c_{i \sigma}^{\dagger} c_{i \sigma}^{\dagger} \left( 1 - n_{i \sigma'} \right).
    \end{aligned}
  \end{equation*}
  Above definition of Hubbard operators shows that the no double occupancy condition is satisfied.
}. As there exist one-to-one mapping of the $X^{pq}$ operators to the $su(2)$ coherent state operators $X_{cs}^{pq}=\bra{z, f} X^{pq} \ket{z, f}$ \cite{ferraz_2022_FractionalizationStrongly_PhysRevB,ferraz_2011_EffectiveAction_NuclearPhysicsB,berezin_1987_IntroductionSuperanalysis_} --- here the $f$ and $z$ are the spinless charged fermionic field (holon) and spinful bosonic fields (spinon) respectively --- one can transform the $X^{pq}$ Hamiltonian into the coherent symbol $X^{pq}_{cs}$ Hamiltonian. Finally the path integral approach is used to solve the resulting $X_{cs}^{pq}$ Hamiltonian containing the holon and spinon degrees of freedom. Physically this Hamiltonian represent the interaction of the strongly correlated itinerant electrons with the background spin textures. We re-derived this theory in App. \ref{sec:app-th} for convenience of the readers. The resulting Hamiltonian given by Eq. (\ref{eq:gen-ham}) is quite general, in the sense that arbitrary two dimensional lattice structure and spin texture can be plugged into it to solve the specific systems.

As discussed in Sec. \ref{sec:materials} usually the magnetic atoms in vdW magnets create honeycomb bipartite crystal structure. Hence, at first, we solve the Eq. (\ref{eq:gen-ham}) for bipartite lattice with spin texture Eq. (\ref{eq:spin-config-int}), then we impose the honeycomb lattice structure to the resulting Hamiltonian. A bipartite lattice $L$ can be divided into two sub lattices $A$ and $B$, i.e. $L = A \oplus B$. Therefore the nearest neighbor (NN) hopping is always related to the hopping from one sub-lattice to the another ($A \to B$, $B \to A$), and the next nearest neighbor (NNN) hopping is related to the hopping on the same sub-lattices ($A \to A$, $B \to B$). Defining the NN lattice vector as $a_{n}$, and NNN lattice vector as $b_{n}$, the bipartite Hamiltonian, Eq. (\ref{eq:gen-ham}), in momentum space can be written as [see App. \ref{sec:solut-hamilt-bipartr} for solution]:
$$H(\vec{k}) = \sum\limits_{\vec{k}} \bar{\psi}_{\vec{k}} \mathcal{H}(\vec{k}) \psi_{\vec{k}}.$$
The matrix $\psi_{\vec{k}} = \left[ f_{\vec{k},A} \quad f_{\vec{k},B}\right]^{T}$ contains the holon creation operators of the $\vec{k}$-th momentum on the two sub-lattices $A$ and $B$ of the bipartite lattice. The single mode kernel of the Hamiltonian is $\mathcal{H}(\vec{k})=\mathcal{H}_{0} (\vec{k})\cdot \mathcal{I} + \mathcal{H}_{i}(\vec{k}) \cdot \vec{\sigma}_{i}$. Here, $\mathcal{I}$ is the unit matrix; $\vec{\sigma}_{i}$ are the Pauli matrices, and $\mathcal{H}_{i}(\vec{k})$ are the corresponding kernels. Explicitly
\small
\begin{equation}
  \label{eq:kerner-ham}
  \begin{aligned}
    &\mathcal{H}(\vec{k}) = \mathcal{H}_{0} \mathcal{I} + \mathcal{H}_x (\vec{k}) \sigma_{x} + \mathcal{H}_y (\vec{k}) \sigma_{y} +\mathcal{H}_z (\vec{k}) \sigma_{z};\\
    &\mathcal{H}_{0} = -2 t_{2} \mathscr{w}_{n}^{S} \: \hat{\mathcal{F}}\left[ 1+ \mathscr{g}_{n} \cos 2 \vec{q}_{1}\left( \vec{r}_{i} + \frac{\vec{b}_{n}}{2}\right) \right]^{S} \ast \\
    & \left\{ \cos S \vec{q}_{2} \vec{b}_{n} \cos \vec{k}\vec{b}_{n}  + 2S \sin S\vec{q}_{2}\vec{b}_{n} \sum\limits_{\vec{k}'}\mathscr{p}(\vec{k}') \cos \left( \vec{k} + \vec{k}' \right) \vec{b}_{n} \right\},\\
    &\mathcal{H}_{x} = +t_{1} \mathscr{w}_{n}'^{S} \: \hat{\mathcal{F}}\left[ 1 - \mathscr{g}_{n}' \cos 2 \vec{q}_{1}\left( \vec{r}_{i} + \frac{\vec{a}_{n}}{2}\right) \right]^{S} \ast \cos \vec{k} \vec{a}_{n}\\
    &\mathcal{H}_{y} = +t_{1} \mathscr{w}_{n}'^{S} \: \hat{\mathcal{F}}\left[ 1 - \mathscr{g}_{n}' \cos 2 \vec{q}_{1}\left( \vec{r}_{i} + \frac{\vec{a}_{n}}{2}\right) \right]^{S} \ast \sin \vec{k} \vec{a}_{n}\\
    &\mathcal{H}_{z} = -2 t_{2} \mathscr{w}_{n}^{S} \: \hat{\mathcal{F}}\left[ 1+ \mathscr{g}_{n} \cos 2 \vec{q}_{1}\left( \vec{r}_{i} + \frac{\vec{b}_n}{2}\right) \right]^{S} \ast\\
    & \left\{- \sin S \vec{q}_{2} \vec{b}_n \sin \vec{k}\vec{b}_n + 2S \cos S\vec{q}_{2}\vec{b}_n \sum\limits_{\vec{k}'}\mathscr{p}(\vec{k}') \sin \left( \vec{k} + \vec{k}' \right) \vec{b}_n \right\}.\\
    &\text{Where,}\\
    &\mathscr{w}_{n} \equiv \frac{1}{2}+ \left(\frac{1}{4} + \frac{\cos \vec{q}_{2}\vec{b}_{n}}{4} \right) \cos \vec{q}_{1}\vec{b}_{n};\: \mathscr{g}_{n} \equiv \left[\left( \frac{1}{4} - \frac{\cos \vec{q}_{2}\vec{b}_{n}}{4}\right) \bigg/ \mathscr{w}_{n} \right];\\
   &\mathscr{w}_{n}' \equiv \frac{1}{2} - \left(\frac{1}{4} + \frac{\cos \vec{q}_{2}\vec{a}_{n}}{4} \right) \cos \vec{q}_{1}\vec{a}_{n};\: \mathscr{g}_{n}' \equiv \left[\left( \frac{1}{4} - \frac{\cos \vec{q}_{2}\vec{a}_{n}}{4}\right) \bigg/ \mathscr{w}_{n}' \right].\\      
  \end{aligned}
\end{equation}
\normalsize
Here, $t_{1}$ and $t_{2}$ are the electron NN and NNN hopping factors. $\vec{k}$ and $\vec{k}'$ are the momentum which take values in the first Brillouin zone (BZ). $\mathscr{p}(\vec{k}')$ is the Fourier coefficients of the Fourier transform of the real space function [see Eq. (\ref{eq:approx-atan-first})]:
\begin{equation}
  \label{eq:atan-main-fucntion}
  \atan \left[ \frac{1}{\mathscr{h}(r) \csc \vec{q}_{2}\vec{b}_{n} + \cot \vec{q}_{2}\vec{b}_{n}} \right] = \sum\limits_{k'} \mathscr{p}(\vec{k}') \mathrm{e}^{-i \vec{k}'\vec{r}}.
\end{equation}
$\mathscr{h}(r)$ is the function or $\vec{r}_{i}$ and $\vec{q}_{1}$ as defined in Eq. (\ref{eq:phi-A}). The $\mathscr{p}(\vec{k}')$ is a function of $\vec{q}_{1}$ and $\vec{q}_{2}$ as the L.H.S. of Eq. (\ref{eq:atan-main-fucntion}) depends on $\vec{q}_{1}$ and $\vec{q}_{2}$. However irrespective of the values of the $\vec{q}_{1}$ and $\vec{q}_{2}$ the values of $\mathscr{p}(\vec{k}') \approx 0$ in Brioullion zone for almost all $\vec{k}'$ [see Fig. \ref{fig:p-k}]. The values of $\mathscr{w}_{n}$, $\mathscr{w}_{n}'$, $\mathscr{g}_{n}$, and $\mathscr{g}_{n}'$ also depend on the spin modulation vectors $\vec{q}_{1}$ and $\vec{q}_{2}$ [see Fig. \ref{fig:g_g_dash} and \ref{fig:w_w_dash}]. $\hat{\mathcal{F}}$ represents the Fourier transform operator;

<<$\ast$>> is the convolution operator. $S$ is the effective spin of the magnetic atoms. For the integer spins ($S=1,2,\dots$) the Fourier transform is easy to find. However, for half-integer spins ($S=1/2,3/2,\dots$) analytical expression of Fourier transform can be found only in extreme limits of $\mathscr{g}_{n}$ and $\mathscr{g}_{n}'$. Both these cases along with their topological properties on a honeycomb lattice are discussed in the following sections.

\subsection{Chern number for Integer spin}
\label{sec:integer-spin-1}
\subsubsection{Chern number for $S=1$}
\label{sec:s=1}

\begin{figure}
  \centering
  \subfloat[][]{\includegraphics[width=0.24\textwidth]{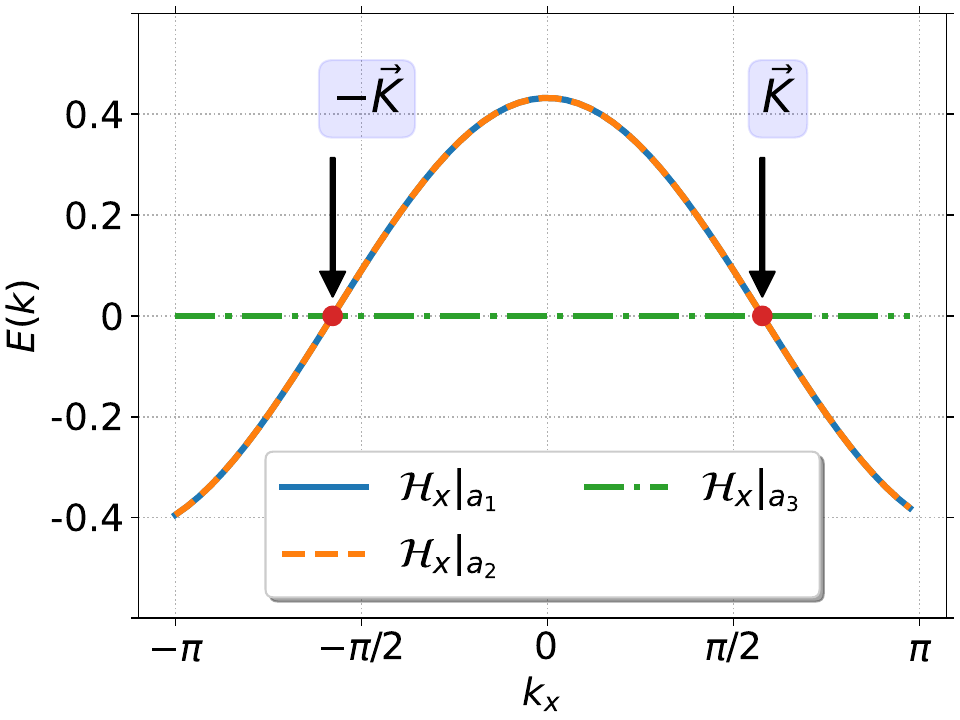}\label{fig:chern_hx}}
  \subfloat[][]{\includegraphics[width=0.24\textwidth]{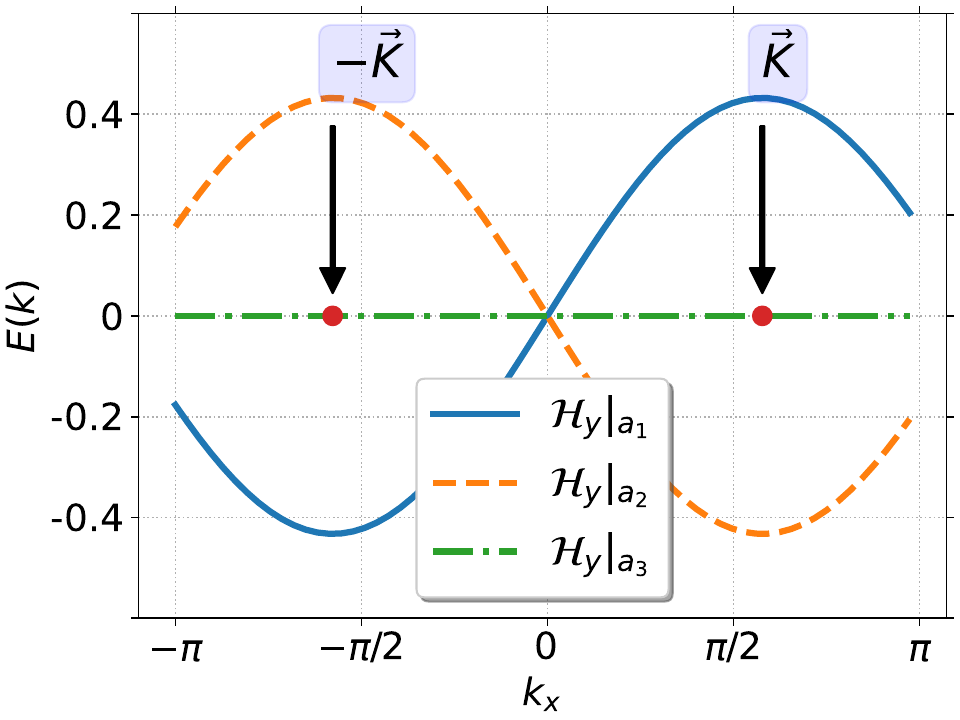}\label{fig:chern_hy}}\\
  \subfloat[][]{\includegraphics[width=0.24\textwidth]{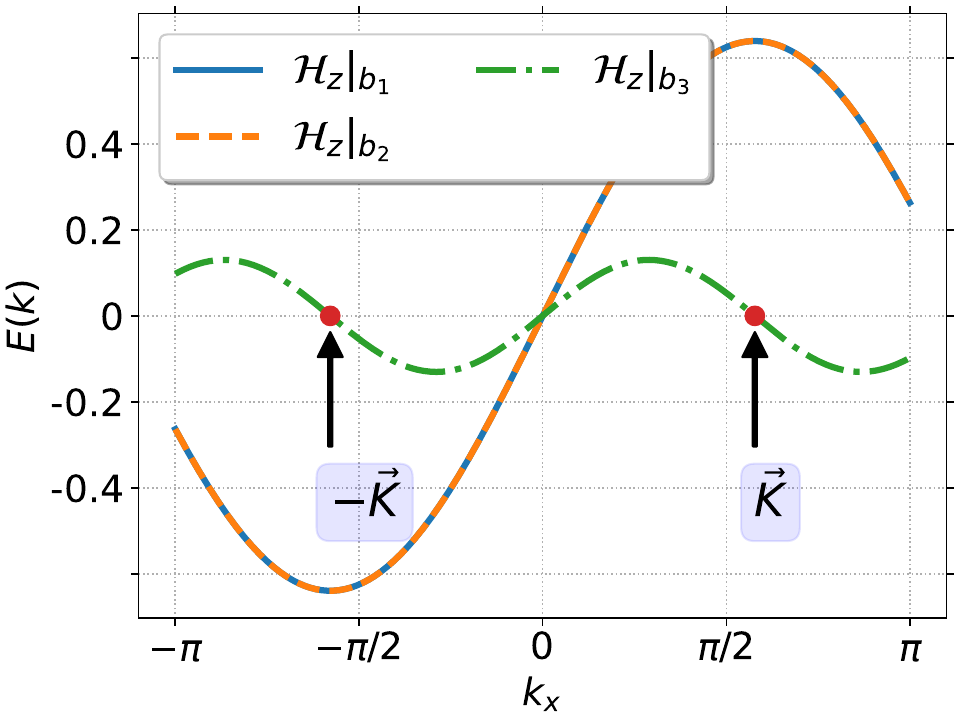}\label{fig:chern_hz}}
  \subfloat[][]{\includegraphics[width=0.24\textwidth]{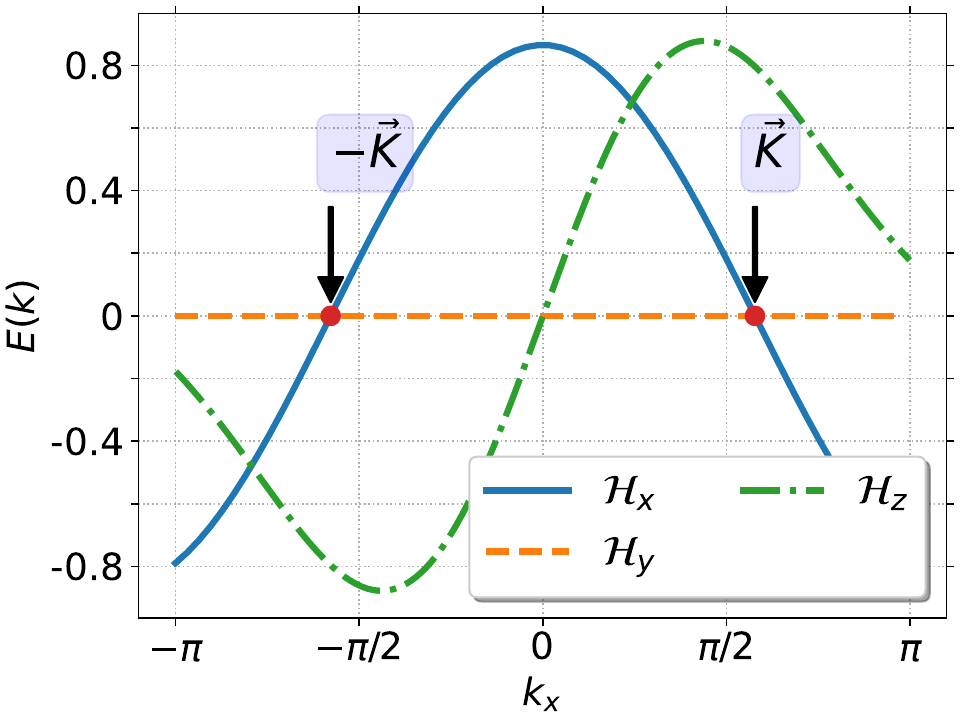}\label{fig:chern_hxyz}}\\  
  \caption{Dependence of (a) $\mathcal{H}_{x}$ and (b) $\mathcal{H}_{y}$ on the momentum vector $k_{x}$ corresponding to the three NN vectors; the value of $k_{y}=0$ everywhere. We have taken $\vec{q}_{1}=\left(\pi/4,0  \right)$ and $\vec{q}_{2}=\left(\pi/4,0  \right)$. The value of $\mathcal{K}=\left( \pm \pi/\sqrt{3},0 \right)$. (c) The same dependence of $\mathcal{H}_{z}$ on corresponding NNN vectors. (d) Summation of the previously plotted three components of the $\mathcal{H}_{x}$, $\mathcal{H}_{y}$, and $\mathcal{H}_{z}$.}
  \label{fig:chern_hx_hy_hz}
\end{figure}
In this section we calculate the topological properties of the Hamiltonian Eq. (\ref{eq:kerner-ham}) for honeycomb lattice. The honeycomb bipartite lattice is shown in Fig. \ref{fig:hc-lat}. It has the following nearest neighbor (NN, $\vec{a}_{n}$) and next nearest neighbor (NNN, $\vec{b}_{n}$) wave vectors:
\begin{equation*}
  \label{eq:honey-lattice-vect}
  \begin{aligned}
    &\vec{a}_{1}= \left( \frac{\sqrt{3}}{2}, \frac{1}{2} \right), && \vec{a}_{2}= \left(- \frac{\sqrt{3}}{2}, \frac{1}{2} \right), &&& \vec{a}_{3}= \left( 0, 1 \right),\\
    &\vec{b}_{1}= \left( -\frac{\sqrt{3}}{2}, \frac{3}{2} \right), && \vec{b}_{2}= \left( -\frac{\sqrt{3}}{2}, -\frac{3}{2} \right), &&& \vec{b}_{3}= \left( \sqrt{3},0 \right).
  \end{aligned}
\end{equation*}
For $S=1$ Eq. (\ref{eq:kerner-ham}) will be:
\small
\begin{equation}
  \label{eq:kerner-ham-k-space}
  \begin{aligned}
    &\mathcal{H}_{0} = -2 t_{2} \sum\limits_{n} \mathscr{w}_{n} \left[ 1+ \frac{\mathscr{g}_{n}}{2} \cos 2 \vec{q}_{1} \vec{b}_{n} \right] \times \\
    & \left\{\cos \vec{q}_{2} \vec{b}_{n} \cos \vec{k}\vec{b}_{n}  + 2 \sin \vec{q}_{2}\vec{b}_{n} \sum\limits_{\vec{k}'}\mathscr{p}(\vec{k}') \cos \left( \vec{k} + \vec{k}' \right) \vec{b}_{n} \right\}\\
    &\mathcal{H}_{x} = +t_{1} \sum\limits_{n} \mathscr{w}_{n}' \left[ 1 - \frac{\mathscr{g}_{n}'}{2} \cos 2 \vec{q}_{1} \vec{a}_{n} \right] \times \cos \vec{k} \vec{a}_{n}\\
    &\mathcal{H}_{y} = +t_{1} \sum\limits_{n} \mathscr{w}_{n}' \left[ 1 - \frac{\mathscr{g}_{n}'}{2} \cos 2 \vec{q}_{1}\vec{a}_{n}\right] \times \sin \vec{k} \vec{a}_{n}\\
    &\mathcal{H}_{z} = -2 t_{2} \sum\limits_{n} \mathscr{w}_{n} \left[ 1+ \frac{\mathscr{g}_{n}}{2} \cos 2 \vec{q}_{1}\vec{b}_{n} \right] \times\\
    & \left\{- \sin \vec{q}_{2} \vec{b}_{n} \sin \vec{k}\vec{b}_{n}  + 2 \cos \vec{q}_{2}\vec{b}_{n} \sum\limits_{\vec{k}'}\mathscr{p}(\vec{k}') \sin \left( \vec{k} + \vec{k}' \right) \vec{b}_{n} \right\}.
     \end{aligned}
\end{equation}
\normalsize
For topological properties to appear the condition $\mathcal{H}_{x}=\mathcal{H}_{y}=0$ and $\mathcal{H}_{z} \neq 0$ should be satisfied simultaneously \cite[ses Sec. 3.5.6 of Ref.][]{fruchart_2013_IntroductionTopological_ComptesRendusPhysique}. In our case if we take $\vec{q}_{1}=\left(2 q_{1x}/\sqrt{3}, 0 \right)$ and $\vec{q}_{2}=\left( 2q_{2x}/\sqrt{3}, 0 \right)$, then at the point $\vec{K}=\left(\pm \pi/\sqrt{3},0 \right)$ the condition is satisfied identically [see App. \ref{sec:chern-numb-calc}]. To show this graphically in Fig. \ref{fig:chern_hx_hy_hz} we plotted the three individual components of $\mathcal{H}_{x}$ (Fig. \ref{fig:chern_hx}), $\mathcal{H}_{y}$ (Fig. \ref{fig:chern_hy}) and $\mathcal{H}_{z}$ (Fig. \ref{fig:chern_hz}) corresponding to the respective NN and NNN vectors for $\vec{q}_{1}=\left( \pi/4 ,0 \right)$ and $\vec{q}_{2}=\left( \pi/4 ,0 \right)$. We observe that all three components of $\mathcal{H}_{x}$ are zero at $\vec{K}$. For $\mathcal{H}_{y}$ the $\vec{a}_{3}$ component is zero at $\vec{K}$; the $\vec{a}_{1}$ and $\vec{a}_{2}$ components are of opposite sign. In result the sum of the all three component of $\mathcal{H}_{y}$ is zero at $\vec{K}$. For $\mathcal{H}_{z}$ the $\vec{a}_{3}$ component is zero at $\vec{K}$, however the $\vec{a}_{2}$ and $\vec{a}_{3}$ components are non zero. Chern number for this case is [see App. \ref{sec:chern-numb-calc}]:
\begin{equation}
  \label{eq:Chern-numb-honey}
  c_{1} = \mathrm{sgn} \left[ \sin \left( q_{2x} \right) \right], \quad -\frac{\sqrt{3}\pi}{2} \leq q_{2x} \leq \frac{\sqrt{3}\pi}{2}.
\end{equation}
Interestingly, the Chern number depends only on the spin wave vectors $\vec{q}_{2}=\left( q_{2x},0 \right)$ --- the azimuthal angle. However, it does not mean that when $\vec{q}_{1}=0$ the topological effect is present; it is absent \footnote{When $\vec{q}_{1}=0$ the $\phi_{ji}=0$ in Eq. (\ref{eq:phi-A}). In this case the time reversal symmetry is conserved.}. Physically, it means for topological effects to occur some inclination of spin vector is necessary, which is related to nonzero effective magnetic field \cite{nagaosa_2013_TopologicalProperties_NatureNanotech}.

\subsubsection{Chern number for $S=2,3,\dots$}
\label{sec:s=2-3-dots}
\begin{figure}
  \centering
  \includegraphics[width=0.44\textwidth]{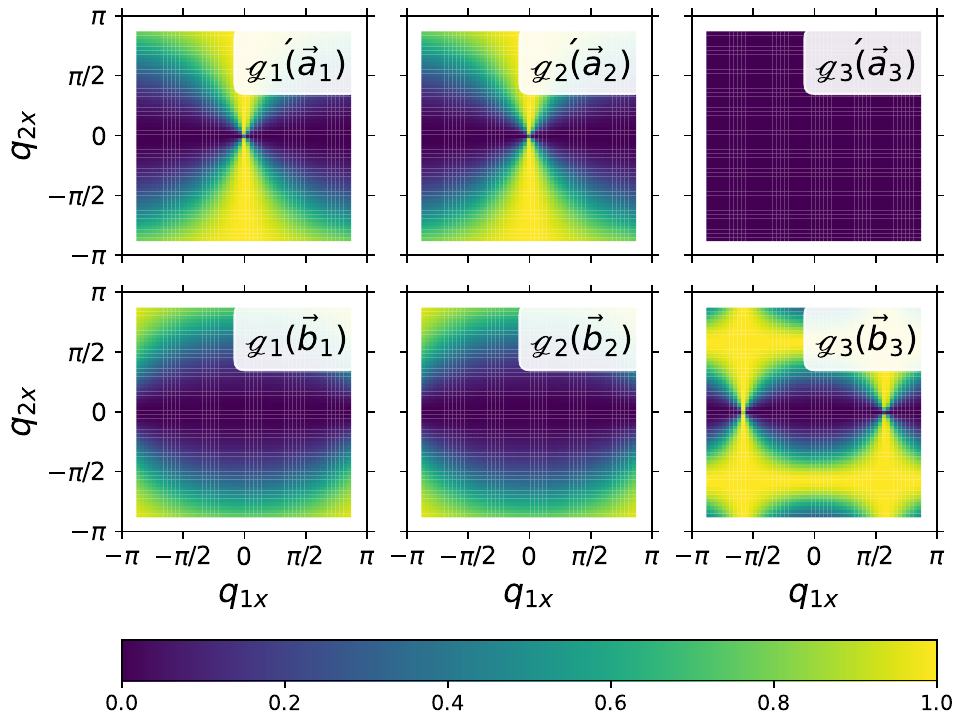}
  \caption{Plot of the dependence of the $\mathscr{g}_{n}'$ and $\mathscr{g}_{n}$ on the wave vectors $\vec{q}_{1}=\left(2 q_{1x}/\sqrt{3}, 0 \right)$ and $\vec{q}_{2}=\left( 2q_{2x}/\sqrt{3}, 0 \right)$ on a honeycomb bipartite lattice calculated using Eq. (\ref{eq:kerner-ham}). The values of $q_{1x}$ and $q_{2x}$ lies in the range $\left( -\sqrt{3}\pi/2, \sqrt{3}\pi/2 \right)$.}
  \label{fig:g_g_dash}
\end{figure}
For higher spin $S=2,3,\dots$ one will need to expand the terms containing $\mathscr{g}_{n}$ and $ \mathscr{g}_{n}'$ using binomial theorem [see App. \ref{sec:hamilt-high-integ}]. In Fig. \ref{fig:g_g_dash} we plotted the dependence of $\mathscr{g}_{n}$ and $ \mathscr{g}_{n}'$ on $q_{1x}$ and $q_{2x}$. We observe that the value of $\mathscr{g}_{n}$ and $\mathscr{g}_{n}'$ are always positive and less than unity for most part of the phase space $q_{1x}$ and $q_{2x}$. Hence, one can neglect the higher powers of $\mathscr{g}_{n}$ and $\mathscr{g}_{n}'$ and keeping terms only upto first power in the binomial expansion. The resulting Hamiltonian is [see App. \ref{sec:hamilt-high-integ} for derivation]:
\begin{equation}
  \label{eq:kerner-ham-k-space-large-s}
  \begin{aligned}
    &\mathcal{H}_{0} \approx -2 t_{2} \sum\limits_{n} \mathscr{w}_{n}^{S} \left[ 1+ \frac{S\mathscr{g}_{n}}{2} \cos 2 \vec{q}_{1} \vec{b}_{n} \right] \times \\
    & \left\{\cos S\vec{q}_{2} \vec{b}_{n} \cos \vec{k}\vec{b}_{n}  + 2S \sin S\vec{q}_{2}\vec{b}_{n} \sum\limits_{\vec{k}'}\mathscr{p}(\vec{k}') \cos \left( \vec{k} + \vec{k}' \right) \vec{b}_{n} \right\}\\
    &\mathcal{H}_{x} \approx +t_{1} \sum\limits_{n} \mathscr{w}_{n}'^{S} \left[ 1 - \frac{S\mathscr{g}_{n}'}{2} \cos 2 \vec{q}_{1} \vec{a}_{n} \right] \times \cos \vec{k} \vec{a}_{n}\\
    &\mathcal{H}_{y} \approx +t_{1} \sum\limits_{n} \mathscr{w}_{n}'^{S} \left[ 1 - \frac{S\mathscr{g}_{n}'}{2} \cos 2 \vec{q}_{1}\vec{a}_{n}\right] \times \sin \vec{k} \vec{a}_{n}\\
    &\mathcal{H}_{z} \approx -2 t_{2} \sum\limits_{n} \mathscr{w}_{n}^{S} \left[ 1+ \frac{S\mathscr{g}_{n}}{2} \cos 2 \vec{q}_{1}\vec{b}_{n} \right] \times\\
    & \left\{- \sin S\vec{q}_{2} \vec{b}_{n} \sin \vec{k}\vec{b}_{n}  + 2S \cos S\vec{q}_{2}\vec{b}_{n} \sum\limits_{\vec{k}'}\mathscr{p}(\vec{k}') \sin \left( \vec{k} + \vec{k}' \right) \vec{b}_{n} \right\}.
     \end{aligned}
\end{equation}
Comparing Eq. (\ref{eq:kerner-ham-k-space}) and Eq. (\ref{eq:kerner-ham-k-space-large-s}) one can see the extra factor of $S$ appears in different terms of the Hamiltonian. For Eq. (\ref{eq:kerner-ham-k-space-large-s}) the topological condition, $\mathcal{H}_{x}=\mathcal{H}_{y}=0$ and $\mathcal{H}_{z} \neq 0$, is satisfied simultaneously at $\vec{K}=\left(\pm \pi/\sqrt{3},0 \right)$ for $\vec{q}_{1}=\left(2 q_{1x}/\sqrt{3}, 0 \right)$ and $\vec{q}_{2}=\left( 2q_{2x}/\sqrt{3}, 0 \right)$. The Chern number is [see App. \ref{sec:spin-s=2-3} for derivation]:
\begin{equation}
  \label{eq:chern-large-S}
  \begin{aligned}
    c_{1} = \text{sign}\left[ \left( 1 + \frac{S\mathscr{g}_{2}}{2} \cos 2q_{1x}\right) \left( \sin Sq_{2x} - 2S\epsilon \cos Sq_{2x}  \right) \right]
  \end{aligned}
\end{equation}
Here, we introduced the symbol $\epsilon \equiv \sum\limits_{k'}\mathscr{p}(k') \sin \left( \pi/2 + \vec{k}' \vec{b}_{1}\right)$; it's value is always $\ll 1$ over whole $q_{1x}$ and $q_{2x}$ phase space [see Fig. \ref{p_k_dash_sin}]. One can see that, now the Chern number depends also on the $q_{1x}$ --- the polar angle; for $S=1$ it was absent. In fact, one can recover the Chern number for $S=1$, Eq. (\ref{eq:Chern-numb-honey}), from Eq. (\ref{eq:chern-large-S}). It can be seen as follows. For $S=1$ the $\left| \mathscr{g}_{2} \cos 2q_{1x}/2  \right|$ will always be less than unity (as $0 \leq \mathscr{g}_{2} \leq 1$).In result, the whole term $\left( 1 + \frac{S\mathscr{g}_{2}}{2} \cos 2q_{1x}\right)$ will always be positive irrespective of $q_{1x}$. Therefore this term will not have any affect on the $c_{1}$, and one can drop this from Eq. (\ref{eq:chern-large-S}). As mentioned before the $\epsilon \ll 1$; the only reason for including it in Eq. (\ref{eq:chern-large-S}), is that for large $S$ the term, $2S\epsilon \cos Sq_{2x}$, might be larger than $\sin Sq_{2x}$. However, for $S=1$ the $2\epsilon \cos Sq_{2x}$ will never be greater than $\sin Sq_{2x}$, apart from some narrow regions of $q_{1x}$-$q_{2x}$ phase space. Hence, we will drop the $2\epsilon \cos Sq_{2x}$ term also in Eq. (\ref{eq:chern-large-S}). Therefore we recovered the Eq. (\ref{eq:Chern-numb-honey}).

\subsection{Chern number for Half-Integer Spins}
\label{sec:half-integer-spins}

\begin{table}[tbh]\label{tab:Ham}
  \caption{Hamiltonians for spin $S=1/2$ in extreme limits.}
  \begin{ruledtabular}
    \begin{tabular}{cccc}
      Case &$\mathscr{g}_{n}'$ & $\mathscr{g}_{n}$ &Hamiltonian\\
      \hline
      I & $\ll 1$  & $\propto 1$  & Eq. (\ref{eq:ham-half-int-app-g-1-g-dash-l-1})\\
      II & $\ll 1$  & $\ll 1$ & Eq. (\ref{eq:ham-half-int-app-g-l1-g-dash-l1})\\
      III & $\propto 1$  & $\propto 1$  & Eq. (\ref{eq:ham-half-int-app-g-1-g-dash-1})\\
      IV & $\propto 1$  & $\ll 1$  & Eq. (\ref{eq:ham-half-int-app-g-l-1-g-dash-1})\\
    \end{tabular}
  \end{ruledtabular}
\end{table}
Finding a general analytical expression of Chern number for $S=1/2$ is little bit tricky, as Fourier transform of the terms,
\begin{equation}
  \label{eq:fourier-half-int}
  \begin{aligned}
    &\hat{\mathcal{F}}\left[ 1- \mathscr{g}_{n}' \cos 2 \vec{q}_{1}\left( \vec{r}_{i} + \frac{\vec{a}_{n}}{2}\right) \right]^{1/2},\\
    &\hat{\mathcal{F}}\left[ 1+ \mathscr{g}_{n} \cos 2 \vec{q}_{1}\left( \vec{r}_{i} + \frac{\vec{b}_{n}}{2}\right) \right]^{1/2},
  \end{aligned}
\end{equation}
are not available. Therefore, one need to find the Hamiltonian in the limiting cases of $\mathscr{g}_{n}$ and $\mathscr{g}_{n}'$, and then make the necessary conclusions. We can have four limiting case of $\mathscr{g}_{n}$ and $\mathscr{g}_{n}'$ in different combinations. The four cases and corresponding Hamiltonian are given in Tab. \ref{tab:Ham}. Observing these Hamiltonian we will see that they are analogous to the Eq. (\ref{eq:kerner-ham-k-space}). Meaning, the term containing $\vec{k}$ remains same. The only difference is the factors involving $\mathscr{g}_{n}$ and $\mathscr{g}_{n}'$. Hence, their Chern number will be given by Eq. (\ref{eq:Chern-numb-honey}).

The calculation for Hamiltonian for large half-integer $S$  is given in App. \ref{sec:hamilt-spin-s=32}. The main idea is to represent:
\small
\begin{equation}
  \label{eq:fourier-half-int-large-S}
  \begin{aligned}
    &\left[ 1- \mathscr{g}_{n}' \cos 2 \vec{q}_{1}\left( \vec{r}_{i} + \frac{\vec{a}_{n}}{2}\right) \right]^{1/2} \approx 
     \left[ 1- \mathscr{g}_{n}' \cos 2 \vec{q}_{1}\left( \vec{r}_{i} + \frac{\vec{a}_{n}}{2}\right) \right]^{1/2} \\ & \qquad \times \left[ 1- \mathscr{g}_{n}' \cos 2 \vec{q}_{1}\left( \vec{r}_{i} + \frac{\vec{a}_{n}}{2}\right) \right]^{(S-1/2)}.
  \end{aligned}
\end{equation}
\normalsize
Then use the approximations for $S=1/2$ (App. \ref{sec:hamilt-spin-s=12}) and higher integer $S$ (App. \ref{sec:s=2-3-dots}) to find the Fourier transform of the resulting expression. The calculation procedure is described in App. \ref{sec:hamilt-spin-s=32}. The resulting Hamiltonians will also be analogous to the Eq. (\ref{eq:kerner-ham-k-space-large-s}). Therefore the Chern number will be given by Eq. (\ref{eq:chern-large-S}).

\section{Discussion}
\label{sec:discussion}

\subsection{General formula for Chern number}
\label{sec:gener-form-chern}
\begin{figure}[tbh]
  \centering
  \includegraphics[width=0.48\textwidth]{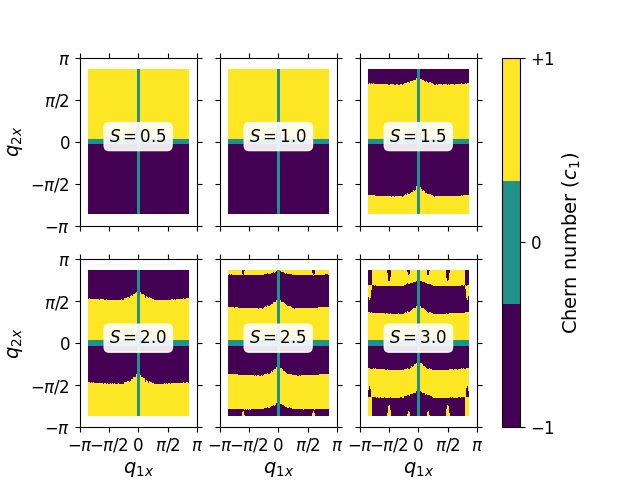}
  \caption{Chern number calculated using Eq. (\ref{eq:chern-large-S}).}
  \label{fig:chern-numb-all-spin}
\end{figure}
From the above discussions we can conclude that Eq. (\ref{eq:chern-large-S}) gives the Chern number for both integer or half-integer spins. In Fig. \ref{fig:chern-numb-all-spin} we plotted the Chern number dependence on $q_{1x}$ and $q_{2x}$ for spin upto $S=3$. Near $q_{1x}\approx 0$ and $q_{2x}$ the Chern number is not defined. The $q_{1x}=0$ physically represent the spin vector pointing towards the north pole, and $q_{2x}=0$ represent the rotation of the spin only along some fixed longitude of Bloch sphere; both these cases have ill defined Chern number. One can observe that apart from some small discrepancies in most part of the figure the Chern number depends only on the azimuthal modulating vector $q_{2x}$. Hence Eq. (\ref{eq:Chern-numb-honey}) can be used as an approximated formula for Chern number. Physically, the dependence of Chern number on $Sq_{2x}$ can be understood as the amount of effective magnetic field threading through three lattice points locally \cite{ohgushi_2000_SpinAnisotropy_PhysRevB}. The effective local magnetic field through three lattice points ($i,j,k$) is $\text{B}_{eff}= \vec{S}_{i}\cdot \left[ \vec{S}_{j} \times \vec{S}_{k} \right]$ is proportional to the $S^{3}$, hence, with increase in $S$ the threaded magnetic field also increases.

\subsection{Energy band gap closing}
\label{sec:energy-band-gap}

\begin{figure}[tbh]
  \centering
  \includegraphics[width=0.48\textwidth]{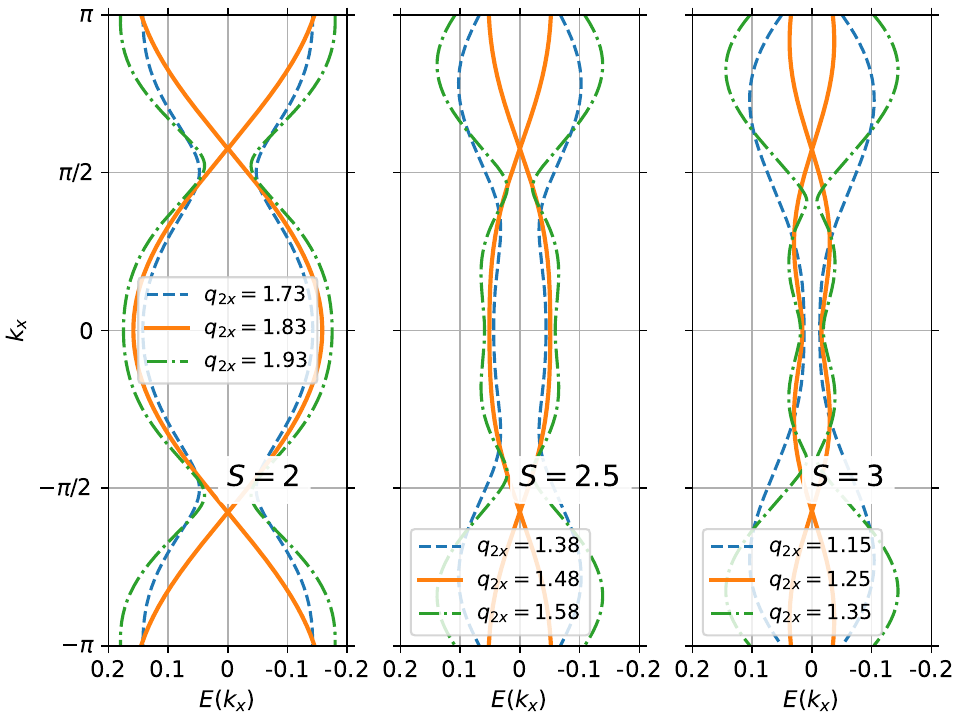}
  \caption{Dependence of energy $E(k_{x}) = \sqrt{\mathcal{H}_{x}+\mathcal{H}_{y}+\mathcal{H}_{z}}$ on $k_{x}\in \left( -\pi, +\pi \right)$ for different $q_{2x}$, and $S=2,5/2,3$.}
  \label{fig:energy-gap}
\end{figure}
From the energy band point of view valence and conduction band are represented by the $\mathcal{H}_{0} \pm \sqrt{\mathcal{H}_{x}^{2} + \mathcal{H}_{y}^{2} + \mathcal{H}_{z}^{2}}$. The wave vector $q_{2x}$ controls the gap of the bands. At specific values of the $q_{2x}$ the gap closes at $\pm\vec{K}$, and there is a change of the Chern number in the corresponding band. Explicitly the values of $q_{2x}$ where the gap closes is found as:
\small
\begin{equation}
  \label{eq:value-q2x-gap}
  q_{2x} = \frac{2}{\sqrt{3}S}\arcsin \left[ 2S \cos \left( \frac{Sq_{2x}\sqrt{3}}{2} \right) \sum\limits_{\vec{k}'} \mathscr{p}(\vec{k}') \sin \left( \frac{\pi}{2} + \vec{k}'\vec{b}_{2}\right) \right].
\end{equation}
\normalsize
In Fig. \ref{fig:energy-gap} we show the values of $q_{2x}$ where the gap closes for different $S2,2.5,3$. The energy dispersion at the points at which the gap closes can be represented through massive Dirac Hamiltonian. The Berry phase is concentrated around these points, hence, the Chern number can be calculated numerically by integrating around these points.

\subsection{Comparison with Haldane model}
\label{sec:comp-with-hald}
\begin{figure}[tbh]
  \centering
  \includegraphics[width=0.48\textwidth]{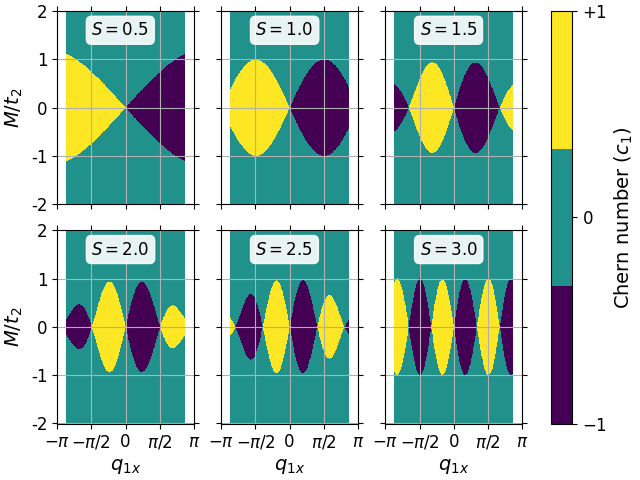}
  \caption{Chern number dependence on sub lattice potential $M/t_{2}$ and $q_{2x}$ for $q_{1x}$ =0 given by Eq. (\ref{eq:chern-haldane}).}
  \label{fig:chern-num-haldane}
\end{figure}
If compared to the Haldane model \cite{haldane_1988_ModelQuantum_PhysRevLett}, then the topological property of the Haldane model was controlled by the sub-lattice symmetry breaking onsite potential ($M$) and the phase accumulation of electrons due to NNN hopping (defined as $\phi$ in Ref. \cite{haldane_1988_ModelQuantum_PhysRevLett}). In our case the factor $Sq_{2x}$ plays the analogous phase accumulation role due to NNN hopping. In fact we can phenomenologically insert additional sublattice hopping potential $M$ into the $\mathcal{H}_{z}$ of Eq. (\ref{eq:kerner-ham}). The Chern number in this case is:
\small
\begin{equation}
  \label{eq:chern-haldane}
  \begin{aligned}
    &c_{1} = \\
    & \frac{\text{sign}\left[M -2t_{2}\left( 1 + \frac{S\mathscr{g}_{2}}{2} \cos 2q_{1x}\right) \left( \sin Sq_{2x} - 2S\epsilon \cos Sq_{2x}  \right) \right]}{2} \\
    & - \frac{\text{sign}\left[M +2t_{2}\left( 1 + \frac{S\mathscr{g}_{2}}{2} \cos 2q_{1x}\right) \left( \sin Sq_{2x} - 2S\epsilon \cos Sq_{2x}  \right) \right]}{2}.
  \end{aligned}
\end{equation}
\normalsize
In Fig. \ref{fig:chern-num-haldane} we plotted the Chern number dependence on $M/t_{2}$ and $q_{2x}$ for $S=3$; the $q_{1x}=\pi/4$ is kept constant.

\subsection{Minimum Energy}
\label{sec:minimum-energy}
\begin{figure}[tbh]
  \centering
  \includegraphics[width=0.44\textwidth]{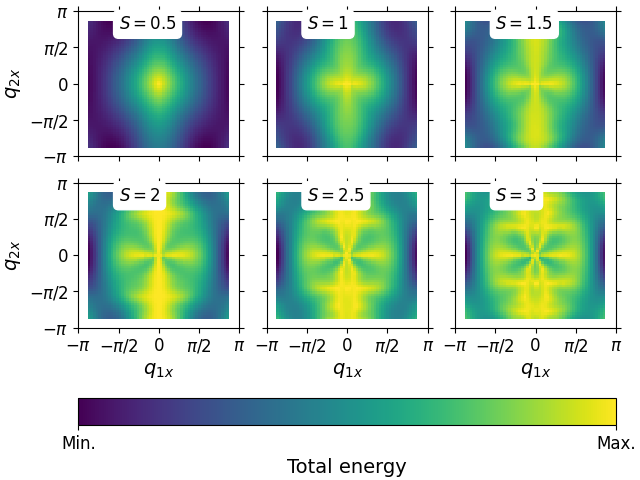}
  \caption{Dependence of total energy on the wave vectors $q_{1x}$ and $q_{2x}$. The total energy is calculated by integrating Eq. (\ref{eq:tot-energy-1}) over 1st Brillouin zone of the honeycomb lattice.}
  \label{fig:minq}
\end{figure}

To understand the thermodynamics of the system we calculate the total internal energy of the system through equation:
\begin{equation}
  \label{eq:tot-energy-1}
  U_{\text{internal}} = \sum\limits_{k \in BZ} E(k) f \left[ E(k), \mu\right]
\end{equation}
$E(k)$ is the energy of the lowest band defined as $\mathcal{H}_{0} - \sqrt{\mathcal{H}_{x}^{2}+\mathcal{H}_{y}^{2}+\mathcal{H}_{z}^{2}}$ in Eq. $f \left[ E(k), \mu \right]$ is the Fermi-Dirac distribution; $k_{B}=8.6\times 10 ^{-5}$ eV K$^{-1}$ is the Boltzman constant; $T$ is the temperature. In our calculation we take $T \approx 0$ K. The summation in Eq. (\ref{eq:tot-energy-1}) is taken over the whole Brillouin zone. In Fig. \ref{fig:minq} we plotted the $U_{\text{internal}}$ for different $S$. In most cases the lowest energy has an anti-ferromagnetic configuration, which was expected, as the original Hamiltonian, Eq. (\ref{eq:gen-ham}) will have lowest energy when the dot product of the $\vec{S}_{i} \cdot \vec{S}_{j}$ will have negative sign.

\subsection{Perspective Materials for application of the model}
\label{sec:materials}

\begin{table}[tbh]\label{tab:Mat}
  \caption{Properties of vdW materials to which proposed model can be applied. The spin lattice is the 2D crystal structure formed by the magnetic atoms. Atomic spin is the effective localized magnetic moments of the magnetic atoms.}
  \begin{ruledtabular}
    \begin{tabular}{llll}
      Material &Crystal   &Spin &Atomic \\
               &Structure &Lattice  &Spin  \\
               &          &         &     \\
      \hline
      Fe$_{3}$GeTe$_{2}$ &Hexagonal &Honeycomb  &3  \\
      (F3GT)  \\
      Fe$_{4}$GeTe$_{2}$ &Rhombohedral  &Honeycomb  &7/2   \\
      (F4GT)        \\
      Cr$_{2}$Ge$_{2}$Te$_{6}$  &Hexagonal  &Honeycomb  &3/2 \\
      (C2GT)  \\
      Cr-X$_{3}$  &Rhombohedral  &Honeycomb  &3/2  \\
      X=Cl,Br,I   \\      
    \end{tabular}
  \end{ruledtabular}
\end{table}
In Tab. \ref{tab:Mat} we give a list of widely studied vdW materials. In these materials usually the magnetic layers are sandwiched in between the non magnetic layer in different configurations. For example, in F3GT a single Fe-Fe dumbbell and a single Fe atom are placed alternatively in a honeycomb lattice pattern; at the center of the honeycomb lattice the single Ge atom is placed \cite{seo_2020_NearlyRoom_SciAdva,kim_2021_DrasticChange_SciRep}. This whole structure containing the Fe and Ge atoms are sandwiched in between Te atoms. The F4GT has the same honeycomb lattice structure, however, now instead of single dumbbell two dumbbells of Fe atoms are present; the single Fe atoms are absent \cite{seo_2020_NearlyRoom_SciAdva,kim_2021_DrasticChange_SciRep}. The structure of C2GT is more complex. Single unit cell contains four different layers of magnetic atoms \cite{gong_2017_DiscoveryIntrinsic_Nature}. However, If considered only the magnetic Cr ions then they form an effective honeycomb lattice \cite{gong_2017_DiscoveryIntrinsic_Nature,selter_2020_MagneticAnisotropy_PhysRevB,fumega_2020_ElectronicStructure_JMaterChemC}. In Cr-X$_{3}$ the Cr atoms are sandwiched in between layers of halide (X=Cl, Br, I) atoms. The Cr$^{3+}$ ions form a honeycomb lattice structure. The same honeycomb spin lattice is formed in other transition metal trihalides \cite{tomar_2019_IntrinsicMagnetism_JournalOfMagnetismAndMagneticMaterials}.

\begin{figure}[tbh]
  \centering
  \subfloat[][Fe$_{3}$GeTe$_{2}$]{\includegraphics[width=0.4\textwidth]{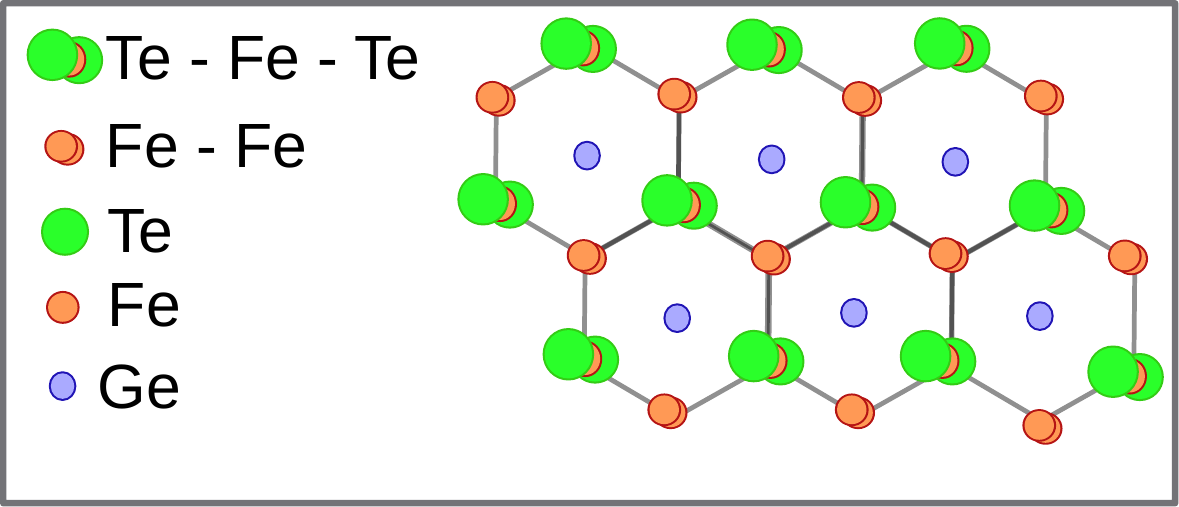}\label{fig:Fe3GeTe2}}\\
  \subfloat[][Fe$_{4}$GeTe$_{2}$]{\includegraphics[width=0.4\textwidth]{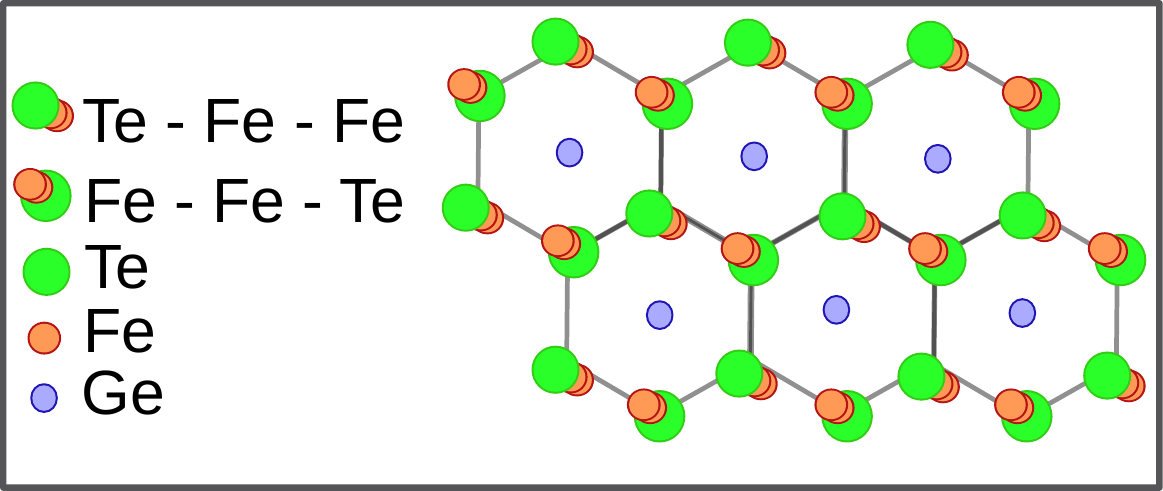}\label{fig:Fe3GeTe2}}\\
  \subfloat[][CrI$_{3}$]{\includegraphics[width=0.4\textwidth]{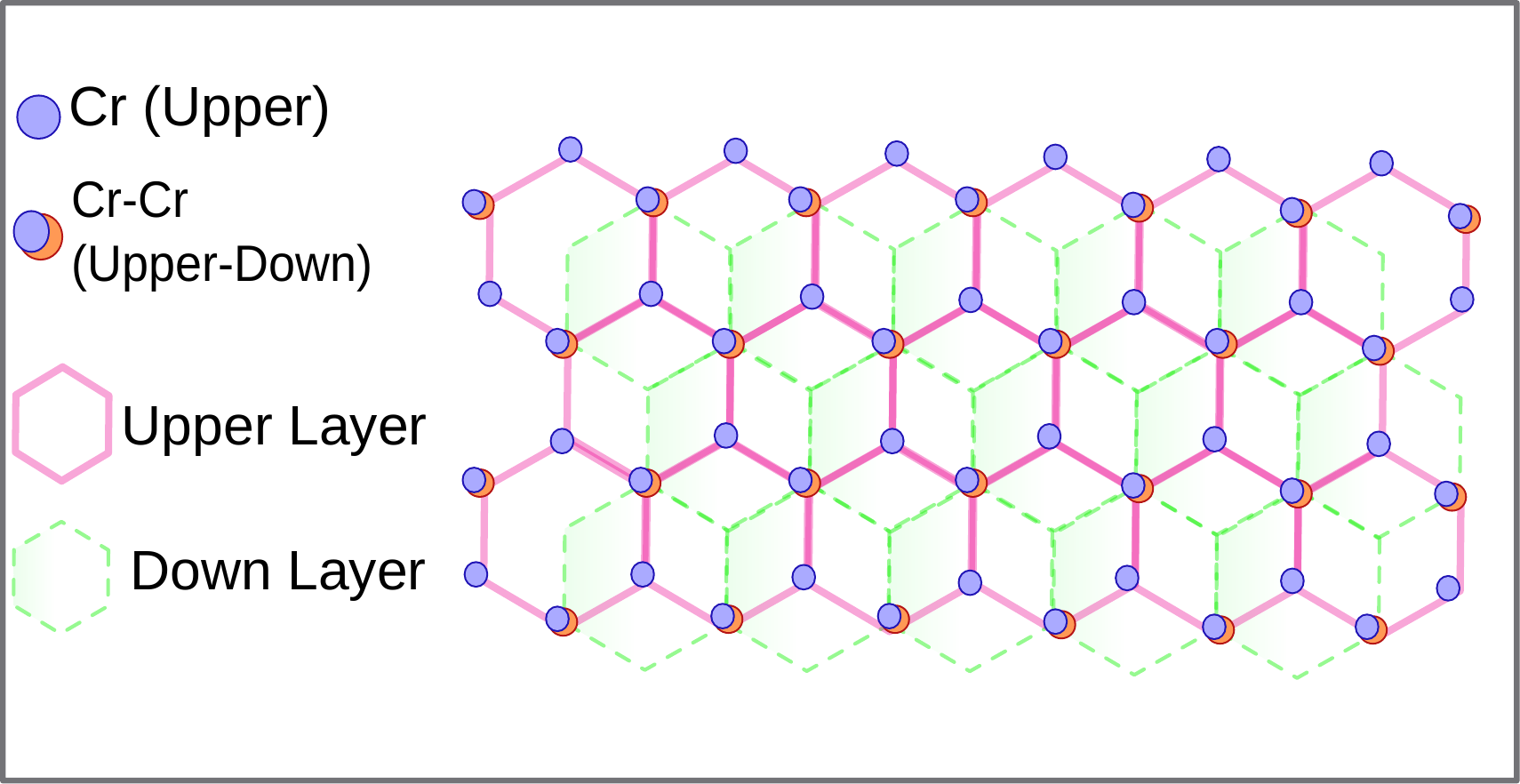}\label{fig:CrI3}}\\
  \subfloat[][CrMnI$_{6}$]{\includegraphics[width=0.4\textwidth]{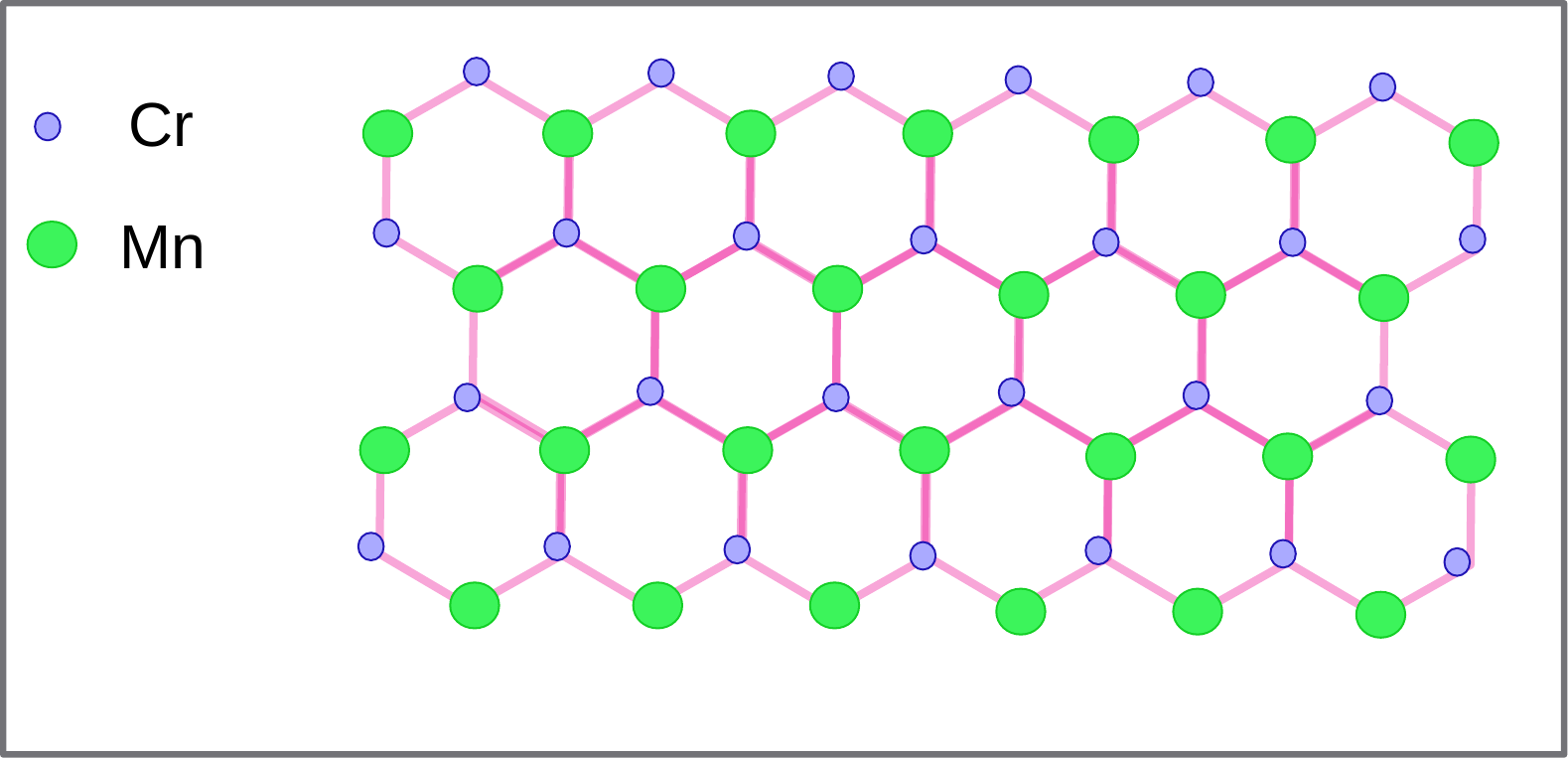}\label{fig:CrMnI6}}\\
  \caption{Generation of magnetic bipartite honeycomb lattice.}
  \label{fig:material-schematics}
\end{figure}
All the aforementioned materials are natural bipartriate lattice or can be treated as effective bipartriate lattice in some special configurations. For example, although monolayer of the Cr-X$_{3}$ can not be considered as bipartite lattice, however, due to stacking configuration of bilayers at low temperature rhombohedral phase, one can consider them as bipartriate lattice \cite{mcguire_2015_CouplingCrystal_ChemMater,jiang_2019_StackingTunable_PhysRevB,soriano_2019_InterplayInterlayer_SolidStateCommunications}. Besides, a sister compound CrMnI$_{6}$, where alternate Cr atoms are replaced by the Mn atoms, is proposed, whose monolayer can be considered as bipartite lattice \cite{zhang_2020_PredictionMonolayered_PhysRevB,zhang_2023_GiantDzyaloshinskiiMoriya_NpjQuantumMater}. In F3GT the Fe-Fe dumbbell and Fe atoms corresponds to the two sub-lattices of the bipartriate lattice. For F4GT the bonding between Fe-Fe dumbbell and Te atoms creates an effective bipartite lattice, i.e. the one of the sublattice contains the dumbbell with upper Fe atom of the dumbbell bonded to the Te atom, and the other sublattice contains the lower Fe atom of the dumbbell bonded to the Te atom. In C2GT the strong hybridization of the Cr $3d$, Ge $4p$ and Te $5p$ electrons gives rise to the bipartite lattice \cite{wang_2019_GiantContribution_PhysChemChemPhys}.

Two fundamental requirement of the proposed model is the localized spin momentum and strong electron correlation. In F3GT the electron correlation is around $U=5$ to $5.5$ eV \cite{zhu_2016_ElectronicCorrelation_PhysRevB,ghosh_2023_UnravelingEffects_NpjComputMater}, which is comparable to the $U=6.2$ eV for the prototypical heavy Fermion compound CeIn$_{3}$. Regarding localized spin momentum, recently, it was shown that 3$d$ electrons of the Fe ions in F3GT have a strong localized character \cite{yamagami_2021_ItinerantFerromagnetism_PhysRevB,xu_2020_SignatureNonStoner_PhysRevB}. The itinerant ferromagnetism observed before \cite{fei_2018_TwodimensionalItinerant_NatureMater}, is in fact due to the delocalized Ge/Te $p$ electrons \cite{yamagami_2021_ItinerantFerromagnetism_PhysRevB}. In case of Cr-X$_{3}$, the Cr sites usually have localized spin $S=3/2$ \cite{craco_2020_MottLocalization_PhysRevB,besbes_2019_MicroscopicOrigin_PhysRevB}, and the correlation between 3$d$ electrons is $U \approx 3$ eV \cite{kvashnin_2022_DynamicalCorrelations_PhysRevB}. The same is also true for C2GT with $U \approx 4$ eV \cite{menichetti_2019_ElectronicStructure_2DMater}.

\subsection{Experiments}
\label{sec:experiments}
\begin{figure}[tbh]
  \centering
  \includegraphics[width=0.5\textwidth]{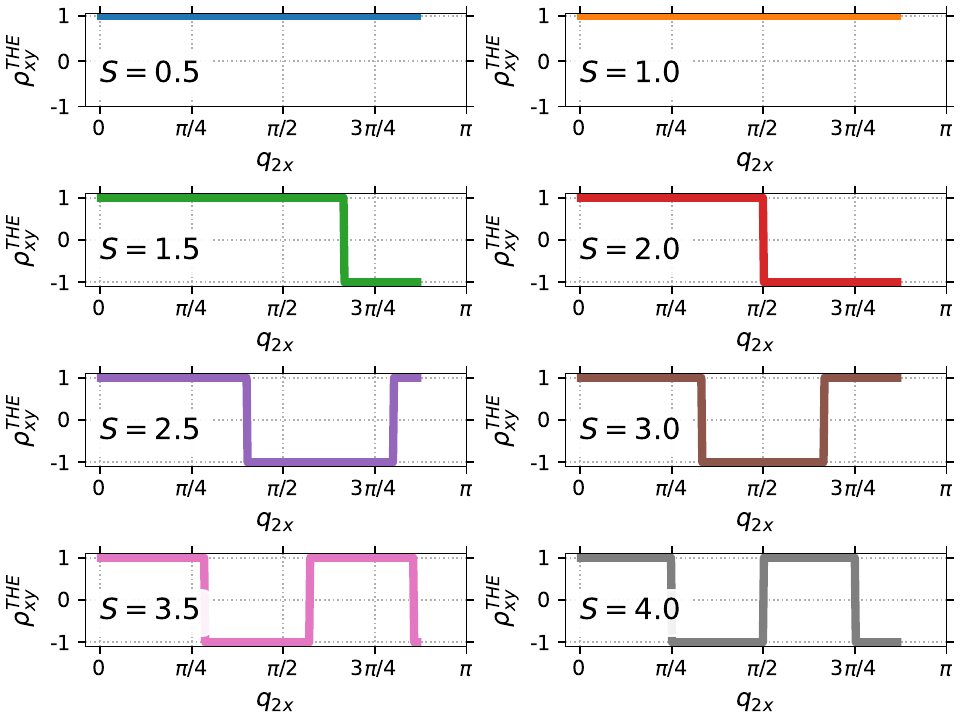}
  \caption{Dependence of the topological Hall resistivity $\rho_{xy}^{THE}$ on the azimuthal spin modulation vector $q_{2x}$. The $\rho_{xy}^{THE}$ is represented in units of $2 \pi \hbar/e^{2}$. The polar angle $q_{1x}$ is kept constant.}
  \label{fig:chern-spin}
\end{figure}
Naturally, the question arises how the above effect can be physically verified? One of the indirect way of confirming this effect is through the hall resistivity measurement. The resistivity due to topological Hall effect is proportional to the Chern number \cite{bruno_2004_TopologicalHall_PhysRevLett,ohgushi_2000_SpinAnisotropy_PhysRevB,kimbell_2022_ChallengesIdentifying_CommunMater}: $\rho_{xy}^{THE} = - \frac{2 \pi \hbar}{e^{2}c_{1}}$. The Hall resistivity measurements are usually done under applied varying magentic field where the THE appears as additional resistivity in some magnetic field range, as in this field range the chiral spin texture is stabilized. In our case, from a theoretical point of view, after removing the ordinary Hall effect and anomalous Hall effect \cite{kimbell_2022_ChallengesIdentifying_CommunMater,wang_2022_TopologicalHall_ProgressInMaterialsScience} we will see the sign change of the $\rho_{xy}^{THE}$ with changing $q_{2x}$. In Fig. \ref{fig:chern-spin} we plotted the dependence of $\rho_{xy}^{THE}$ on the azimuthal angle $q_{2x}$, while keeping $q_{1x}=\pi/4$ constant. One can observe that for $S=1/2$, $1$ there is no change in the sign of the $\rho_{xy}^{THE}$, hence, from Hall resistivity experiment it is not possible to detect the predicted effect. However for higher $S=5/2, 3, 7/2$ there are multiple sign change with increasing $q_{2x}$. Therefore for materials with high magnetic moments (F3GT, F4GT) are good experimental platforms to observe this effect.

\begin{figure}[tbh]
  \centering
  \includegraphics[width=0.4\textwidth]{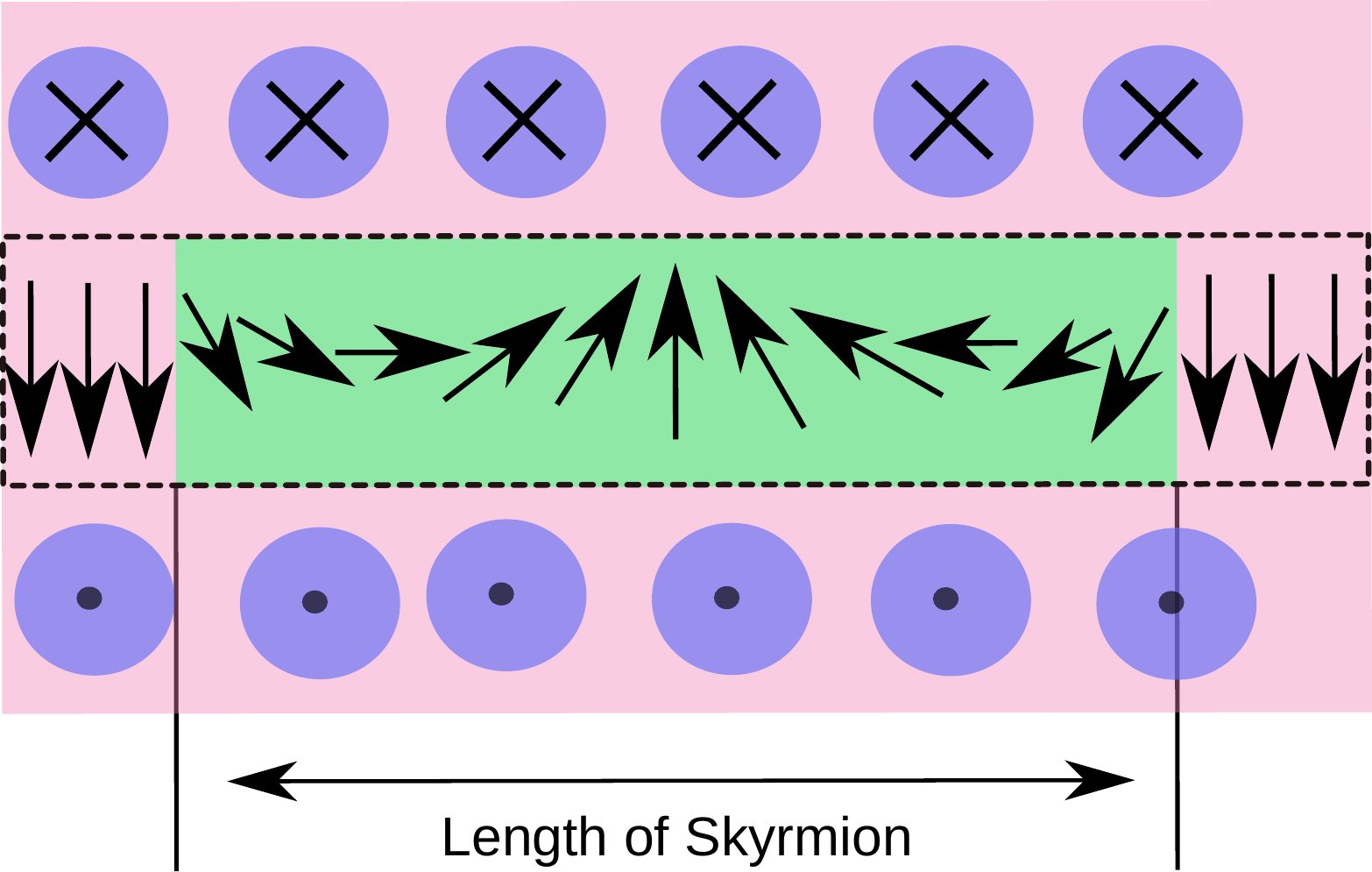}
  \caption{domain-wall Skyrmion spin texture. The $z$ projection of the spin on the walls are opposite to each other. The $x$ and $y$ projection of the spin along the length of the Skyrmions is defined by Eq. (\ref{eq:spin-config-int}) \cite[see Eq. (2) of Ref.][]{cheng_2019_MagneticDomain_PhysRevB}. The length of the Skyrmion which is proportional to the spin modulation vector $\vec{q}_{2}$ is controlled by DMI and PMA \cite{cheng_2019_MagneticDomain_PhysRevB}.} 
  \label{fig:dw-skx}
\end{figure}
Now the question arises, how one can control the modulation of the azimuthal vector $\vec{q}_{2}$. To generate such spin textures one need to use materials or hetero-structures where different magnetic energy scales, e.g. exchange interaction, perpendicular magnetic anisotropy (PMA), DMI, competes with each other; the competition of these energy scales gives rise to different exotic chiral spin textures \cite{jiang_2017_SkyrmionsMagnetic_PhysicsReports}. One of the promising way for verification of the Chern number dependence on $q_{2x}$ is through domain-wall Skyrmions \cite{cheng_2019_MagneticDomain_PhysRevB}. In these spin texture on the two sides of the domain-walls the $z$-component of the spin is directed along opposite direction, while inside the domain-wall the spin rotates on the $xy$ and $yz$ plane. The spin texture is shown in Fig. \ref{fig:dw-skx}. Inside the domain-walls the spin texture (only along the length of the Skyrmion) can be given by Eq. (\ref{eq:spin-config-int}) \cite[see Eq. (2) of Ref.][]{cheng_2019_MagneticDomain_PhysRevB}. In real materials the domain-wall Skyrmion spin texture has already been predicted in Janus monolayer of Cr-X$_{3}$ \cite{xu_2020_TopologicalSpin_PhysRevB}, Mn$_{3}$Sn \cite{li_2019_ChiralDomain_NatCommun}, chiral magnets \cite{nagase_2021_ObservationDomain_NatCommun,amari_2023_DomainwallSkyrmion_ArXivE-prints}. One can control the length of the Skyrmions --- which is proportional to the $\vec{q}_{2}$ --- by tuning the PMA and DMI in these materials \cite{cheng_2019_MagneticDomain_PhysRevB,lemesh_2017_AccurateModel_PhysRevB}. Several extrinsic and intrinsic methods are available for modification of the PMA and DMI \cite{kammerbauer_2023_DzyaloshinskiiMoriya_JPhysSocJpn,kurebayashi_2022_MagnetismSymmetry_NatRevPhys,rana_2019_MagnonicDevices_CommunPhys}. We propose a hetero-structure in which upper layer consists of Janus Cr-X$_{3}$ monolayer on the substrate for which the induced DMI can be controlled by extrinsic means of strain, voltage, or electric current \cite{kammerbauer_2023_DzyaloshinskiiMoriya_JPhysSocJpn}.

Another idea to control $\vec{q}_{2}$ is to have heterostructures with two perpendicular DMI vectors: (i) bulk DMI, (ii) interfacial DMI. The bulk DMI arises due to intrinsic broken bulk inversion symmetry ($r \to -r$), and the interfacial DMI occurs due to cosmetically broken mirror symmetry ($z \to -z$) at the interface of the heterostructures \cite{ahmed_2018_ChiralBobbers_PhysRevMater,rowland_2016_SkyrmionsChiral_PhysRevB}. The structure specific bulk DMI is hard to control, however, the interfacial DMI can be controlled by the extrinsic means \cite{schott_2021_ElectricField_JournalOfMagnetismAndMagneticMaterials,dai_2022_ReviewVoltagecontrolled_JournalOfMagnetismAndMagneticMaterials,nawaoka_2015_VoltageInduction_ApplPhysExpress,ma_2016_InterfacialControl_PhysRevB,yang_2018_ControllingDzyaloshinskiiMoriya_SciRep}. One can use the Janus vdW magnets, where the bulk inversion symmetry is broken naturally, as the upper layer \cite{jiang_2023_TwodimensionalMagnetic_MaterHoriz}. For substrate one can use the materials with strong spin orbit coupling, where electrical control of DMI is possible \cite{manchon_2019_CurrentinducedSpinorbit_RevModPhys}.

\section{conclusion}
\label{sec:conclusion}
In this work we analyzed the topological properties of the spin texture Eq. (\ref{eq:spin-config-int}) on a honeycomb lattice and strongly correlated materials. We showed that, the Chern number depends strongly on the azimuthal angle of the spin texture, on spin of the magnetic atoms $S$, and weakly on polar spin modulation vector. The model can be applied to the vdW magnets shown in Tab. \ref{tab:Mat}. We discuss some experimental setups for observing the effects.

\begin{acknowledgments}
The author will like to thank E. A. Kochetov and P. A. Maksimov for important discussions on the problem. The author acknowledges the financial support from the JINR grant for young scientists and specialists, the Foundation for the Advancement of Theoretical Physics and Mathematics ”Basis” for grant \# 23-1-4-63-1, RFBR grant No. 21-52-12027.
\end{acknowledgments}

\appendix

\section{Theory}
\label{sec:app-th}
We will work in the strong correlation regime in which the underlying Hilbert space is modified as the double occupancy is prohibited. It results in constrained electrons operators which are now isomorphic to the Hubbard operators \cite{wiegmann_1988_SuperconductivityStrongly_PhysRevLett}. Those operators appear as a generators of the \emph{su(2|1)} superalgebras. As a result charge and spin degrees of freedom can be represented as product of the \emph{SU(2|1)} supergroup. This is applicable for particles described by the $S=1/2$, however, for arbitrary spin $S>1/2$ one will use the $su(2)$ algebra. In the $SU(2)$ formalism the required theory is constructed under the condition that the background spin field affects the fermion hopping without breaking the global symmetry. The $su(2)$ coherent states (CS) have these properties, as the electron hopping factor is affected only due to CS overlap factors \cite{shankar_1990_HolesQuantum_NuclearPhysicsB,zhang_1990_CoherentStates_RevModPhys}. In condensed matter it is analogous to the Peierls phase factor generated in an external magnetic field. It is the vector potential generated by the non-collinear chiral spin textures \cite{nagaosa_2013_TopologicalProperties_NatureNanotech}. Physically it can be thought of fictitious magnetic field through a plaquete. In field theory it is the same as the emergent artificial gauge field generated by the $U(1)$ local connection one-form of the spin $U(1)$ complex line bundle. It provides a covariant (geometric) quantization of a spin \cite{stone_1989_SupersymmetryQuantum_NuclearPhysicsB}. In this approach the underlying base space appears as a classical spin phase space. It is a two sphere $S^{2}$ which can be mapped to a complex projective space $CP^{1}$, endowed with a set of local coordinates $\left( z,\bar{z} \right)$. In this case quantum spin is represented as the section $\ket{z}$ of the principle (monopole) line bundle $P \left( CP^{1},U(1) \right)$. The local connection of the bundle is $a^{(0)} = i \bra{z}d\ket{z}$; $d$ is the exterior derivative.

Physically, we start from the lattice Kondo-model:
\begin{equation}
  \label{eq:ham-kondo}
  \begin{aligned}
    H=-\sum_{ij\sigma}&[t_{ij}+J \: S(S+1)\:\delta_{ij}]c^{\dagger}_{i\sigma}c_{j\sigma}\\
                      &+J\sum_i \hat{\mathbf{S}}_i\cdot(c^{\dagger}_{i\sigma}\vec{\sigma}_{\sigma\sigma'}c_{i\sigma'}).
  \end{aligned}
\end{equation}
Here $c^{\dagger}_{i\sigma}$ $\left( c_{i\sigma} \right)$ is the electron creation (annihilation) operator with the spin $\sigma$ on site $i$; $J>0$ is the exchange coupling constant; $\vec{\sigma}$ is the vector of the \emph{Pauli} spin matrices; $\hat{\mathbf{S}}_{i}$ is the nuclear spin operator at \emph{i}-th site. The extra $J$ dependent term $ J \: S(S+1) \:\delta_{ij} $ introduced in the hopping parameter is to make sure a finite $J \to \infty$ limit \cite{ivantsov_2022_StrongCorrelation_AnnalsOfPhysics}. Under the mean field approximation one can represent the nuclear spin operator as product of the localized spin magnitude ($S$) and their direction ($\vec{n}_{i}$):  $\left\langle \mathbf{\hat{S}}_{i} \right\rangle = S \cdot \vec{n}_{i}$. In the large Kondo limit $J \to \infty$ the Hubbard model goes into the $U \to \infty$ limit (strongly correlated electronic system) \cite{ivantsov_2022_StrongCorrelation_AnnalsOfPhysics}:
\begin{equation}
  \label{eq:Ham-U-infty}
  H \approx - \sum\limits_{i,j,\sigma}t_{ij} \bar{c}_{i\sigma}^{\dagger} \bar{c}_{j\sigma}.
\end{equation}
Here $\bar{c}_{i\sigma} = c_{i \sigma} (1-n_{i \bar{\sigma}})$ is the constrained electron operator; $n_{i \bar{\sigma}} = c_{i \bar{\sigma}}^{\dagger} c_{i \bar{\sigma}}$ is the number operator of the complementary spin. The constraint operator as explained above can be dynamically factorized into the spinless charge fermionic field $f_{i}$ (holons) and spinfull bosonic $z_{i}$ fields (spinons) \cite{ferraz_2022_FractionalizationStrongly_PhysRevB,kesharpu_2023_TopologicalHall_PhysRevB}. It can be seen that as long as the fermionic field satisfy the condition $f_{i}^{2} \equiv 0$, the local no double occupancy of strongly correlated electron is satisfied rigorously. Here, the holons acquire the band structure of their own; it is usual behavior for fractionalized electrons \cite{maciejko_2015_FractionalizedTopological_NaturePhys}. The spinons are handled by mean-field treatment.

The necessary theory was given by the authors recently \cite{kesharpu_2023_TopologicalHall_PhysRevB}, hence, here we briefly derive the required Hamiltonian. The high spin CS theory is constructed from the fundamental $S=1/2$ representation:
\begin{equation}
  \label{eq:z-section}
  \ket{z} = \left( 1 + \left| z \right|^{2} \right) ^{-S} \mathrm{e}^{z \hat{S}^{-}} \ket{S}.
\end{equation}
Here $\ket{S}$ is the highest spin-\emph{S} $su(2)$ state; $\hat{S}^{-}$ is the spin lowering operator. The \emph{S}-dependent partition function will be:
\begin{equation}
  \label{eq:part-func}
  Z = \int D \mu \left( z, f \right) \:\: \exp \mathcal{A}.
\end{equation}
The measure $D \mu \left( z, f \right)$ is:
\begin{equation}
  \label{eq:measure}
  D \mu \left( z, f \right) = \prod\limits_{i,t} \frac{S}{\pi i } \frac{d \bar{z}_{i}\left( t \right) d \bar{z}_{i}\left( t \right)}{\left( 1 + \left| z_{i} \right|^{2} \right)^{2}} d\bar{f}_{i}(t) d f_{i}(t).
\end{equation}
Here $z_{i}$ keeps track of the spin and complex, while $f_{i}$ keeps track of charge and is a \emph{Grassman} variable. The effective action $\mathcal{A}$ is defined as:
\begin{equation}
  \label{eq:effective-action}
  \mathcal{A} = \sum\limits_{i} \int\limits_0^{\beta} \left[ i a_{i}^{(0)} - \bar{f}_{i} \left( \partial_{t} + i a_{i}^{(0)} \right) f_{i}  \right] dt - \int\limits_0^{\beta} H \: dt.
\end{equation}
Here, $i a_{i}^{(0)}$ is the $u(1)$-valued connection one-form of the magnetic monopole bundle as a \emph{spin kinetic} term:
\begin{equation}
  \label{eq:spin-kin-term}
  i a_{i}^{(0)} = - \bra{z} \partial_{t} \ket{z} = S \frac{\dot{\bar{z}} z - \bar{z} \dot{z}}{1 + \left| z \right|^{2}}.
\end{equation}
It is analogous to the Berry connection. The Hamiltonian in $\mathcal{A}$ can be written as:
\begin{equation}
  \label{eq:hamiltonian-effect-action}
  H = - \sum\limits_{ij} t_{ij} \bar{f}_{i} f_{j} \mathrm{e}^{i a_{ji}} + H.c. + \mu \sum\limits_{i} \bar{f}_{i} f_{i},
\end{equation}
where,
\begin{equation}
  \label{eq:a-ij}
  a_{ij} = -i \log \bra{z_{i}}\ket{z_{j}}, \: \bra{z_{i}}\ket{z_{j}} = \frac{\left( 1 + \bar{z}_{i} z_{j} \right)^{2S}}{\left( 1 + \left|z_{j} \right|^{2} \right)^{S} \left( 1 + \left|z_{i} \right|^{2} \right)^{S}} \: .
\end{equation}
Under a \emph{global SU(2)} rotation
\begin{equation}
  \label{eq:su-2-rot}
  z_{i} \to \frac{\quad u z_{i} + v }{- \bar{v} z_{i} + \bar{u}}
\end{equation}
the phase will be:
\begin{equation}
  \label{eq:phase}
  a_{i}^{(0)} \to a_{i}^{(0)} - \partial_{t} \theta_{i}, \quad a_{ij} \to a_{ij} + \theta_{j} - \theta_{i}.
\end{equation}
Here,
\begin{equation}
  \label{eq:theta-su-2}
  \theta_{i} = i S \log \left( \frac{v \bar{z}_{i} + u}{\bar{v} z_{i} + \bar{u}} \right); \quad
  \begin{bmatrix}
    \:\: \: u & v \\
    -\bar{v}  &\bar{u}
  \end{bmatrix}
  \in SU(2).
\end{equation}
The effective action $\mathcal{A}$ remains unchanged under $SU(2)$ transformation of the $z_{i}$ in conjunction with $U(1)$ transformation of the $f_{i} \to \mathrm{e}^{i \theta_{i}} f_{i}$. The real and imaginary part of the $a_{ji}$ is defined as:
\begin{equation}
  \label{eq:a-ji-re-im}
  a_{ji} = \phi_{ji} + i \chi_{ji}; \: \phi_{ji}=\bar{\phi}_{ji}; \: \chi_{ji}=\bar{\chi}_{ji}.
\end{equation}
The $\phi_{ji}$ and $\chi_{ji}$ are defined as
\begin{equation}
  \label{eq:phi-chi-ji}
  \begin{aligned}
    &\phi_{ji} &&=iS \log \frac{1 + \bar{z}_{i} z_{j}}{1+ \bar{z}_{j} z_{i}}\\
    &          &&=iS \log \frac{\left( S+S_{i}^{z} \right)\left( S+S_{j}^{z} \right)+S_{i}^{-} S_{j}^{+}}{\left( S+S_{i}^{z} \right)\left( S+S_{j}^{z} \right)+S_{j}^{-} S_{i}^{+}};\\
    &\chi_{ji} &&= -S \log \frac{\left( 1+\bar{z}_{i}z_{j} \right)\left( 1+\bar{z}_{j}z_{i} \right)}{\left( 1+|z_{i}|^{2} \right) \left( 1+|z_{j}|^{2} \right)}\\
    &   &&=- S \log \left( \frac{\vec{S}_{i} \cdot \vec{S}_{j}}{2S^{2}} + \frac{1}{2} \right).
  \end{aligned}
\end{equation}
Here, $\vec{S}_{i}$ stands for the CS symbols of the $su(2)$ generators \cite{ferraz_2022_FractionalizationStrongly_PhysRevB}. The corresponding values are:
\begin{equation}
  \label{eq:CS-symb}
  \begin{aligned}
    &S_{i}^{z} = \frac{1}{2} \frac{1-|z|^{2}}{1+|z|^{2}}, \: S_{i}^{+} = \frac{z}{1+|z|^{2}}, \: S_{i}^{-} = \frac{\bar{z}}{1+|z|^{2}}.
  \end{aligned}
\end{equation}
There is a one-to-one correspondence between the $su(2)$ generators and their CS symbols~\cite{berezin_1987_IntroductionSuperanalysis_}. Under $SU(2)$ \emph{global} rotation $\chi_{ji}$ remains intact, however, $\phi_{ji}$ transforms as:
\begin{equation}
  \label{eq:su-2-trans-phi}
  \phi_{ji} \to \phi_{ji} + \theta_{i} - \theta_{j}.
\end{equation}
This transformation is analogous to gauge fixing by choosing a specific rotational covariant frame. The dynamical fluxes do not depend on the chosen covariant frame. Substituting Eq. \eqref{eq:phi-chi-ji} in the dynamical Hamiltonian Eq. \eqref{eq:hamiltonian-effect-action} we get:
\begin{equation}
  \label{eq:gen-ham}
  \begin{aligned}
    H = - \sum\limits_{ i,j } t_{ij} \bar{f}_{i} f_{j} \mathrm{e}^{i \phi_{ji}} \left( \frac{\vec{S}_{i} \cdot \vec{S_{j}}}{2 S^{2}} + \frac{1}{2} \right)^{S} + \mu \sum\limits_{i} \bar{f}_{i} f_{i}.
  \end{aligned}
\end{equation}
Physically, this represents the interaction of the underlying spin field and itinerant spinless fermions.

\section{Solution of Hamiltonian on a bipartite lattice}
\label{sec:solut-hamilt-bipartr}
We use the Hamiltonian Eq. (\ref{eq:gen-ham}) with the spin texture Eq. (\ref{eq:spin-config-int}) to analyze the topological properties of the two band materials, as they are the simplest one. We take a bipartite lattice \emph{L} with two sub-lattices \emph{A} and \emph{B}: $L=A \oplus B$. The charge and spin degrees of freedom on \emph{A} sub-lattice are $f_{i}$ and $z_{i}$ respectively. For convenience on sub-lattice \emph{B} we can define:
\begin{equation}
  \label{eq:deg-of-free-B}
  f_{i} \to f_{i} \mathrm{e}^{i \theta_{i}^{(0)}}, \quad z_{i} \to - \frac{1}{\bar{z}_{i}}; \qquad i \in B.
\end{equation}
Where $\theta_{i}^{(0)} \equiv \theta_{i} |_{u=0,v=1}$. Under these transformations the $\chi_{ji}$ remains unchanged, while the $\phi_{ji} \to \phi_{ij} + \theta_{j}^{(0)} - \theta_{i}^{(0)}$. This transformation makes the calculation easier, while conserving the form of $a_{ij}$ in Eq. (\ref{eq:phase}) under \emph{global} $SU(2)$ rotation. The coherent states symbols in Eq. (\ref{eq:CS-symb}) also changes from $\vec{S}_{i} \to - \vec{S}_{i}$. For bipartite lattice the total Hamiltonian will contain three parts depending on hopping of electrons: (i) $A \to A$, (ii) $B \to B$, (iii) $A \to B$. For $A \to A$ ($i,j \in A$) the $\chi_{ji}$ is:
\begin{equation}
  \label{eq:chi-A}
  \begin{aligned}
    \mathrm{e}^{\chi_{ij}} &= \mathscr{w}_{ij}^{S}\left[ 1 + \mathscr{g}_{ij} \cos \vec{q}_{1}\left( 2\vec{r}_{i} + \vec{r}_{ij}\right)\right]^{S};\\
    \\
    \mathscr{w}_{ij} &\equiv \frac{1}{2}+ \left(\frac{1}{4} + \frac{\cos \vec{q}_{2}\vec{r}_{ij}}{4} \right) \cos \vec{q}_{1}\vec{r},\\
    \mathscr{g}_{ij} &\equiv \left[\left( \frac{1}{4} - \frac{\cos \vec{q}_{2}\vec{r}_{ij}}{4}\right) \bigg/ \mathscr{w}_{ij} \right].\\
  \end{aligned}
\end{equation}
Here we defined $\vec{r}_{ij} = \vec{r}_{j} - \vec{r}_{i}$. The values of $\mathscr{w}_{ij}$ and $\mathscr{g}_{ij}$ are bounded $\in \left[ 0,1 \right]$, and constant for a given $\vec{q}_{1}$ and $\vec{q}_{2}$. The $\phi_{ji}$ is:
\begin{equation}
  \label{eq:phi-A}
  \begin{aligned}
    &\phi_{ji} = 2S \atan \left[ \frac{1}{\mathscr{h}(r) \csc \vec{q}_{2}\vec{r}_{ij} + \cot \vec{q}_{2}\vec{r}_{ij}} \right];
    \\
    & \mathscr{h} (r) \equiv\\
    & \frac{ 2 + \cos \vec{q}_{1}\vec{r}_{ij} + \left[4 \cos \frac{\vec{q}_{1} \vec{r}_{ij}}{2} + 1 \right] \: \cos 2 \vec{q}_{1} \left(\vec{r}_{i} + \frac{\vec{r}_{ij}}{2}\right) }{\cos \vec{q}_{1}\vec{r}_{ij} - \cos 2\vec{q}_{1}\left(\vec{r}_{i} + \frac{\vec{r}_{ij}}{2}\right)}.
  \end{aligned}
\end{equation}
The $\phi_{ji}$ is a periodic, but bounded function. $\mathscr{h}(r)$ is also a periodic function which depends only on the polar angle $\vec{q}_{1}$. For $B \to B$ the $\chi_{ij}$ remains same. However, the $\phi_{ji}\big|_{i,j \in B} \to -\phi_{ji} \big|_{i,j \in A} + 2S\vec{q}_{2}\vec{r}_{ij}$ \footnote{
  On \emph{B} sublattice $S_{i} \to -S_{i}$. Under this transformation the spin part of the $\phi_{ji}$ in Eq. (\ref{eq:phi-chi-ji}) can be written as:
  \begin{equation*}
    \label{eq:fn:phi}
    \begin{aligned}
      &\frac{\left( S - S_{i}^{z} \right)\left( S - S_{j}^{z} \right)+S_{i}^{-} S_{j}^{+}}{\left( S - S_{i}^{z} \right)\left( S - S_{j}^{z} \right)+S_{j}^{-} S_{i}^{+}}\\
      & \qquad= \frac{\left( S+S_{i}^{z} \right)\left( S+S_{j}^{z} \right)+S_{j}^{-} S_{i}^{+}}{\left( S+S_{i}^{z} \right)\left( S+S_{j}^{z} \right)+S_{i}^{-} S_{j}^{+}} \cdot \frac{S_{j}^{+} S_{i}^{-}}{S_{i}^{+}S_{j}^{-}}.
    \end{aligned}
  \end{equation*}
}. Explicitly,
\begin{equation}
  \label{eq:phi-B}
  \begin{aligned}
    &\phi_{ji} = 2S \atan \left[ \frac{1}{\mathscr{h}(r) \csc \vec{q}_{2}\vec{r}_{ij} + \cot \vec{q}_{2}\vec{r}_{ij}} \right] + 2S \vec{q}_{2} \vec{r}_{ij}.
  \end{aligned}
\end{equation}
For $A \to B$ the $\chi_{ji}$ is:
\begin{equation}
  \label{eq:chi-AB}
  \begin{aligned}
    \mathrm{e}^{\chi_{ij}} &= \mathscr{w}_{ij}'^{S}\left[ 1 - \mathscr{g}_{ij}' \cos 2\vec{q}_{1}\left(\vec{r}_{i} + \frac{\vec{r}_{ij}}{2}\right)\right]^{S};\\
    \mathscr{w}_{ij}' &\equiv \frac{1}{2} - \left(\frac{1}{4} + \frac{\cos \vec{q}_{2}\vec{r}_{ij}}{4} \right) \cos \vec{q}_{1}\vec{r}_{ij};
    \\
    \mathscr{g}_{ij}' &\equiv \left[\left( \frac{1}{4} - \frac{\cos \vec{q}_{2}\vec{r}_{ij}}{4}\right) \bigg / \mathscr{w}'_{ij} \right],\\
  \end{aligned}
\end{equation}
and the phase 
\begin{equation}
  \label{eq:phi-AB}
  \begin{aligned}
    \mathrm{e}^{i \phi_{ji}} = e^{iS \left( \vec{q_{2}} \: \vec{r}_{ij}  - \pi\right)}.
  \end{aligned}
\end{equation}
Substituting the above derived $\phi_{ji}$ and $\chi_{ij}$ in Eq. (\ref{eq:gen-ham}), and assuming the inter-lattice hopping parameter ($t_{1}$) and intra-lattice hopping parameter ($t_{2}$) we find the total Hamiltonian:
\begin{equation}
  \label{eq:gen-Ham-skyrm}
  \begin{aligned}
    &H = \\
    &-t_{2} \sum\limits_{i,j \in A} \bar{f}_{i} f_{j} \mathscr{w}_{ij}^{S}
      \left[ 1 + \mathscr{g}_{ij} \cos 2 \vec{q}_{1}\left(\vec{r}_{i} + \frac{\vec{r}_{ij}}{2}\right) \right]^{S}\\
    &\qquad \exp \left\{+ iS \left[  2 \atan \left[ \frac{1}{\mathscr{h}(r) \csc \vec{q}_{2}\vec{r}_{ij} + \cot \vec{q}_{2}\vec{r}_{ij}} \right]  -\vec{q}_{2}\vec{r}_{ij}\right] \right\}\\
    &-t_{2} \sum\limits_{i,j \in B} \bar{f}_{i} f_{j}  \mathscr{w}_{ij}^{S}
      \left[ 1 + \mathscr{g}_{ij} \cos 2\vec{q}_{1}\left( \vec{r}_{i} + \frac{\vec{r}_{ij}}{2}\right) \right]^{S}\\
    &\qquad \exp \left\{- iS \left[  2 \atan \left[ \frac{1}{\mathscr{h}(r) \csc \vec{q}_{2}\vec{r}_{ij} + \cot \vec{q}_{2}\vec{r}_{ij}} \right]  -\vec{q}_{2}\vec{r}_{ij}\right] \right\}\\
    &+t_{1} \sum\limits_{\substack{i \in A \\ j \in B}} \bar{f}_{i} f_{j}  \mathscr{w}_{ij}'^{S} \left[ 1 - \mathscr{g}_{ij}' \cos 2 \vec{q}_{1} \left(\vec{r}_{i} + \frac{\vec{r}_{ij}}{2}\right) \right]^{S}.
  \end{aligned}
\end{equation}
The first two terms --- corresponding to the hopping either only on sub-lattice \emph{A}, or on sub-lattice \emph{B} respectively --- are complex conjugate of each other. The third term corresponding to the hopping between sub-lattice \emph{A} and \emph{B} does not contain the imaginary part. Hence, the Hamiltonian and its complex conjugate are not identical to each other, which breaks the time reversal symmetry. At $\vec{q}_{1}=0$ or at $\vec{q}_{2}=0$ the $\phi_{ji}$ in Eq. (\ref{eq:phi-A}) and (\ref{eq:phi-B}) collapses. In this case there won't be any topological properties. It was expected as planar spin textures does not show any THE. 

\begin{figure}[tbh]
  \centering
  \subfloat[][]{\includegraphics[width=0.25\textwidth]{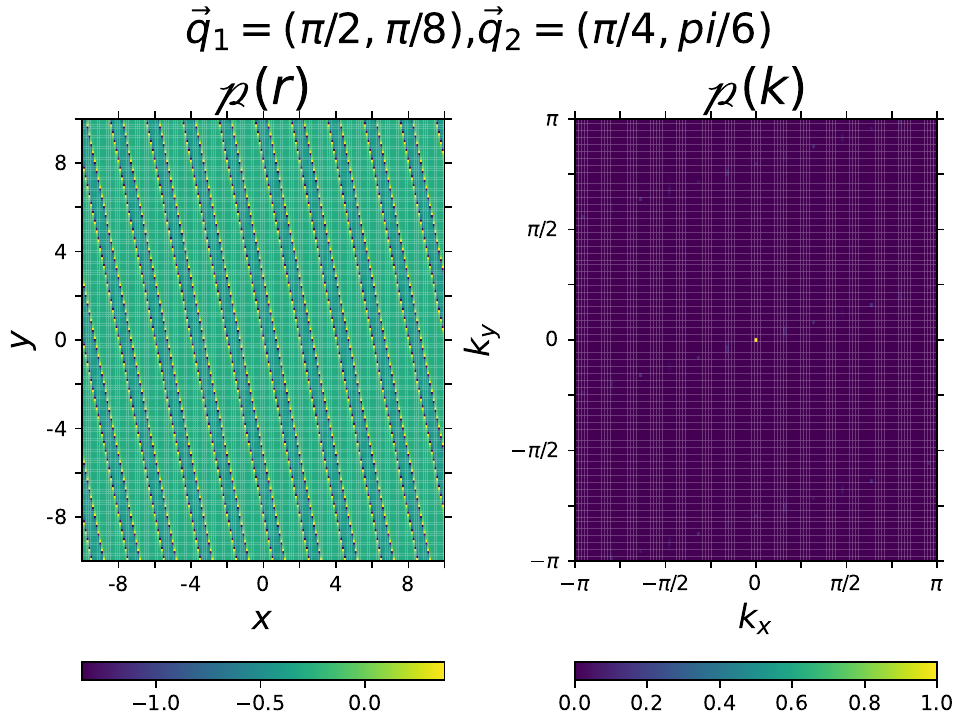}\label{fig:pk1}}
  \subfloat[][]{\includegraphics[width=0.25\textwidth]{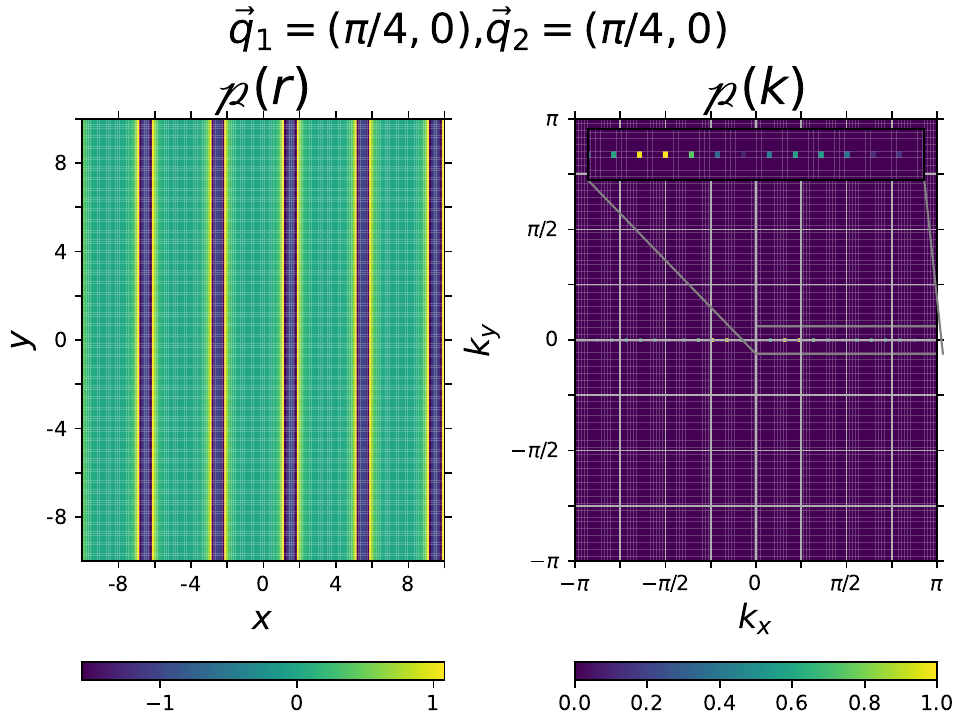}\label{fig:pk2}}\\
  \subfloat[][]{\includegraphics[width=0.25\textwidth]{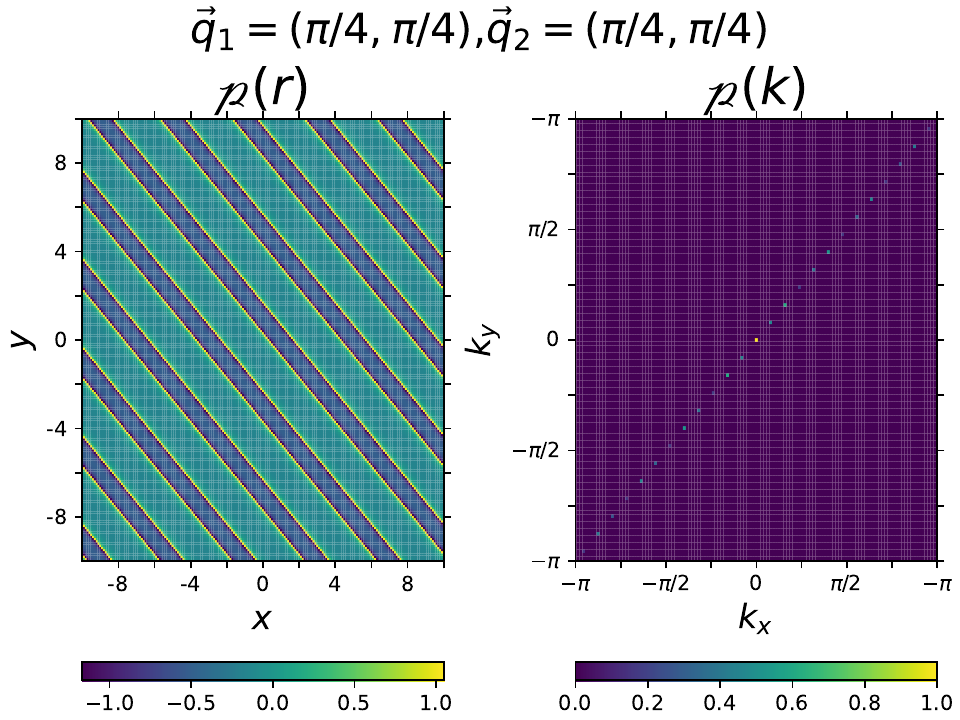}\label{fig:pk3}}
  \subfloat[][]{\includegraphics[width=0.25\textwidth]{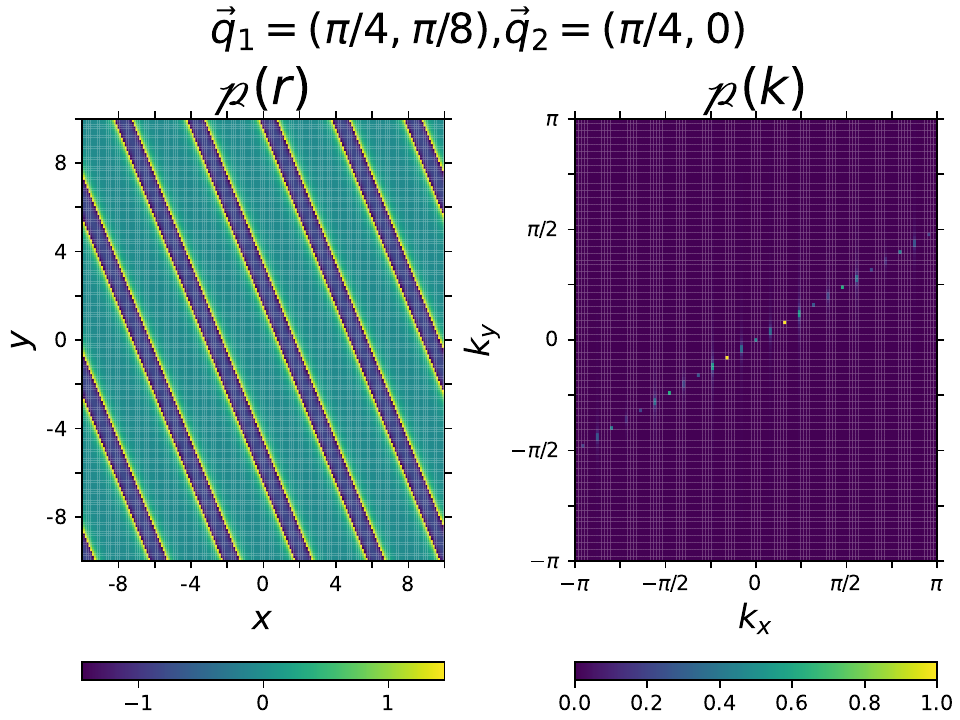}\label{fig:pk4}}
  \caption{Real space spin texture generated from Eq. (\ref{eq:spin-config-int}) for different $\vec{q}_{1}$ and $\vec{q}_{2}$, and their corresponding Fourier transformation.}
  \label{fig:p-k}
\end{figure}
In momentum space the two band Hamiltonian can be written as:
\begin{equation}
  \label{eq:cas-1-ham-k-symb}
  H(\vec{k}) = \sum\limits_{\vec{k}} \bar{\psi}_{\vec{k}} \mathcal{H}(\vec{k}) \psi_{\vec{k}}. 
\end{equation}
The wave vector $\vec{k}$ is taken over the first Brillouin zone. The matrix $\psi_{\vec{k}} = \left[ f_{k,A} f_{k,B}\right]$ contains the creation operators of the $\vec{k}$-th momentum on the \emph{A} and \emph{B} sub-lattices. The single mode kernel of the Hamiltonian is $\mathcal{H}(\vec{k})=\mathcal{H}_{0} (\vec{k})\cdot \mathcal{I} + \mathcal{H}_{i}(\vec{k}) \cdot \vec{\sigma}_{i}$. Here, $\mathcal{I}$ is the unit matrix; $\vec{\sigma}_{i}$ are the Pauli matrices, and $\mathcal{H}_{i}(\vec{k})$ are the corresponding kernels
\footnote{$\mathcal{H}(\vec{k})$ is a $2 \times 2$ matrix. In terms of \emph{Pauli matrices} it is represented as
  \begin{equation*}
    \label{eq:pauli-mat}
    \begin{aligned}
      &\mathcal{H}(\vec{k}) = \mathcal{H}_{0} \mathcal{I} + \mathcal{H}_x (\vec{k}) \sigma_{x} + \mathcal{H}_y (\vec{k}) \sigma_{y} +\mathcal{H}_z (\vec{k}) \sigma_{z},\\
      &\text{where,}\\
      &\mathcal{H}_{0}(\vec{k}) = \frac{H_{i,j \in A} + H_{i,j \in B}}{2}, \mathcal{H}_{x}(\vec{k}) = \Re \left[ H_{i \in A, j \in B} \right],\\
      &\mathcal{H}_{z}(\vec{k})=\frac{H_{i,j \in A} - H_{i,j \in B}}{2},  \mathcal{H}_{y}(\vec{k}) = \Im \left[ H_{i \in A, j \in B} \right].
    \end{aligned}
  \end{equation*}
  Here, $\mathcal{I}$ is the $2 \times 2$ unit matrix; $\sigma_{x}$, $\sigma_{y}$, and $\sigma_{z}$ are the Pauli matrices.
}.
To find an analytical formulation of $\mathcal{H}(\vec{k})$ we first represent:
\begin{equation}
  \label{eq:approx-atan-first}
  \atan \left[ \frac{1}{\mathscr{h}(r) \csc \vec{q}_{2}\vec{r}_{ij} + \cot \vec{q}_{2}\vec{r}_{ij}} \right] = \sum\limits_{k'} \mathscr{p}(k') \mathrm{e}^{-i k'r};
\end{equation}
it is the Fourier series representation. As the value of $\mathscr{p}(k') \ll 1$ for most of the $\vec{q}_{1},\vec{q}_2$ as shown in Fig. \ref{fig:p-k}, we can approximate
\begin{equation}
  \label{eq:approx-atan}
  \begin{aligned}
    \exp& \left\{i \: 2S \atan \left[ \frac{1}{\mathscr{h}(r) \csc \vec{q}_{2}\vec{r}_{ij} + \cot \vec{q}_{2}\vec{r}_{ij}} \right] \right\}\\
        & \qquad \approx 1 + i 2 S \sum\limits_{k'} \mathscr{p}(k') \mathrm{e}^{-i k' r}.
  \end{aligned}
\end{equation}
Using this the kenel $\mathcal{H}(\vec{k})$ can be written as:
\small
\begin{equation}
  \label{eq:kerner-ham-app}
  \begin{aligned}
    &\mathcal{H}_{0} = -2 t_{2} \mathscr{w}_{ij}^{S} \: \hat{\mathcal{F}}\left[ 1+ \mathscr{g}_{ij} \cos 2 q_{1}\left( \vec{r}_{i} + \frac{\vec{r}_{ij}}{2}\right) \right]^{S} \ast \\
    & \left\{ \cos S \vec{q}_{2} \vec{r}_{ij} \cos \vec{k}\vec{r}_{ij}  + 2S \sin S\vec{q}_{2}\vec{r}_{ij} \sum\limits_{k'}\mathscr{p}(k') \cos \left( \vec{k} + \vec{k}' \right) \vec{r}_{ij} \right\}\\
    &\mathcal{H}_{x} = +t_{1} \mathscr{w}_{ij}'^{S} \: \hat{\mathcal{F}}\left[ 1 - \mathscr{g}_{ij}' \cos 2 q_{1}\left( \vec{r}_{i} + \frac{\vec{r}_{ij}}{2}\right) \right]^{S} \ast \cos \vec{k} \vec{r}_{ij}\\
    &\mathcal{H}_{y} = +t_{1} \mathscr{w}_{ij}'^{S} \: \hat{\mathcal{F}}\left[ 1 - \mathscr{g}_{ij}' \cos 2 q_{1}\left( \vec{r}_{i} + \frac{\vec{r}_{ij}}{2}\right) \right]^{S} \ast \sin \vec{k} \vec{r}_{ij}\\
    &\mathcal{H}_{z} = -2 t_{2} \mathscr{w}_{ij}^{S} \: \hat{\mathcal{F}}\left[ 1+ \mathscr{g}_{ij} \cos 2 q_{1}\left( \vec{r}_{i} + \frac{\vec{r}_{ij}}{2}\right) \right]^{S} \ast\\
    & \left\{- \sin S \vec{q}_{2} \vec{r}_{ij} \sin \vec{k}\vec{r}_{ij} + 2S \cos S\vec{q}_{2}\vec{r}_{ij} \sum\limits_{k'}\mathscr{p}(k') \sin \left( \vec{k} + \vec{k}' \right) \vec{r}_{ij} \right\}.
  \end{aligned}
\end{equation}
\normalsize
Here, $\hat{\mathcal{F}}$ represents the Fourier transform operator; <<$\ast$>> is the convolution operator.

\section{Hamiltonian for higher integer spins $S=2,3,4,\dots$}
\label{sec:hamilt-high-integ}

\begin{figure}[tbh]
  \centering
  \includegraphics[width=0.44\textwidth]{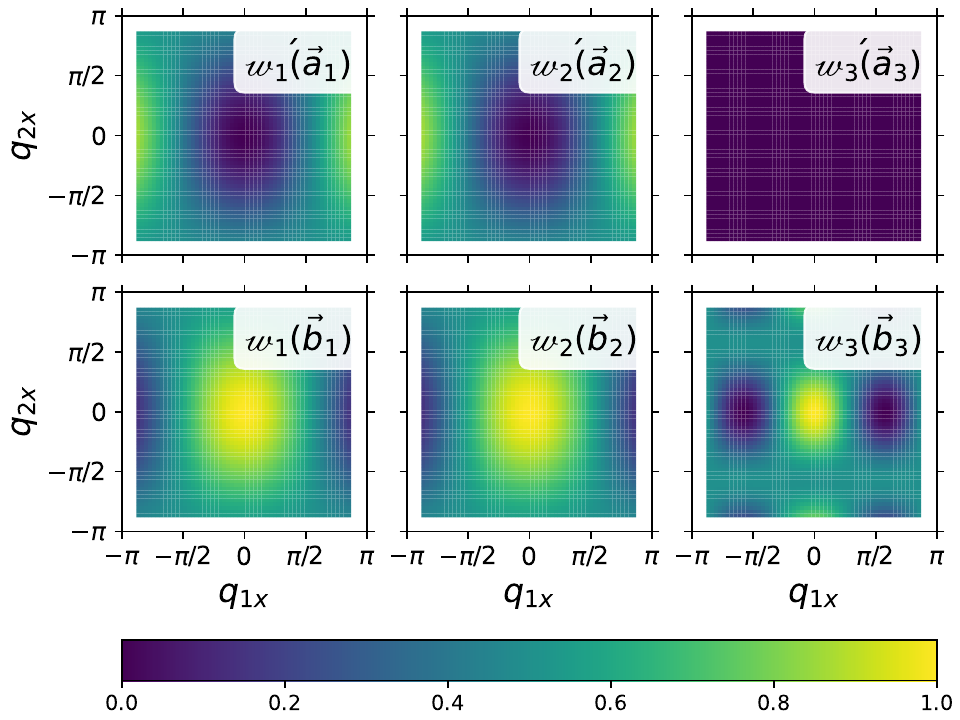}
  \caption{Dependence of the $\mathscr{w}_{n}'$ and $\mathscr{w}_{n}$ on the wave vectors $\vec{q}_{1}=\left(2 q_{1x}/\sqrt{3}, 0 \right)$ and $\vec{q}_{2}=\left( 2q_{2x}/\sqrt{3}, 0 \right)$ on a honeycomb bipartite lattice. The value of $q_{1x}$ and $q_{2x}$ changes from $\sqrt{3}\pi/2$ to $-\sqrt{3}\pi/2$.}
  \label{fig:w_w_dash}
\end{figure}
Here we analyze the Hamiltonian, Eq. (\ref{eq:kerner-ham-app}) (in main text Eq. (\ref{eq:kerner-ham}), for higher integer spins $S=2,3,\dots$. In Eq. (\ref{eq:kerner-ham-app}) only four terms depends on spin:
\small
\begin{equation}
  \label{eq:spin-dep-gn}
  \begin{aligned}
    \left[ 1+ \mathscr{g}_{n} \cos 2 \vec{q}_{1}\left( \vec{r}_{i} + \frac{\vec{b}_{n}}{2}\right) \right]^{S},
    \left[ 1- \mathscr{g}_{n}' \cos 2 \vec{q}_{1}\left( \vec{r}_{i} + \frac{\vec{a}_{n}}{2}\right) \right]^{S},\\
  \end{aligned}
\end{equation}
\normalsize
$\mathscr{w}_{n}^{S}$, and $\mathscr{w}_{n}'^{S}$. The values of $\mathscr{w}_{n}$ and $\mathscr{w}_{n}'$ depend only on $\vec{q}_{1}$ and $\vec{q}_{2}$. In Fig. \ref{fig:w_w_dash} we plot the dependence of the $\mathscr{w}_{n}$ and $\mathscr{w}_{n}'$ on $\vec{q}_{1}=\left( q_{1x},0 \right)$ and $\vec{q}_{2}=\left( q_{2x},0 \right)$ for honeycomb lattice NN and NNN vectors; here $q_{1x}=q_{2x} \in \left( -\sqrt{3}\pi/2, \sqrt{3}\pi/2\right)$. The same dependence for $\mathscr{g}_{n}$ and $\mathscr{g}_{n}'$ is shown in Fig. \ref{fig:g_g_dash}. We observe that for all the cases the values of $\mathscr{w}_{n}$, $\mathscr{w}_{n}'$, $\mathscr{g}_{n}$, and $\mathscr{g}_{n}'$ are always smaller than unity for most values of $q_{1x}$ and $q_{2x}$. It is equal to unity only at some specific values of $q_{1x}$ and $q_{2x}$.

As a first step, to find the expression for Hamiltonian, one needs to expand the terms containing power of $S$ in Eq. (\ref{eq:spin-dep-gn}) using binomial theorem:
\small
\begin{equation}
  \label{eq:bin-thm}
  \begin{aligned}
    &\left[ 1+ \mathscr{g}_{n} \cos 2 \vec{q}_{1}\left( \vec{r}_{i} + \frac{\vec{b}_{n}}{2}\right) \right]^{S} =\\
    & 1 + \begin{pmatrix} S \\ 1 \end{pmatrix} \mathscr{g}_{n} \cos 2 \vec{q}_{1}\left( \vec{r}_{i} +\frac{\vec{b}_{n}}{2}\right) + \begin{pmatrix} S \\ 2 \end{pmatrix} \mathscr{g}_{n}^{2} \cos^{2} 2 \vec{q}_{1}\left( \vec{r}_{i} + \frac{\vec{b}_{n}}{2}\right) + \dots;\\
    \\
    &\left[ 1- \mathscr{g}_{n}' \cos 2 \vec{q}_{1}\left( \vec{r}_{i} + \frac{\vec{a}_{n}}{2}\right) \right]^{S}=\\
    & 1 - \begin{pmatrix} S \\ 1 \end{pmatrix} \mathscr{g}_{n}' \cos 2 \vec{q}_{1}\left( \vec{r}_{i} +\frac{\vec{b}_{n}}{2}\right) + \begin{pmatrix} S \\ 2 \end{pmatrix} \mathscr{g}_{n}'^{2} \cos^{2} 2 \vec{q}_{1}\left( \vec{r}_{i} + \frac{\vec{b}_{n}}{2}\right) - \dots.\\    
  \end{aligned}
\end{equation}
\normalsize
We observe that the expansion contains the higher powers of $\mathscr{g}_{n}'$ and $\mathscr{g}_{n}$. As the values of $\mathscr{g}_{n}'$ and $\mathscr{g}_{n}$ are never greater than unity [see Fig. \ref{fig:g_g_dash}], we can neglect their higher powers. Keeping this in mind one can approximate, Eq. (\ref{eq:kerner-ham}) to first order in $\mathscr{g}_{n}$ and $\mathscr{g}_{n}'$. The Fourier transform of the approximated Hamiltonian is:
\begin{equation}
  \label{eq:kerner-ham-k-space-large-s-app}
  \begin{aligned}
    &\mathcal{H}_{0} \approx -2 t_{2} \sum\limits_{n} \mathscr{w}_{n}^{S} \left[ 1+ \frac{S\mathscr{g}_{n}}{2} \cos 2 \vec{q}_{1} \vec{b}_{n} \right] \times \\
    & \left\{\cos S\vec{q}_{2} \vec{b}_{n} \cos \vec{k}\vec{b}_{n}  + 2S \sin \vec{q}_{2}\vec{b}_{n} \sum\limits_{\vec{k}'}\mathscr{p}(\vec{k}') \cos \left( \vec{k} + \vec{k}' \right) \vec{b}_{n} \right\}\\
    &\mathcal{H}_{x} \approx +t_{1} \sum\limits_{n} \mathscr{w}_{n}'^{S} \left[ 1 - \frac{S\mathscr{g}_{n}'}{2} \cos 2 \vec{q}_{1} \vec{a}_{n} \right] \times \cos \vec{k} \vec{a}_{n}\\
    &\mathcal{H}_{y} \approx +t_{1} \sum\limits_{n} \mathscr{w}_{n}'^{S} \left[ 1 - \frac{S\mathscr{g}_{n}'}{2} \cos 2 \vec{q}_{1}\vec{a}_{n}\right] \times \sin \vec{k} \vec{a}_{n}\\
    &\mathcal{H}_{z} \approx -2 t_{2} \sum\limits_{n} \mathscr{w}_{n}^{S} \left[ 1+ \frac{S\mathscr{g}_{n}}{2} \cos 2 \vec{q}_{1}\vec{b}_{n} \right] \times\\
    & \left\{- \sin S\vec{q}_{2} \vec{b}_{n} \sin \vec{k}\vec{b}_{n}  + 2S \cos S\vec{q}_{2}\vec{b}_{n} \sum\limits_{\vec{k}'}\mathscr{p}(\vec{k}') \sin \left( \vec{k} + \vec{k}' \right) \vec{b}_{n} \right\}.
     \end{aligned}
\end{equation}
The only difference between Eq. (\ref{eq:kerner-ham-k-space-large-s-app}) with Eq. (\ref{eq:kerner-ham-k-space}) is the power of $\mathscr{w}_{n}$ and $\mathscr{w}_{n}'$, and the pre-factor of $S$ in the term containing $\mathscr{g}_{n}$ and $\mathscr{g}_{n}'$. We should mention that, even if we have kept full expansion given in Eq. (\ref{eq:bin-thm}) still Eq. (\ref{eq:kerner-ham-k-space-large-s-app}) would have the analogous form.

\section{Hamiltonians for half-integer spins}
\label{sec:hamilt-half-integ}

\subsection{Hamiltonian For spin $S=1/2$}
\label{sec:hamilt-spin-s=12}

\begin{figure}[tbh]
  \centering
  \includegraphics[width=0.35\textwidth]{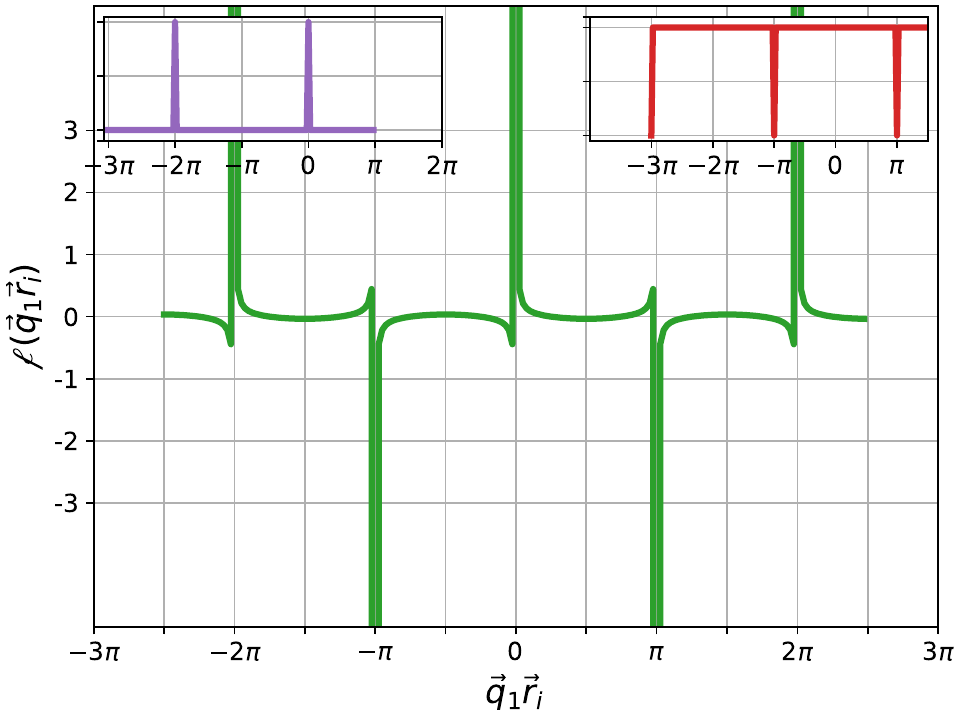}
  \caption{Plot of the function $\mathscr{f}(\vec{q}_{1}\vec{r}_{i})$ given in Eq. (\ref{eq:fourier-half-gn-eq-1}). It is applicable in the limit $g_{n} \approx 1$. (inset) The function $\mathscr{f}(\vec{q}_{1}\vec{r}_{i})$ can be approximated by combining two dirac combs shifted by a phase $\pi$ w.r.t each other.}
  \label{fig:fk-dash}
\end{figure}
We first find the Hamiltonian, Eq. (\ref{eq:kerner-ham}), for $S=1/2$. As there is no Fourier transform for the functions
\begin{equation}
  \label{eq:fourier-half-int-app}
  \begin{aligned}
    &\hat{\mathcal{F}}\left[ 1- \mathscr{g}_{n}' \cos 2 \vec{q}_{1}\left( \vec{r}_{i} + \frac{\vec{a}_{n}}{2}\right) \right]^{1/2}\\
    &\hat{\mathcal{F}}\left[ 1+ \mathscr{g}_{n} \cos 2 \vec{q}_{1}\left( \vec{r}_{i} + \frac{\vec{b}_{n}}{2}\right) \right]^{1/2},
  \end{aligned}
\end{equation}
we will find the Fourier transform for the limiting case of $\mathscr{g}_{n}$ and $\mathscr{g}_{n}'$. When $\mathscr{g}_{n} \ll 1$ one can approximate:
\begin{equation}
  \label{eq:fourier-half-gn-ll-1}
  \begin{aligned}
    &\hat{\mathcal{F}}\left[ 1+ \mathscr{g}_{n} \cos 2 \vec{q}_{1}\left( \vec{r}_{i} + \frac{\vec{b}_{n}}{2}\right) \right]^{1/2}\\
    &\approx \hat{\mathcal{F}}\left[ 1+ \frac{\mathscr{g}_{n}}{2} \cos 2 \vec{q}_{1}\left( \vec{r}_{i} + \frac{\vec{b}_{n}}{2}\right) \right].
  \end{aligned}
\end{equation}
When $\mathscr{g}_{n}' \ll 1$ one can approximate:
\begin{equation}
  \label{eq:fourier-half-gn-dash-ll-1}
  \begin{aligned}
    &\hat{\mathcal{F}}\left[ 1 - \mathscr{g}_{n}' \cos 2 \vec{q}_{1}\left( \vec{r}_{i} + \frac{\vec{a}_{n}}{2}\right) \right]^{1/2}\\
    &\approx \hat{\mathcal{F}}\left[ 1 - \frac{\mathscr{g}_{n}'}{2} \cos 2 \vec{q}_{1}\left( \vec{r}_{i} + \frac{\vec{a}_{n}}{2}\right) \right].
  \end{aligned}
\end{equation}
Hence, when $\mathscr{g}_{n} \ll 1$ and $\mathscr{g}_{n}' \ll 1$ the complete Hamiltonian can be approximated as:
\small
\begin{equation}
  \label{eq:ham-half-int-app-g-l1-g-dash-l1}
  \begin{aligned}
    &\mathcal{H}_{0} \approx -2 t_{2} \sum\limits_{n} \mathscr{w}_{n} \left[ 1+ \frac{\mathscr{g}_{n}}{4} \cos 2 \vec{q}_{1} \vec{b}_{n} \right] \times \\
    & \left\{\cos \frac{\vec{q}_{2} \vec{b}_{n}}{2} \cos \vec{k}\vec{b}_{n}  + \sin \frac{\vec{q}_{2}\vec{b}_{n}}{2} \sum\limits_{\vec{k}'}\mathscr{p}(\vec{k}') \cos \left( \vec{k} + \vec{k}' \right) \vec{b}_{n} \right\}\\
    &\mathcal{H}_{x} \approx +t_{1} \sum\limits_{n} \mathscr{w}_{n}' \left[ 1 - \frac{\mathscr{g}_{n}'}{4} \cos 2 \vec{q}_{1} \vec{a}_{n} \right] \times \cos \vec{k} \vec{a}_{n}\\
    &\mathcal{H}_{y} \approx +t_{1} \sum\limits_{n} \mathscr{w}_{n}' \left[ 1 - \frac{\mathscr{g}_{n}'}{4} \cos 2 \vec{q}_{1}\vec{a}_{n}\right] \times \sin \vec{k} \vec{a}_{n}\\
    &\mathcal{H}_{z} \approx -2 t_{2} \sum\limits_{n} \mathscr{w}_{n} \left[ 1+ \frac{\mathscr{g}_{n}}{4} \cos 2 \vec{q}_{1}\vec{b}_{n} \right] \times\\
    & \left\{- \sin \frac{\vec{q}_{2} \vec{b}_{n}}{2} \sin \vec{k}\vec{b}_{n}  +  \cos \frac{\vec{q}_{2}\vec{b}_{n} }{2} \sum\limits_{\vec{k}'}\mathscr{p}(\vec{k}') \sin \left( \vec{k} + \vec{k}' \right) \vec{b}_{n} \right\}.
     \end{aligned}
\end{equation}
\normalsize

When $\mathscr{g}_{n} \approx 1$ one first need to approximate $\mathscr{g}_{n} = 1-\gamma_{n}$; where $\gamma_{n}$ is a small quantity. Further substituting this in Eq. (\ref{eq:fourier-half-int-app}) we will get:
\small
\begin{equation}
  \label{eq:fourier-half-gn-eq-1}
  \begin{aligned}
    &\hat{\mathcal{F}}\left[ 1+ \mathscr{g}_{n} \cos 2 \vec{q}_{1}\left( \vec{r}_{i} + \frac{\vec{b}_{n}}{2}\right) \right]^{1/2}\\
    &\approx \hat{\mathcal{F}}\left[ 1 - \frac{\gamma_{n}}{2 \sqrt{2}} \mathscr{f}(\vec{q}_{1}\vec{r}_{i}) \right]
      =\hat{\mathcal{F}}\left[ 1 - \frac{1-\mathscr{g}_{n}}{2 \sqrt{2}} \mathscr{f}(\vec{q}_{1}\vec{r}_{i}) \right];\\
    & \mathscr{f}(\vec{q}_{1}\vec{r}_{i}) \equiv \frac{\cos 2\vec{q}_{1} \left( \vec{r}_{i} + \frac{\vec{b}_{n}}{2}\right)}{\cos \vec{q}_{1} \left( \vec{r}_{i} + \frac{\vec{b}_{n}}{2}\right)}.
  \end{aligned}
\end{equation}
\normalsize
Eq. (\ref{eq:fourier-half-gn-eq-1}) is applicable only when the condition $1+\cos 2\vec{q}_{1} \left( \vec{r}_{i} + \frac{\vec{b}_{n}}{2}\right) \gg \gamma$ is satisfied. Besides, the function $\mathscr{f}(\vec{q}_{1}\vec{r}_{i})$ is defined everywhere except in the vicinity of $\vec{q}_{1}\vec{r}_{i} \approx \pi/2-(\vec{q}_{1}\vec{b}_{n}/2)$. In Fig. \ref{fig:fk-dash} we plot the function $\mathscr{f}(\vec{q}_{1}\vec{r}_{i})$ for $\gamma=0.1$; with increase in $\gamma$ the shape of the function does not change much. In fact one can approximate this function using Dirac comb, i.e. we can write the function as summation of the two opposite valued Dirac combs shifted by $\pi$ compared to each other:
\begin{equation}
  \label{eq:dirac-comb}
   \mathscr{f}(\vec{q}_{1}\vec{r}_{i}) \approx \sum\limits_{n} \delta(\vec{q}_{1}\vec{r}_{i}-2n\pi) - \sum\limits_{n} \delta(\vec{q}_{1}\vec{r}_{i}-(2n+1)\pi)
\end{equation}
Here, $\delta$ is the Dirac delta function. Hence, explicitly we will have:
\small
\begin{equation}
  \label{eq:f_k_2}
  \begin{aligned}
    &\hat{\mathcal{F}}\left[ 1 - \frac{\gamma}{2 \sqrt{2}} \mathscr{f}(\vec{q}_{1}\vec{r}_{i}) \right]\\
    & = \hat{\mathcal{F}}\left[ 1 - \frac{\gamma}{2 \sqrt{2}} \left\{ \sum\limits_{n} \delta(\vec{q}_{1}\vec{r}_{i}-2n\pi) - \sum\limits_{n} \delta(\vec{q}_{1}\vec{r}_{i}-(2n+1)\pi)  \right\} \right]
  \end{aligned}
\end{equation}
\normalsize
The fourier transform of Dirac comb is known. The Fourier transform of the first part of the Eq. (\ref{eq:f_k_2}):
\begin{equation}
  \label{eq:four-dirac-comb}
  \mathcal{F} \left[ \sum\limits_{n} \delta(\vec{q}_{1}\vec{r}_{i}-2n\pi) \right]= \sum\limits_{n} 2 \cos \left( 2n\pi \right)
\end{equation}
Similarly we can write the Fourier transform of the second part of the Eq. (\ref{eq:f_k_2}). When $\mathscr{g}_{n}' \approx 1$ we proceed the same way; defining  $\mathscr{g}_{n}' = 1-\gamma_{n}'$ we will have:
\small
\begin{equation}
  \label{eq:fourier-half-gn-dash-eq-2}
  \begin{aligned}
    &\hat{\mathcal{F}}\left[ 1- \mathscr{g}_{n}' \cos 2 \vec{q}_{1}\left( \vec{r}_{i} + \frac{\vec{a}_{n}}{2}\right) \right]^{1/2}\\
    &\approx \hat{\mathcal{F}}\left[ 1 + \frac{\gamma_{n}'}{2 \sqrt{2}} \mathscr{f}(\vec{q}_{1}\vec{r}_{i}) \right]
      =\hat{\mathcal{F}}\left[ 1 + \frac{1-\mathscr{g}_{n}'}{2 \sqrt{2}} \mathscr{f}'(\vec{q}_{1}\vec{r}_{i}) \right];\\
    & \mathscr{f}'(\vec{q}_{1}\vec{r}_{i}) \equiv \frac{\cos 2\vec{q}_{1} \left( \vec{r}_{i} + \frac{\vec{a}_{n}}{2}\right)}{\sin \vec{q}_{1} \left( \vec{r}_{i} + \frac{\vec{a}_{n}}{2}\right)}.
  \end{aligned}
\end{equation}
\normalsize
As in Eq. (\ref{eq:fourier-half-gn-eq-1}), Eq. (\ref{eq:fourier-half-gn-dash-eq-2}) is applicable when the condition $1+\cos 2\vec{q}_{1} \left( \vec{r}_{i} - \frac{\vec{a}_{n}}{2}\right) \gg \gamma$ is satisfied. The function $\mathscr{f}'(\vec{q}_{1}\vec{r}_{i})$ is defined everywhere except in the vicinity of $\vec{q}_{1}\vec{r}_{i} \approx \pi/2-(\vec{q}_{1}\vec{a}_{n}/2)$. The Fourier transform of Eq. (\ref{eq:fourier-half-gn-dash-eq-2}) is found by the same way as in Eq. (\ref{eq:f_k_2}). Substituting the Fourier transforms of Eq. (\ref{eq:f_k_2}) and Eq. (\ref{eq:fourier-half-gn-dash-eq-2}) in Eq. (\ref{eq:kerner-ham}) we will get the Hamiltonian:
\small
\begin{equation}
  \label{eq:ham-half-int-app-g-1-g-dash-1}
  \begin{aligned}
    &\mathcal{H}_{0} = -2 t_{2} \sum\limits_{n} \mathscr{w}_{n} \left[ 1 - \frac{1-\mathscr{g}_{n}}{\sqrt{2}} \left\{ \sum\limits_{n} \cos (2 n \pi) - \sum\limits_{n} \cos (2 n \pi) \right\} \right] \times \\
    & \left\{\cos \frac{\vec{q}_{2} \vec{b}_{n}}{2} \cos \vec{k}\vec{b}_{n}  + \sin \frac{\vec{q}_{2}\vec{b}_{n}}{2} \sum\limits_{\vec{k}'}\mathscr{p}(\vec{k}') \cos \left( \vec{k} + \vec{k}' \right) \vec{b}_{n} \right\}\\
    &\mathcal{H}_{x} = +t_{1} \sum\limits_{n} \mathscr{w}_{n}' \left[ 1 + \frac{1-\mathscr{g}_{n}'}{\sqrt{2}}  \left\{ \sum\limits_{n} \cos (2 n \pi) - \sum\limits_{n} \cos (2 n \pi) \right\} \right] \times \cos \vec{k} \vec{a}_{n}\\
    &\mathcal{H}_{x} = +t_{1} \sum\limits_{n} \mathscr{w}_{n}' \left[ 1 + \frac{1-\mathscr{g}_{n}'}{\sqrt{2}}  \left\{ \sum\limits_{n} \cos (2 n \pi) - \sum\limits_{n} \cos (2 n \pi) \right\} \right] \times \sin \vec{k} \vec{a}_{n}\\    
    &\mathcal{H}_{z} =  -2 t_{2} \sum\limits_{n} \mathscr{w}_{n} \left[ 1 - \frac{1-\mathscr{g}_{n}}{\sqrt{2}} \left\{ \sum\limits_{n} \cos (2 n \pi) - \sum\limits_{n} \cos (2 n \pi) \right\} \right] \times \\
    & \left\{- \sin \frac{\vec{q}_{2} \vec{b}_{n}}{2} \sin \vec{k}\vec{b}_{n}  +  \cos \frac{\vec{q}_{2}\vec{b}_{n} }{2} \sum\limits_{\vec{k}'}\mathscr{p}(\vec{k}') \sin \left( \vec{k} + \vec{k}' \right) \vec{b}_{n} \right\}.
     \end{aligned}
\end{equation}
\normalsize
The Hamiltonian for $\mathscr{g}' \ll 1$ and $\mathscr{g} \approx 1$ is found by using above mentioned approximation. Specifically one will combine the $\mathcal{H}_{0}$ and $\mathcal{H}_{z}$ from Eq. (\ref{eq:ham-half-int-app-g-1-g-dash-1}) and $\mathcal{H}_{x}$ and $\mathcal{H}_{y}$ from Eq. (\ref{eq:ham-half-int-app-g-l1-g-dash-l1}). Explicitly:
\small
\begin{equation}
  \label{eq:ham-half-int-app-g-1-g-dash-l-1}
  \begin{aligned}
    &\mathcal{H}_{0} = -2 t_{2} \sum\limits_{n} \mathscr{w}_{n} \left[ 1 - \frac{1-\mathscr{g}_{n}}{\sqrt{2}} \left\{ \sum\limits_{n} \cos (2 n \pi) - \sum\limits_{n} \cos (2 n \pi) \right\} \right] \times \\
    & \left\{\cos \frac{\vec{q}_{2} \vec{b}_{n}}{2} \cos \vec{k}\vec{b}_{n}  + \sin \frac{\vec{q}_{2}\vec{b}_{n}}{2} \sum\limits_{\vec{k}'}\mathscr{p}(\vec{k}') \cos \left( \vec{k} + \vec{k}' \right) \vec{b}_{n} \right\}\\
    &\mathcal{H}_{x} = +t_{1} \sum\limits_{n} \mathscr{w}_{n}' \left[ 1 - \frac{\mathscr{g}_{n}'}{4} \cos 2 \vec{q}_{1} \vec{a}_{n} \right] \times \cos \vec{k} \vec{a}_{n}\\
    &\mathcal{H}_{y} = +t_{1} \sum\limits_{n} \mathscr{w}_{n}' \left[ 1 - \frac{\mathscr{g}_{n}'}{4} \cos 2 \vec{q}_{1}\vec{a}_{n}\right] \times \sin \vec{k} \vec{a}_{n}\\
    &\mathcal{H}_{z} =  -2 t_{2} \sum\limits_{n} \mathscr{w}_{n} \left[ 1 - \frac{1-\mathscr{g}_{n}}{\sqrt{2}} \left\{ \sum\limits_{n} \cos (2 n \pi) - \sum\limits_{n} \cos (2 n \pi) \right\} \right] \times \\
    & \left\{- \sin \frac{\vec{q}_{2} \vec{b}_{n}}{2} \sin \vec{k}\vec{b}_{n}  +  \cos \frac{\vec{q}_{2}\vec{b}_{n} }{2} \sum\limits_{\vec{k}'}\mathscr{p}(\vec{k}') \sin \left( \vec{k} + \vec{k}' \right) \vec{b}_{n} \right\}.
     \end{aligned}
\end{equation}
\normalsize
Similarly, for $\mathscr{g}' \ll 1$ and $\mathscr{g} \approx 1$ the Hamiltonian is found by combining the $\mathcal{H}_{x}$ and $\mathcal{H}_{y}$ from Eq. (\ref{eq:ham-half-int-app-g-1-g-dash-1}) and $\mathcal{H}_{0}$ and $\mathcal{H}_{z}$ from Eq. (\ref{eq:ham-half-int-app-g-l1-g-dash-l1}).
\small
\begin{equation}
  \label{eq:ham-half-int-app-g-l-1-g-dash-1}
  \begin{aligned}
    &\mathcal{H}_{0} = -2 t_{2} \sum\limits_{n} \mathscr{w}_{n} \left[ 1+ \frac{\mathscr{g}_{n}}{4} \cos 2 \vec{q}_{1} \vec{b}_{n} \right] \times \\
    & \left\{\cos \frac{\vec{q}_{2} \vec{b}_{n}}{2} \cos \vec{k}\vec{b}_{n}  + \sin \frac{\vec{q}_{2}\vec{b}_{n}}{2} \sum\limits_{\vec{k}'}\mathscr{p}(\vec{k}') \cos \left( \vec{k} + \vec{k}' \right) \vec{b}_{n} \right\}\\    
    &\mathcal{H}_{x} = +t_{1} \sum\limits_{n} \mathscr{w}_{n}' \left[ 1 + \frac{1-\mathscr{g}_{n}'}{\sqrt{2}}  \left\{ \sum\limits_{n} \cos (2 n \pi) - \sum\limits_{n} \cos (2 n \pi) \right\} \right] \times \cos \vec{k} \vec{a}_{n}\\
    &\mathcal{H}_{x} = +t_{1} \sum\limits_{n} \mathscr{w}_{n}' \left[ 1 + \frac{1-\mathscr{g}_{n}'}{\sqrt{2}}  \left\{ \sum\limits_{n} \cos (2 n \pi) - \sum\limits_{n} \cos (2 n \pi) \right\} \right] \times \sin \vec{k} \vec{a}_{n}\\    
    &\mathcal{H}_{z} = -2 t_{2} \sum\limits_{n} \mathscr{w}_{n} \left[ 1+ \frac{\mathscr{g}_{n}}{4} \cos 2 \vec{q}_{1}\vec{b}_{n} \right] \times\\
    & \left\{- \sin \frac{\vec{q}_{2} \vec{b}_{n}}{2} \sin \vec{k}\vec{b}_{n}  +  \cos \frac{\vec{q}_{2}\vec{b}_{n} }{2} \sum\limits_{\vec{k}'}\mathscr{p}(\vec{k}') \sin \left( \vec{k} + \vec{k}' \right) \vec{b}_{n} \right\}.
     \end{aligned}
\end{equation}
\normalsize
\subsection{Hamiltonian For spin $S=3/2, 5/2, \dots$}
\label{sec:hamilt-spin-s=32}
Hamiltonian for higher half-integer spin can be found by combining the aproximation made for $S=1/2$ and for $S=2,3,\dots$ in App. \ref{sec:hamilt-high-integ}. For example for $S=5/2$ one can write the term:
\begin{equation}
  \label{eq:g-high-half-int-app}
  \begin{aligned}
    &\left[ 1- \mathscr{g}_{n}' \cos 2 \vec{q}_{1}\left( \vec{r}_{i} + \frac{\vec{a}_{n}}{2}\right) \right]^{5/2}\\
    &\: = \left[ 1- \mathscr{g}_{n}' \cos 2 \vec{q}_{1}\left( \vec{r}_{i} + \frac{\vec{a}_{n}}{2}\right) \right]^{2}
      \left[ 1- \mathscr{g}_{n}' \cos 2 \vec{q}_{1}\left( \vec{r}_{i} + \frac{\vec{a}_{n}}{2}\right) \right]^{1/2}.
  \end{aligned}
\end{equation}
The integer power term (square) can be approximated as [see App. \ref{sec:hamilt-high-integ}]:
\begin{equation}
  \label{eq:g-high-half-int-app-pow-2}
  \begin{aligned}
    &\left[ 1- \mathscr{g}_{n}' \cos 2 \vec{q}_{1}\left( \vec{r}_{i} + \frac{\vec{a}_{n}}{2}\right) \right]^{2}
    \approx 1 - 2 \mathscr{g}_{n}'\cos 2 \vec{q}_{1}\left( \vec{r}_{i} + \frac{\vec{a}_{n}}{2}\right).
  \end{aligned}
\end{equation}
For $\mathscr{g}_{n}' \ll 1$ the square root term can be:
\begin{equation}
  \label{eq:g-high-half-int-app-pow-1-2}
  \begin{aligned}
    &\left[ 1- \mathscr{g}_{n}' \cos 2 \vec{q}_{1}\left( \vec{r}_{i} + \frac{\vec{a}_{n}}{2}\right) \right]^{1/2}
    \approx 1 - \frac{\mathscr{g}_{n}'}{2}\cos 2 \vec{q}_{1}\left( \vec{r}_{i} + \frac{\vec{a}_{n}}{2}\right).
  \end{aligned}
\end{equation}
Multiplying Eq. (\ref{eq:g-high-half-int-app-pow-2}) and (\ref{eq:g-high-half-int-app-pow-1-2}) and keeping the terms only first order in $\mathscr{g}_{n}'$ we will have:
\begin{equation}
  \label{eq:g-high-half-int-app-pow-tot}
  \begin{aligned}
    &\left[ 1- \mathscr{g}_{n}' \cos 2 \vec{q}_{1}\left( \vec{r}_{i} + \frac{\vec{a}_{n}}{2}\right) \right]^{5/2}\\
    &\approx 1 - \frac{5 \mathscr{g}_{n}'}{2}\cos 2 \vec{q}_{1}\left( \vec{r}_{i} + \frac{\vec{a}_{n}}{2}\right).
  \end{aligned}
\end{equation}
For arbitrary half-integer $S$ Eq. (\ref{eq:g-high-half-int-app-pow-tot}) can be written as:
\begin{equation}
  \label{eq:g-high-half-int-app-pow-S}
  \begin{aligned}
    &\left[ 1- \mathscr{g}_{n}' \cos 2 \vec{q}_{1}\left( \vec{r}_{i} + \frac{\vec{a}_{n}}{2}\right) \right]^{(S-1/2)/2 + 1/2}\\
    &\approx 1 - S \mathscr{g}_{n}' \cos 2 \vec{q}_{1}\left( \vec{r}_{i} + \frac{\vec{a}_{n}}{2}\right).
  \end{aligned}
\end{equation}
The same trick can be applied for finding the Hamiltonian for other cases of $\mathscr{g}_{n}$ and $\mathscr{g}_{n}'$. The total Hamiltonian will be analogous to the Eqs. (\ref{eq:ham-half-int-app-g-l1-g-dash-l1}), (\ref{eq:ham-half-int-app-g-1-g-dash-1}), (\ref{eq:ham-half-int-app-g-1-g-dash-l-1}), and (\ref{eq:ham-half-int-app-g-l-1-g-dash-1}).

\section{Chern number Calculations for Honeycomb bipartriate lattice}
\label{sec:chern-numb-calc}
\subsection{Spin $S=1$}
\label{sec:spin-s=1}
\begin{table}
  \label{app-chern-tab}
  \caption{Table showing dot product between lattice vectors and vectors $\vec{q}_{1}$, $\vec{q}_{2}$, and $\vec{K}$.}
\begin{ruledtabular}  
  \centering
  \begin{tabular}{lccc}
    Lat. Vect.    &$\vec{q}_{1}=\left( \frac{2q_{1x}}{\sqrt{3}},0 \right)$  &$\vec{q}_{2}=\left( \frac{2q_{2x}}{\sqrt{3}},0 \right)$  &$\vec{K}=\left( \pm\frac{\pi}{\sqrt{3}}, 0 \right)$ \\
    \hline
    $\vec{a}_{1}=\left( \frac{\sqrt{3}}{2}, \frac{1}{2} \right)$  &$q_{1x}$  &$q_{2x}$  &$\pm \frac{\pi}{2}$\\
    $\vec{a}_{2}=\left( \frac{-\sqrt{3}}{2}, \frac{1}{2} \right)$  &$-q_{1x}$  &$-q_{2x}$  &$\mp \frac{\pi}{2}$\\
    $\vec{a}_{3}=\left( 0,1 \right)$  &$0$  &$0$  &$0$\\
    $\vec{b}_{1}=\left( \frac{-\sqrt{3}}{2}, \frac{3}{2} \right)$  &$-q_{1x}$  &$-q_{2x}$  &$\mp \frac{\pi}{2}$\\
    $\vec{b}_{2}=\left( \frac{-\sqrt{3}}{2}, \frac{3}{2} \right)$  &$-q_{1x}$  &$-q_{2x}$  &$\mp \frac{\pi}{2}$\\
    $\vec{b}_{3}=\left( \sqrt{3},0 \right)$  &$2q_{1x}$  &$2q_{2x}$  &$\pm \pi$\\
  \end{tabular}
\end{ruledtabular}  
\end{table}
We first calculate the Chern number for $S=1$ using the Hamiltonian Eq. (\ref{eq:kerner-ham-k-space}). As mentioned before the Chern number is calculated at values of momentum $\pm\vec{K}$ when the condition $\mathcal{H}_{x}=\mathcal{H}_{y}=0$ and $\mathcal{H}_{z} \neq 0$ is satisfied simultaneously \cite[see Sec. 3.5.6 of Ref.][]{fruchart_2013_IntroductionTopological_ComptesRendusPhysique}. If we take $\vec{q}_{1}=\left(2 q_{1x}/\sqrt{3}, 0 \right)$ and $\vec{q}_{2}=\left( 2q_{2x}/\sqrt{3}, 0 \right)$, then at the point $\vec{K}=\left(\pm \pi/\sqrt{3},0 \right)$ the condition is satisfied identically. To see this, defining the three NN ($a_{n}$) and NNN ($b_{n}$) lattice vectors of the honeycomb lattice, in Eq. (\ref{eq:kerner-ham-k-space}) for spin $S=1$ we will get:
\begin{equation}
  \label{eq:app-chern-ham-hx}
  \begin{aligned}
    \mathcal{H}_{x}=&+t_{1} \mathscr{w}_{1}' \left[ 1 - \frac{\mathscr{g}_{1}'}{2} \cos 2\vec{q}_{1}\vec{a}_{1} \right] \times \cos \vec{k}\vec{a}_{1}\\
                    & \quad +t_{1} \mathscr{w}_{2}' \left[ 1 - \frac{\mathscr{g}_{2}'}{2} \cos 2\vec{q}_{1}\vec{a}_{2} \right] \times \cos \vec{k}\vec{a}_{2}\\
                    & \qquad +t_{1} \mathscr{w}_{3}' \left[ 1 - \frac{\mathscr{g}_{3}'}{2} \cos 2\vec{q}_{1}\vec{a}_{3} \right] \times \cos \vec{k}\vec{a}_{3}\\
  \end{aligned}
\end{equation}
\begin{equation}
  \label{eq:app-chern-ham-hy}
  \begin{aligned}
    \mathcal{H}_{y}=&+t_{1} \mathscr{w}_{1}' \left[ 1 - \frac{\mathscr{g}_{1}'}{2} \cos 2\vec{q}_{1}\vec{a}_{1} \right] \times \sin \vec{k}\vec{a}_{1}\\
                    & \quad +t_{1} \mathscr{w}_{2}' \left[ 1 - \frac{\mathscr{g}_{2}'}{2} \cos 2\vec{q}_{1}\vec{a}_{2} \right] \times \sin \vec{k}\vec{a}_{2}\\
                    & \qquad +t_{1} \mathscr{w}_{3}' \left[ 1 - \frac{\mathscr{g}_{3}'}{2} \cos 2\vec{q}_{1}\vec{a}_{3} \right] \times \sin \vec{k}\vec{a}_{3}\\
  \end{aligned}
\end{equation}
\begin{equation}
  \label{eq:app-chern-ham-hz}
  \begin{aligned}
    &\mathcal{H}_{z}=\\
    &-2t_{2} \mathscr{w}_{1} \left[ 1 + \frac{\mathscr{g}_{1}}{2} \cos 2\vec{q}_{1}\vec{b}_{1} \right] \times \\
    &\left\{- \sin \vec{q}_{2}\vec{b}_{1}\sin \vec{k}\vec{b}_{1} + 2 \cos \vec{q}_{2}\vec{b}_{1} \sum\limits_{k'} \mathscr{p}\left( k' \right) \sin \left( \vec{k} + \vec{k}' \right) \vec{b}_{1}\right\}\\
    &-2t_{2} \mathscr{w}_{2} \left[ 1 + \frac{\mathscr{g}_{2}}{2} \cos 2\vec{q}_{1}\vec{b}_{2} \right] \times \\
    &\left\{- \sin \vec{q}_{2}\vec{b}_{2}\sin \vec{k}\vec{b}_{2} + 2 \cos \vec{q}_{2}\vec{b}_{2} \sum\limits_{k'} \mathscr{p}\left( k' \right) \sin \left( \vec{k} + \vec{k}' \right) \vec{b}_{2}\right\}\\
    &-2t_{2} \mathscr{w}_{3} \left[ 1 + \frac{\mathscr{g}_{3}}{2} \cos 2\vec{q}_{1}\vec{b}_{3} \right] \times \\
    &\left\{- \sin \vec{q}_{2}\vec{b}_{3}\sin \vec{k}\vec{b}_{3} + 2 \cos \vec{q}_{2}\vec{b}_{3} \sum\limits_{k'} \mathscr{p}\left( k' \right) \sin \left( \vec{k} + \vec{k}' \right) \vec{b}_{3}\right\}.
  \end{aligned}
\end{equation}
In Eqs. (\ref{eq:app-chern-ham-hx}), (\ref{eq:app-chern-ham-hy}), and (\ref{eq:app-chern-ham-hz}) we explicitly wrote the all three components of the $\mathcal{H}_{x}$, $\mathcal{H}_{y}$ and $H_{z}$. We dropped the $\mathcal{H}_{0}$ component as it does not play any role in determining the Chern number. Further one need to substitute the values of $\vec{a}_{n}$, $\vec{b}_{n}$, $\vec{q}_{1}=\left(2 q_{1x}/\sqrt{3}, 0 \right)$, $\vec{q}_{2}=\left( 2q_{2x}/\sqrt{3}, 0 \right)$, and $\vec{K}=\left(\pm \pi/\sqrt{3},0 \right)$ in above equations. In Tab. \ref{app-chern-tab} we have given the dot product of these values. Substituting these in Eqs. (\ref{eq:app-chern-ham-hx})(\ref{eq:app-chern-ham-hy})(\ref{eq:app-chern-ham-hz}) we will get:
\begin{equation}
  \label{eq:app-chern-hx-sub}
  \mathcal{H}_{x}=0,
\end{equation}
\begin{equation}
  \label{eq:app-chern-hy-sub}
  \begin{aligned}
    &\mathcal{H}_{y}=\\&\pm t_{1} \mathscr{w}_{1}' \left[ 1 - \frac{\mathscr{g}_{1}'}{2} \cos 2\vec{q}_{1x}\right]
                     \mp t_{1} \mathscr{w}_{2}' \left[ 1 - \frac{\mathscr{g}_{2}'}{2} \cos 2\vec{q}_{1x}\right]=0,
  \end{aligned}  
\end{equation}
\begin{equation}
  \label{eq:app-chern-hz-sub}
  \begin{aligned}
    &\mathcal{H}_{z}=\\
    &-2t_{2} \mathscr{w}_{1} \left[ 1 + \frac{\mathscr{g}_{1}}{2} \cos 2q_{1x} \right] \times \\
    &\left\{\pm \sin \vec{q}_{2x} + 2 \cos q_{2x}\sum\limits_{k'} \mathscr{p}\left( k' \right) \sin \left( \pm \pi/2 + \vec{k}' \vec{b}_{1} \right) \right\}\\
    &-2t_{2} \mathscr{w}_{2} \left[ 1 + \frac{\mathscr{g}_{2}}{2} \cos 2q_{1x} \right] \times \\
    &\left\{\pm \sin \vec{q}_{2x} + 2 \cos q_{2x}\sum\limits_{k'} \mathscr{p}\left( k' \right) \sin \left( \pm \pi/2 + \vec{k}' \vec{b}_{2} \right) \right\}\\
    &-2t_{2} \mathscr{w}_{3} \left[ 1 + \frac{\mathscr{g}_{3}}{2} \cos 2q_{1x} \right] \times \\
    &\left\{+ 2 \cos q_{2x}\sum\limits_{k'} \mathscr{p}\left( k' \right) \sin \left( \pm \pi + \vec{k}' \vec{b}_{3} \right) \right\}\\    
  \end{aligned}
\end{equation}
We observe that all the terms of $\mathcal{H}_{x}$ are identically zero; the term involving $\vec{a}_{1}$ and $\vec{a}_{2}$ are zero as $\cos \vec{k}\vec{a}_{1} = \cos \pi/2 =0$ and $\cos \vec{k} \vec{a}_{2}= \cos \pi/2 =0$; the term involving $a_{3}$ is zero as $\mathscr{w}_{3}=0$. For $\mathcal{H}_{y}$ the summation of all three terms are zero; the term involving $\vec{a}_{3}$ is zero as $\sin \vec{k}a_{3}= \sin 0 = 0$; the terms involving $\vec{a}_{1}$ and $\vec{a}_{2}$ are opposite of each other, hence they cancel. The three terms of the $\mathcal{H}_{z}$ are not zero at $\vec{K}$.

\begin{figure}
  \centering
  \includegraphics[width=0.44\textwidth]{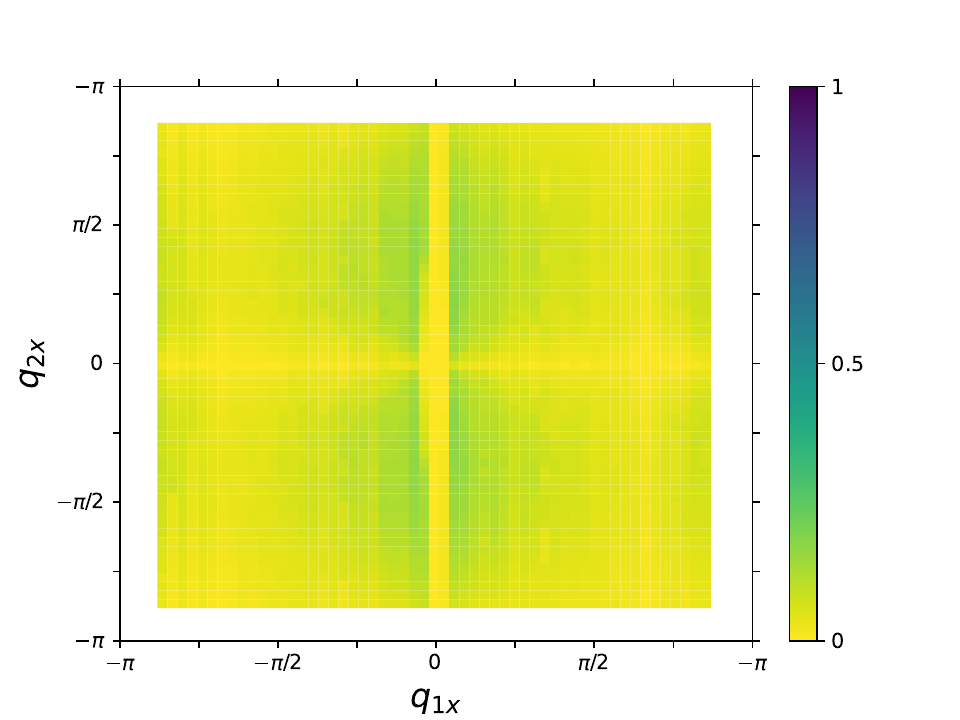}
  \caption{Dependence of $\sum\limits_{k'}\mathscr{p}(k') \sin \left( \pi/2 + k'b_{1} \right)$ on $\vec{q}_{1}= \left( q_{1x},0 \right)$ and $\vec{q}_{2}= \left( q_{2x},0 \right)$. In the figure only the $q_{1x}$ and $q_{2x}$ value changes from $-\sqrt{3}\pi/2$ to $\sqrt{3}\pi/2$. The same is applicable for $\vec{b}_{2}$.}
  \label{p_k_dash_sin}
\end{figure}
For $\mathcal{H}_{z}$ the term containing $\vec{b}_{3}$ one can observe that both $\mathscr{p}(k') \sim 0$ [see Fig. \ref{fig:p-k}] and the term $\sin \left( \pm \pi + \vec{k}' \vec{b}_{3} \right) \sim 0$; hence, we can neglect this term due to at least an order of smallness compared to other two terms. The terms containing $\vec{b}_{1}$ and $\vec{b}_{2}$ controls the Chern number. In these, usually the terms containing summation over $\vec{k}'$ are negligible for most values of $q_{1x}$ and $q_{2x}$; in Fig. \ref{p_k_dash_sin} we showed the summation $\sum\limits_{k'}\mathscr{p}(k') \sin \left( \pi/2 + k'b_{1} \right) \ll 1$ at $\vec{K}= \left(\pi/\sqrt{3},0 \right)$ for arbitrary values of the $q_{1x}$ and $q_{2x}$. It should be kept in mind that at $\vec{K}= \left(-\pi/\sqrt{3},0 \right)$ the summation $\sum\limits_{k'}\mathscr{p}(k') \sin \left( \pi/2 + k'b_{1} \right) \ll 1$ will be negative. Due to smallness of the summation, the term containing the same, won't have any effect on the Chern number, except at values where $\sin q_{2x} \lesssim 2 \left| \cos q_{2x}  \right| \sum\limits_{k'}\mathscr{p}(k') \sin \left( \pi/2 + k'b_{1} \right)$, which is in the vicinity $q_{2x} \sim 0$ and $q_{2x} \sim \pm \pi/ \sqrt{3}$. Away from these areas one can easily find the Chern number analytically. We will neglect the term containing the $\mathscr{p}(k')$.

The values of $\mathscr{g}_{n}'$ and $\mathscr{w}_{n}$ are always positive and less than unity. Hence, they also won't have any effect on Chern number. Therefore the Chern number will be defined by only the term $\sin \left( q_{2x} \right)$. The first Chern number is calculated by using the formula \cite[See Eq. (42) of Ref. ][]{fruchart_2013_IntroductionTopological_ComptesRendusPhysique}:
\begin{equation}
  \label{eq:chern-explanation}
  \begin{aligned}
    &c_{1}=\\
      &\frac{\text{sgn} \left\{ \mathcal{H}_{z} \left( \vec{K} =( \pi/\sqrt{3},0) \right) \right\}  - \text{sgn} \left\{ \mathcal{H}_{z} \left( \vec{K} =(-\pi/\sqrt{3},0) \right) \right\}}{2}.
  \end{aligned}
\end{equation}
Explicitly the Chern number will be:
\begin{equation}
  \label{eq:chern-number-app}
  \begin{aligned}
    &c_{1}= \text{sgn} \left[ \sin \left( q_{2x} \right) \right].
  \end{aligned}
\end{equation}

\subsection{Spin $S=2,3,\dots$}
\label{sec:spin-s=2-3}

The approximate Hamiltonian for higher spins is given in Eq. (\ref{eq:kerner-ham-k-space-large-s-app}). If one compares the Hamiltonian Eq. (\ref{eq:kerner-ham-k-space-large-s-app}) and (\ref{eq:kerner-ham-k-space}) they almost look identical, except for extra $S$ factors occurring in different places. Comparing $\mathcal{H}_{x}$ in Eq. (\ref{eq:kerner-ham-k-space-large-s-app}) and (\ref{eq:kerner-ham-k-space}) one will observe two changes: (i) $\mathscr{w}_{n}'$ has a power of $S$; (ii) $\mathscr{g}_{n}'$ has an extra coefficient $S$. However, the terms involving momentum $\vec{k}$ remains same. The similar is true for $\mathcal{H}_{y}$. For $\mathcal{H}_{z}$ one can observe five changes: (i) $\mathscr{w}_{n}$ has a power of $S$; (ii) $\mathscr{g}_{n}$ has an extra coefficient $S$; (iii) the angle of $\sin \vec{q}_{2}\vec{b}_{n}$ has now an $S$ coefficient: $\sin \left( S \vec{q}_{2}\vec{b}_{n} \right)$; (iv) the angle of $\cos \vec{q}_{2}\vec{b}_{n}$ has now an $S$ coefficient: $\cos \left( S \vec{q}_{2}\vec{b}_{n} \right)$; (v) the last term has a coefficient $2S$. However, as in $\mathcal{H}_{x}$ and $\mathcal{H}_{y}$ in $\mathcal{H}_{z}$ also the term containing momentum $\vec{k}$ remains same.

The topological properties is found from the same condition $\mathcal{H}_{x}=\mathcal{H}_{y}=0$ and $\mathcal{H}_{z} \neq 0$. The $\mathcal{H}_{x}$ and $\mathcal{H}_{z}$ are zero at $\vec{K}= \left( \pm \pi/\sqrt{3},0 \right)$, $\vec{q}_{1}=\left(2 q_{1x}/\sqrt{3}, 0 \right)$ and $\vec{q}_{2}=\left( 2q_{2x}/\sqrt{3}, 0 \right)$ due to the same reasons as explained for the $S=1$ case. For $\mathcal{H}_{z}$ the term containing $\vec{b}_{3}$ can be neglected due to two order of smallness as explained for $S=1$ case. Therefore the terms containing $\vec{b}_{1}$ and $\vec{b}_{2}$ which are non zero at $\vec{K}$ controls the Chern number. For $S=2$ case the Chern number can be described by Eq. (\ref{eq:chern-explanation}): $c_{1}= \text{sgn} \left[ \sin \left( 2q_{2x} \right) \right]$ for most part of the $q_{1x}$ and $q_{2x}$ phase space. However with increasing spin $S$ the expression will not work. Hence, a more general expression for Chern number is needed.

From Eq. (\ref{eq:kerner-ham-k-space-large-s-app}) the terms of $\mathcal{H}_{z}$ corresponding to $\vec{b}_{1}$ and $\vec{b}_{2}$ for arbitrary $S$  is:
\begin{equation}
  \label{eq:app-chern-ham-hz-s-large}
  \begin{aligned}
    &\mathcal{H}_{z} \approx \\
    &-2t_{2} \mathscr{w}_{1}^{S} \left[ 1 + \frac{S\mathscr{g}_{1}}{2} \cos 2\vec{q}_{1}\vec{b}_{1} \right] \times \\
    &\left\{- \sin S\vec{q}_{2}\vec{b}_{1}\sin \vec{k}\vec{b}_{1} + 2S \cos S\vec{q}_{2}\vec{b}_{1} \sum\limits_{k'} \mathscr{p}\left( k' \right) \sin \left( \vec{k} + \vec{k}' \right) \vec{b}_{1}\right\}\\
    &-2t_{2} \mathscr{w}_{2}^{S} \left[ 1 + \frac{S\mathscr{g}_{2}}{2} \cos 2\vec{q}_{1}\vec{b}_{2} \right] \times \\
    &\left\{- \sin S\vec{q}_{2}\vec{b}_{2}\sin \vec{k}\vec{b}_{2} + 2S \cos S\vec{q}_{2}\vec{b}_{2} \sum\limits_{k'} \mathscr{p}\left( k' \right) \sin \left( \vec{k} + \vec{k}' \right) \vec{b}_{2}\right\}.
  \end{aligned}
\end{equation}
Using $\vec{K}= \left( \pm \pi/\sqrt{3},0 \right)$, $\vec{q}_{1}=\left(2 q_{1x}/\sqrt{3}, 0 \right)$ and $\vec{q}_{2}=\left( 2q_{2x}/\sqrt{3}, 0 \right)$ and substituting the corresponding value from Tab. \ref{app-chern-tab} in Eq. (\ref{eq:app-chern-ham-hz-s-large}) we will get:
\begin{equation}
  \label{eq:app-chern-ham-hz-s-large-sub-val}
  \begin{aligned}
    &\mathcal{H}_{z} \approx \\
    &-2t_{2} \mathscr{w}_{1}^{S} \left[ 1 + \frac{S\mathscr{g}_{1}}{2} \cos 2q_{1x} \right] \times \\
    &\left\{- \sin Sq_{2x} + 2S \cos Sq_{2x}\sum\limits_{k'} \mathscr{p}\left( k' \right) \sin \left( \pi/2 + \vec{k}'\vec{b}_{1} \right) \right\}\\
    &-2t_{2} \mathscr{w}_{2}^{S} \left[ 1 + \frac{S\mathscr{g}_{2}}{2} \cos 2q_{1x}\right] \times \\
    &\left\{- \sin Sq_{2x} + 2S \cos Sq_{2x}\sum\limits_{k'} \mathscr{p}\left( k' \right) \sin \left( \pi/2 + \vec{k}' \vec{b}_{2} \right) \right\}.
  \end{aligned}
\end{equation}
Although the summation over $\mathscr{p}(k')$ is always small as shown in Fig. \ref{p_k_dash_sin}, however, the whole term, $2S \cos Sq_{2x}\sum\limits_{k'} \mathscr{p}\left( k' \right) \sin \left( \pi/2 + \vec{k}' \vec{b}_{1} \right)$, might not be small compared to the previous term $\sin Sq_{2x}$, due to the coefficient $2S$ present in it. Hence now the effect of term containing $\mathscr{p}(k')$ should also be included; for $S=1$ it was neglected due to smallness. For ease of analysis we approximate the whole term with a small term $\epsilon$ independent of $q_{1x}$ and $q_{2x}$:
\begin{equation}
   \label{eq:epsilong-defn}
    \epsilon \equiv \sum\limits_{k'} \mathscr{p}\left( k' \right) \sin \left( \pi/2 + \vec{k}' \vec{b}_{1}\right).
  \end{equation}
Explicitly we will write:
\begin{equation}
  \label{eq:epsilong-defn}
  \begin{aligned}
    &2S \cos Sq_{2x}\sum\limits_{k'} \mathscr{p}\left( k' \right) \sin \left( \pi/2 + \vec{k}' \vec{b}_{1}\right)
     \equiv 2S \epsilon \cos Sq_{2x}.
  \end{aligned}
\end{equation}
Value of $\epsilon$ is positve for $+\vec{K}$ and negative for $- \vec{K}$. The value of $\mathscr{w}_{1}$ is always positive and less than unity [see Fig. \ref{fig:w_w_dash}], hence they wont affect the Chern number. Similarly, $\mathscr{g}_{n}$ is always positive and less than unity. However, its coefficients: (i) $S$ may be large, (ii) $\cos 2q_{1x}$ can take positive and negative values. Hence, now the term $\left[ 1 + \frac{S\mathscr{g}_{1}}{2} \cos 2q_{1x} \right]$ can also affect the Chern number depending on $S$ and $q_{1x}$. All the above analysis is true for $\vec{b}_{2}$ also. The Chern number using Eq. (\ref{eq:chern-explanation}) will be:
\begin{equation}
  \label{eq:chern-large-S-app}
  \begin{aligned}
    c_{1} = \text{sign}\left[ \left( 1 + \frac{S\mathscr{g}_{2}}{2} \cos 2q_{1x}\right) \left( \sin Sq_{2x} - 2S\epsilon \cos Sq_{2x}  \right) \right]
  \end{aligned}
\end{equation}

\section{Neel type Skyrmion}
\label{sec:equiv-skyrm}
\begin{figure}[tbh]
  \centering
  \includegraphics[width=0.44\textwidth]{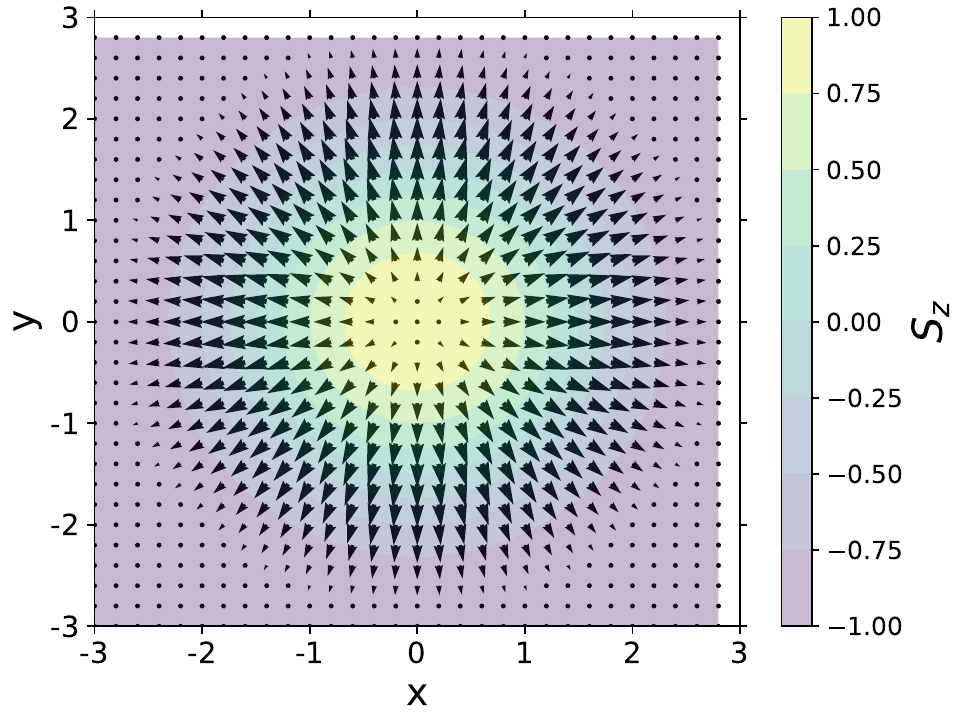}
  \caption{Generation of single Neel type Skyrmion from Eq. (\ref{eq:spin-config-int}).}
  \label{fig:skykrmion-plot}
\end{figure}
Interestingly one can easily show the equivalence of the magnetic Skyrmions (Skx) to the spin texture given in Eq. (\ref{eq:spin-config-int}) in specific limits. The magnetic Skx is represented as \cite[see Eq. (8) of Ref.][]{gobel_2021_SkyrmionsReview_PhysicsReports}:
\begin{equation}
  \label{eq:skyrmions}
  \vec{S}_{i,\text{Skx}} =
  \begin{bmatrix}
    S_{x}\\
    S_{y}\\
    S_{z}\\
  \end{bmatrix}
  = S
  \begin{bmatrix}
     \sin \left( \frac{\pi}{\left| r_{0} \right|} \left| \vec{r}_{i}  \right|  \right) \left( \frac{x_{i}}{\left| \vec{r}_{i} \right|} \cos \gamma  - \omega \frac{y_{i}}{\left| \vec{r}_{i} \right|} {\sin \gamma}\right)\\
     \sin \left( \frac{\pi}{\left| r_{0} \right|} \left| \vec{r}_{i}  \right|  \right)\left( \frac{x_{i}}{\left| \vec{r}_{i} \right|} \sin \gamma  + \omega \frac{y_{i}}{\left| \vec{r}_{i} \right|}{ \sin \gamma}\right)\\
    \rho \cos \left( \frac{\pi}{\left| r_{0}  \right|} \left| \vec{r}_{i}  \right|\right)
  \end{bmatrix}.
\end{equation}
Here, $\left| \vec{r}_{i}  \right|$ is the length of the position vector at $i$-th site; $x_{i}$ ($y_{i}$) is the $x$ ($y$) component of the position vector $\vec{r}_{i}$; $r_{0}$ is the maximum length of the Skx. $p$ is the polarity which represent the out of plane of the magnetization of the Skx; its value changes from $+1$ ($-1$) at the center to the $-1$ ($+1$) at the boundary. $\omega$ is the vorticity which represent rotation of the magnetization around the $z$-axis; it takes multiples of $2\pi$, $\omega=0,\pm 1, \pm 2, \dots$. $\gamma$ is the helicity which represent the rotation of the magnetization around the $i$-th site. If we substitute $\gamma=0$, $\omega=1$ and $\rho=1$ in Eq. (\ref{eq:skyrmions}); and substituting $x_{i}=\left| \vec{r}_{i} \right| \cos \phi$, $y_{i}=\left| \vec{r}_{i} \right| \sin \phi$ we will get:
\begin{equation}
  \label{eq:skyrmions-mod}
  \vec{S}_{i,\text{Skx}} =
  \begin{bmatrix}
    S_{x}\\
    S_{y}\\
    S_{z}\\
  \end{bmatrix}
  = S
  \begin{bmatrix}
    \sin \left( \frac{\pi}{\left| r_{0} \right|} \left| \vec{r}_{i}  \right|  \right)\cos \phi\\
    \sin \left( \frac{\pi}{\left| r_{0} \right|} \left| \vec{r}_{i}  \right|  \right) \sin \phi\\
     \cos \left( \frac{\pi}{\left| r_{0}  \right|} \left| \vec{r}_{i}  \right|\right)
  \end{bmatrix}.
\end{equation}
Physically it represents the Neel-type Skx \cite{gobel_2021_SkyrmionsReview_PhysicsReports}. In Eq. (\ref{eq:skyrmions-mod}) one needs to map $\phi \to q_{2} \atan2(r_{i})$ and $\frac{\pi}{\left| r_{0} \right|} \left| \vec{r}_{i} \right| \to q_{1} |\vec{r}_{i}|$ to see the similarity with Eq. (\ref{eq:spin-config-int}). In contrast to Eq. (\ref{eq:spin-config-int}) now the spin modulating vectors become scalars. Besides for a single Skx (not Skx lattice) the $q_{1}$ and $q_{2}$ can not be chosen arbitrary, they are fixed:
\begin{equation}
  \label{eq:q1-q2-skyrmions}
  q_{2} = 1, \quad  q_{1} = \frac{\pi}{r_{0}}.  
\end{equation}
In Fig. \ref{fig:skykrmion-plot} we plotted the Skx spin texture using Eq. (\ref{eq:skyrmions-mod}).


\providecommand{\noopsort}[1]{}

\end{document}